\documentclass[fleqn,usenatbib]{mnras}

\usepackage{newtxtext,newtxmath}

\usepackage[T1]{fontenc}
\usepackage{ae,aecompl}

\usepackage{booktabs}
\usepackage{subfigure}

\usepackage{graphicx}	
\usepackage{amsmath}	
\usepackage{amssymb}	
\usepackage[flushleft]{threeparttable}

\usepackage{color, colortbl}

\definecolor{LightCyan}{rgb}{0.88,1,1}
\definecolor{Gray}{gray}{0.9}

\newcommand{\hst}{\textit{HST}}
\newcommand{\vlt}{\textit{VLT}}
\newcommand{\muse}{MUSE}
\newcommand{\1}{\,{\sc i}}
\newcommand{\2}{\,{\sc ii}}
\newcommand{\3}{\,{\sc iii}}

\title[Candidate LBV stars in galaxy NGC~7793]{Candidate LBV stars in galaxy NGC~7793 found via HST photometry + MUSE spectroscopy}

\author[Aida Wofford et al.]
{\parbox{\textwidth}{Aida Wofford,$^{1}$\thanks{E-mail: awofford@astro.unam.mx}
Vanesa Ram{\'i}rez,$^{2}$
Janice C. Lee,$^{3}$
David A. Thilker,$^{4}$
Lorenza Della Bruna,$^{5}$
Angela Adamo,$^{5}$
Schuyler D. Van Dyk,$^{3}$
Artemio Herrero,$^{6,7}$
Hwihyun Kim,$^{8}$
Alessandra Aloisi,$^{9}$
Daniela Calzetti,$^{10}$
Rupali Chandar,$^{11}$
Daniel A. Dale,$^{12}$
Selma E. de Mink,$^{13,14}$
John S. Gallagher,$^{15}$
Dimitrios A. Gouliermis,$^{16,17}$
Kathryn Grasha,$^{18,19}$
Eva K. Grebel,$^{20}$
E. Sacchi,$^{9,21}$
Linda J. Smith,$^{9,22}$
Leonardo {\'U}beda,$^{9}$
Rene A. M. Walterbos,$^{23}$
Stephen Hannon,$^{3,24}$
Matteo Messa,$^{10}$}
\\\\
\parbox{\textwidth}{$^{1}$Instituto de Astronom\'ia, Universidad Nacional Aut\'onoma de M\'exico, Unidad Acad\'emica en Ensenada, Km 103 Carr. Tijuana$-$Ensenada, Ensenada 22860, M\'exico\\
$^{2}$Instituto de F\'isica - FCEN, Universidad de Antioquia Calle 70 No. 52-21, Medell\'in, Colombia\\
$^{3}$Caltech/IPAC, Pasadena, CA 91125\\
$^{4}$Dept. of Physics and Astronomy, The Johns Hopkins University, Baltimore, MD\\
$^{5}$Dept. of Astronomy, The Oskar Klein Centre, Stockholm University, Stockholm, Sweden\\
$^{6}$Instituto de Astrof\'isica de Canarias, La Laguna, Tenerife, Spain\\
$^{7}$Departamento de Astrof\'isica, Universidad de La Laguna, Tenerife, Spain\\
$^{8}$Gemini Observatory, Casilla 603, La Serena, Chile\\
$^{9}$Space Telescope Science Institute, 3700 San Martin Drive, Baltimore, MD, USA\\
$^{10}$Dept. of Astronomy, University of Massachusetts -- Amherst, Amherst, MA 01003, USA\\
$^{11}$Dept. of Physics and Astronomy, University of Toledo, Toledo, OH, USA\\
$^{12}$Dept. of Physics and Astronomy, University of Wyoming, Laramie, WY, USA\\
$^{13}$Anton Pannenkoek Institute for Astronomy, University of Amsterdam, NL-1090 GE Amsterdam, the Netherlands 3GRAPPA, University of 
$^{14}$Amsterdam, Science Park 904, 1098 XH Amsterdam, The Netherlands
$^{15}$Dept. of Astronomy, University of Wisconsin--Madison, Madison, WI, USA\\
$^{16}$Zentrum f\"ur Astronomie der Universit\"at Heidelberg, Institut f\"ur Theoretische Astrophysik, Albert-Ueberle-Str.\,2, 69120 Heidelberg, Germany\\
$^{17}$Max Planck Institute for Astronomy,  K\"{o}nigstuhl\,17, 69117 Heidelberg, Germany\\
$^{18}$Research School of Astronomy and Astrophysics, Australian National University, Canberra, ACT 2611, Australia\\
$^{19}$ARC Centre of Excellence for All Sky Astrophysics in 3 Dimensions (ASTRO 3D), Australia\\
$^{20}$Astronomisches Rechen-Institut, Zentrum f\"ur Astronomie der
Universit\"at Heidelberg, M\"onchhofstr.\ 12--14, 69120 Heidelberg, Germany\\
$^{21}$INAF -- OAS Osservatorio di Astrofisica e Scienza dello Spazio, Bologna, Italy\\
$^{22}$European Space Agency\\
$^{23}$Dept. of Astronomy, New Mexico State University, Las Cruces, NM, USA\\
$^{24}$Dept. of Physics \& Astronomy, University of California, CA, USA
}}
\date{Accepted XXX. Received YYY; in original form ZZZ}

\pubyear{2015}

\begin{document}
\label{firstpage}
\pagerange{\pageref{firstpage}--\pageref{lastpage}}
\maketitle

\clearpage

\begin{abstract}
Only about 19 Galactic and 25 extra-galactic bona-fide Luminous Blue Variables (LBVs) are known to date. This incomplete census prevents our understanding of this crucial phase of massive star evolution which leads to the formation of heavy binary black holes via the classical channel. With large samples of LBVs one could better determine the duration and maximum stellar luminosity which characterize this phase. We search for candidate LBVs (cLBVs) in a new galaxy, NGC 7793. For this purpose, we combine high spatial resolution images from two {\it Hubble Space Telescope (HST)} programs with optical spectroscopy from the Multi Unit Spectroscopic Explorer (MUSE). By combining PSF-fitting photometry measured on F547M, F657N, and F814W images, with restrictions on point-like appearance (at HST resolution) and H$\alpha$ luminosity, we find 100 potential cLBVs, 36 of which fall in the MUSE fields. Five of the latter 36 sources are promising cLBVs which have $M_{\rm{V}}\leq-7$ and a combination of: H$\alpha$ with a P-Cygni profile; no [O\1]$\,\lambda6300$ emission; weak or no [O\3]$\,\lambda5007$ emission; large [N\2]/H$\alpha$ relative to H\2 regions; and [S\2]$\,\lambda6716$/[S\2]$\,\lambda6731\sim1$. It is not clear if these five cLBVs are isolated from O-type stars, which would favor the binary formation scenario of LBVs.  Our study, which approximately covers one fourth of the optical disc of NGC 7793, demonstrates how by combining the above \hst~surveys with multi-object spectroscopy from 8-m class telescopes, one can efficiently find large samples of cLBVs in nearby galaxies. 

\end{abstract}




\section{Introduction}

Luminous Blue Variables (LBVs), also known as S Doradus variables\footnote{S Doradus is one of the brightest stars in the LMC and one of the most luminous known. Its absolute V-band magnitude ranges from -7.6 (1969) to -10 (1989, \citealt{vanGenderen2001}). Stars with S Dor like variability are called S Doradus variables.}, constitute a rare and poorly-understood phase in the lives of massive stars. This phase is characterized by episodic mass eruptions, very high luminosities\footnote{Log($L/L_\odot$)$\sim$\,5.4 to $\sim$\,6.4 \citep{Vink2012}; extinction-corrected V-band absolute magnitude, $M_{\rm{V}}\leq-7$ mag \citep{Humphreys2014}}, the highest mass loss rates among all observed stars\footnote{For example, for Eta Carinae, $\dot{M}\sim10^{-4}\,M_\odot\,$yr$^{-1}$ during the quiescent phase and $\dot{M}\sim10^{-2}\,M_\odot\,$yr$^{-1}$ during the eruptive phase; \citep{Humphreys1994}.}, and significant photometric and spectroscopic variability. In the Hertzsprung-Russell (H-R) diagram, LBVs are located between the Main Sequence (MS) and the Humphreys-Davidson (H-D) luminosity limit above which only highly unstable objects are found. The lack of Red Supergiant stars of similar luminosity indicates that LBVs cannot evolve to the red, and if they try to, they probably become unstable. Only a limited population of LBVs in outburst phase have been found to the right of the H-D limit. 

In the Milky Way and the Magellanic Clouds, LBVs are thought to be responsible for some extra-galactic non-SN transients \citep{VanDyk2012,Humphreys2017b}. Furthermore, the episodic mass loss of LBVs has been a reference point for interpreting the dense circumstellar material around Type IIn supernovae (SNe, \citealt{Smith2017}). 

Open questions about LBVs concern the total amount of mass lost in a typical eruption and how it scales with initial mass and metallicity, the duty cycle and number of repeated LBV giant eruptions, and whether all stars or only a special subset suffer the LBV giant eruptions \citep{Herrero2010, Smith2017}. 

In the classical ``Conti scenario" \citep{Conti1975}, which assumes single-star evolution, LBVs are a short phase ($10^4-10^5$\,yr, \citealt{Herrero2010}) in the lives of the most massive stars (M$_{\rm{ZAMS}} \ge40\,M_\odot$), which occurs immediately after H-core exhaustion during the transition to He-core burning. In this scenario, it is the mass loss which occurs during the LBV phase which removes the H envelope of the star, transforming the star into an H-deficient Wolf-Rayet star. In this scenario, stars evolve on times scales of $\le3-4$ Myr\footnote{This is the MS lifetime of an M$_{\rm{ZAMS}} \ge40\,M_\odot$ star} and consequently do not have time to move far from their birthplaces. Thus, one would expect to find LBVs near young O-type MS stars, which observationally, are stars that  appear to be clustered \citep{Smith2019}. However, LMC observations show that LBVs are more spatially dispersed than O-type MS stars, which has led \cite{Smith2015} and \cite{Smith2019} to argue that LBVs are likely the products of close binary evolution. This is because the larger spatial dispersion of LBVs can be explained by either SN kicks which move the LBVs far from their birthplace, or rejuvenation via mass transfer and mergers, which produce older massive stars at a time when the single O-type stars have disappeared. In the binary scenario, LBVs would be massive evolved blue stragglers, which in the H-R diagrams of star clusters are MS stars which are more luminous and bluer than stars at the MS turnoff point of the cluster. Note that \cite{Davidson2016} dispute the results of \cite{Smith2015} and argue for the standard view of LBVs.

Not understanding LBV mass loss imprints large uncertainties in massive-star evolution models. In particular, the type of supernova explosion that the star will undergo and the final mass of the remnant black hole or neutron star (if any) cannot be predicted. In fact, the mass loss during the LBV phase is one of the primary uncertainties in the classical formation channel for heavy binary black holes (e.g. \citealt{Belczynski2016}).

The lack of a complete LBV census prevents an accurate determination of the length of the LBV phase and the properties of stars in this unstable regime. Although cLBVs can be identified relatively quickly on the basis of their spectrum or luminosity, the identification of LBVs requires confirmation of the characteristic spectral and photometric variations. LBVs can be ``quiescent'' for decades or centuries during which they are indistinguishable from many other hot luminous stars. A key to understanding the peculiar instability of LBVs is their high observed luminosities, which often depend on uncertain distances. Much effort has been expended to identify new LBVs, with IR observations playing an important role in this regard. In particular, candidates have been identified via NIR spectroscopy and/or MIR imaging of circumstellar ejection nebulae \citep[and references therein]{Clark2005}.  

Table~\ref{tab:known_lbvs} summarizes the number of known Galactic and extra-galactic LBVs and cLBVs. The galaxies in which they are found sample a wide range of ionized-gas O/H ratios, i.e., 0.03 to 1.7 times the solar value. Trends between LBV properties and galaxy properties are difficult to investigate because i) bona-fide LBVs are incompletely characterized and ii) it is difficult to obtain multi-epoch spectroscopy for a sample of galaxies with a wide range of properties, at the spatial resolution which is necessary for such study.


\begin{table*}
\centering
\caption{\textbf{Census of LBVs or cLBVs in order of increasing distance.} }
\label{tab:known_lbvs}
\begin{tabular}{lllllllll}
\hline
{Galaxy} & \multicolumn{3}{c}{Number} & \multicolumn{2}{c}{Distance}  & \multicolumn{2}{c}{(O/H)/(O/H)$_\odot$} & {Ref$^{\rm d}$} \\
\cmidrule(lr){2-4}
\cmidrule(lr){5-6}
\cmidrule(lr){7-8}
\hfill & LBVs & cLBVs & {Ref$^{\rm a}$} & {(Mpc)} & {Ref$^{\rm b}$} & Value & {Ref$^{\rm c}$} & \hfill \\
\hline
Milky Way & 19 & 42 & (1) & - & - & 1 & (1) & (1) \\
LMC  & 8 & 19 & (1) & 0.04  & (1) & (0.4)-0.5 & (2,3) &  (2) \\
SMC & 2 & 2 & (1) & 0.04 & (1) & 0.23-(0.25) & (2,3) &  (3) \\
NGC 6822  & 0  & 1 & (1) & 0.22-0.46   & (2,3) & 0.28 & (3) &  (4)\\
IC 1613  & 0 & 1  & (1) & 0.65   & (3) & 0.12 & (3) &  (5) \\
IC 10  & 0 & 3 & (1) & 0.74   & (4) & 0.37 & (4) &  (4)\\
M31  & 6 & 10-25 & (1, 6) & 0.74   & (5) & 1.05-1.7 & (5) &  (4, 6)\\
M33  & 5 & 28 & (1) & 0.9   & (6) & 0.47-0.79 & (6) &  (4, 6)\\
NGC~55  & 0 & 4 & (2) & 1.99   & (4) & 0.27-0.30 & (7) & (2) \\
NGC~2403 & 2 & 1 & (3) & 2.5   & (4) & 0.61 & (8) & (3) \\
NGC 2366 & 1 & 0 & (1) & 3.34   & (4) & 0.1 & (9) &  (7) \\
M81 & 0 & 2 & (3) & 3.55   & (3) & 0.27-2.5 & (10) &  (8) \\
M101 & 0 & 11 & (4) & 6.4   & (3) & 0.7-1.12 & (11) & (4)  \\
DDO 68 & 1 & 0 & (5) & 12.75   & (7) & 0.028 & (12) & (9) \\
 Total & 44 & 124 & & & & & & \\
 \hline
\end{tabular}
\begin{tablenotes}
      \item $^{\rm a}$Reference for the number of LBVs/cLBVs. Note that the confirmed LBVs are removed from the cLBV column. (1) \cite{Richardson2018}. (2) \cite{Castro2008}. (3) \cite{Humphreys2019}. (4) \cite{Grammer2015}. (5) \cite{Pustilnik2017}. (6) \cite{King1998}.
      \item $^{\rm b}$Reference for the distance to the galaxy. (1) \cite{Groenewegen2013}. (2) \cite{Rich2014}. (3) \cite{Freedman2001}. (4) \cite{Tully2013}. (5) \cite{Wagner-Kaiser2015}. (6) \cite{Pellerin2011}. 
      \item $^{\rm c}$References for ionized-gas oxygen abundances, except for the Milky Way, which gives the ref. for the Solar photosphere value, which is 12+log(O/H)=8.69. (1) \cite{Asplund2009}. (2) LMC/30 Doradus and SMC/N88A regions, \cite{Garnett1995} (values in parenthesis). (3) CEL values, \cite{Dopita2019}. (4) IC 10, \cite{Magrini2009, Polles2019}. (5) Central value of M31 (abundance gradient measured from 3.9 to 16.1 kpc of the centre: $-0.023\pm0.002$ dex/kpc), \cite{Zurita2012}. (6) Inner 2 kpc of M33, \cite{Bresolin2011}. (7) Range of values obtained with different methods in NGC 55 \cite{Magrini2017}. (8) Central value of NGC 2403 (abundance gradient: ($-0.032\pm0.007\times\,$Rg (dex kpc/1)), \cite{Berg2013}. (9) NGC 2366/Mrk 71\cite{Izotov2011}. (10) Range given for M81 by \cite{Garnett1987}. Abundance gradient: $-0.088\pm0.013$ dex kpc/1 \citep{Stanghellini2014}. (11) Central region of M101, \cite{Bresolin2007}. (12) \cite{Pustilnik2017}.
      \item $^{d}$References with examples of photometry and/or spectra of LBVs/cLBVs. (1) \cite{Wolf1982, Miroshnichenko2014}. (2) \cite{Walborn2017}. (3) \cite{Koenigsberger2010}. (4) \cite{Massey2007}. (5) \cite{Herrero2010}. (6) \cite{Humphreys2014}. (7) \cite{Drissen2001}. (8) \cite{Humphreys2019}. (9) \cite{Pustilnik2017}.
\end{tablenotes}
\end{table*}

 Characterizing large samples of LBVs in different environments can provide valuable clues about their mass loss and evolution. Two \hst~programs, LEGUS (Legacy ExtraGalactic UV Survey, PID 13364, \citealt{Calzetti2015}) and H$\alpha$  LEGUS (PID 13773, PI Chandar, Chandar et al., in prep.) offer a unique opportunity to search for cLBVs in new galaxies. 

LEGUS is a Cycle 21 \hst~treasury program which obtained high spatial resolution ($\sim0.07"$) images of portions of 50 nearby ($\le16$ Mpc) galaxies, using the UVIS channel of the Wide Field Camera Three (WFC3), and broad band filters F275W (2704 \AA), F336W (3355 \AA), F438W (4325 \AA), F555W (5308 \AA), and F814W (8024 \AA), which roughly correspond to the photometric bands NUV, U, B, V, and I, respectively. The survey includes galaxies of different morphological types and spans a factor of $\sim10^3$ in both star formation rate (SFR) and specific star formation rate (sSFR), $\sim10^4$ in stellar mass ($\sim10^7-10^{11}\,\rmn{M_\odot}$), and $\sim10^2$ in oxygen abundance ($12+\rmn{log\,O/H}=7.2-9.2$). Some of the targets in the survey have high quality archival images in bandpasses similar to those required by LEGUS, most of them from the Wide Field Channel of \hst's Advanced Camera for Surveys (ACS), and fewer of them, from ACS's High Resolution Channel (HRC). For the latter targets, LEGUS completed the five band coverage. The choice of filters was dictated by the desire to distinguish young massive bright stars from faint star clusters, to derive accurate star formation histories for the stars in the field from their CMDs, and to obtain extinction-corrected estimates of age and mass for the star clusters. Star and star-cluster catalogues have been released for the LEGUS sample and are described in \cite{Sabbi2018} and \cite{Adamo2017}, respectively. We note that the LEGUS data do not have the spatial resolution to visually resolve massive stars in close binary systems. 

H$\alpha$ LEGUS is a Cycle 22 program which obtained narrow-band, H$\alpha$ (F657N) and medium band, continuum (F547M) images for the 25 LEGUS galaxies with the highest star formation rates, using the WFC3. The corresponding H$\alpha$ observations reveal thousands of previously undetected H\2 regions, including those ionized by stellar clusters and ``single" massive stars. 

In this paper, we focus on NGC 7793, the nearest LEGUS galaxy of the Southern hemisphere. This galaxy was also observed by H$\alpha$ LEGUS. Furthermore, we focus on the LEGUS field NGC 7793W, for which Adamo et al. recently secured ESO \vlt~MUSE \citep{Bacon2010} spectroscopy with Adaptive Optics (AO) of two subfields (Della Bruna et al., subm.). We combine LEGUS and H$\alpha$ LEGUS photometry with MUSE spectroscopy to search for cLBVs in NGC 7793W. Candidate LBVs are defined by \cite{Smith2019} as stars with similar luminosities and spectra to LBVs, often with circumstellar shells that indicate a prior outburst, but which have not yet been observed photometrically to undergo LBV eruptions. 
 
 We start by using the \hst~photometry to create a catalogue of potential cLBVs (hereafter, photometric cLBVs). We then identify the photometric cLBVs which are located within the MUSE fields. Hereafter, we refer to the latter as MUSE cLBVs. Note that the latter are still potential cLBVs. This work focuses on the MUSE cLBVs. We proceed to check if the \hst~coordinates of the MUSE cLBVs are coincident with those of sources in other catalogues of the LEGUS collaboration, in particular:  O-type stars (Lee et al. in prep.), compact star clusters \citep{Hannon2019},  star clusters \citep{Adamo2017}, and H\2 regions (Della Bruna et al., subm.). After this, we determine if the cLBV spectra are contaminated with light from nearby star clusters or H\2 regions. Finally, we spectroscopically classify the MUSE cLBVs and identify the most promising cLBVs. 

In Section~\ref{sec:observations}, we describe the galaxy and its imaging and spectroscopic observations. In section~\ref{sec:photometric_selection}, we present our catalogue of photometric cLBVs.  In Section~\ref{sec:other_catalogs}, we check if our sources have matches in other LEGUS catalogues. In Section~\ref{sec:classification}, we spectroscopically classify the MUSE cLBVs. In Section~\ref{sec:discussion}, we discuss our results. Finally, we summarize and conclude in Section~\ref{sec:conclusions}.


\section{SAMPLE AND OBSERVATIONS}\label{sec:observations}

\subsection{NGC 7793}\label{sec:sample}

NGC 7793 is a SAd flocculent spiral, with an inclination of 47 degrees. It is the nearest Southern-hemisphere galaxy in LEGUS, and one of the brightest galaxies in the Sculptor Group. The main properties of the galaxy are: Cepheid-determined distance, 3.44 Mpc \citep{Pietrzynski2010}; color excess  due to the Galaxy, $E(B-V)=0.017$\,mag \citep{Schlafly2011}; stellar mass obtained from the extinction-corrected B-band luminosity and color information, using the method of \citealt{Bothwell2009}, based on the mass-to-light ratio models of \cite{Bell2001}, $M_\star=3\times10^9\,M_\odot$; galaxy-wide star formation rate calculated from the dust-attenuation corrected GALEX far-UV, adopting a distance of 3.44 Mpc, SFR=0.52~$M_\odot$\,yr$^{-1}$ \citep{lee2009}; SFR based on the integration of LEGUS fields NGC 7793W and NGC 7793E and averaged over the Hubble time (hence not directly comparable to the galaxy-wide UV measurement), calculated from CMD-fitting using F555W and F814W, adopting a distance of $\sim$3.7 Mpc \citep{Sabbi2018}, SFR=0.23$\pm$0.02 M$_\odot$/yr \citep{Sacchi2019}; central value of the oxygen abundance in the ionized gas, 12+log(O/H) = 8.50 $\pm$ 0.02; and radial oxygen abundance gradient adopting a distance of 3.4 Mpc, -0.305 $\pm$ 0.048 dex R$_{25}^{-1}$, \citep{Pilyugin2014}.

\subsection{\hst~imaging}\label{sec:imaging}

Two fields of NGC 7793 (E and W), which have a small overlap, were observed as part of the LEGUS and H$\alpha$ LEGUS programs. In this work, we focus on NGC 7793W, for which we recently obtained MUSE spectroscopy. 

We retrieved the individual ACS and WFC3 flc images of NGC 7793W from the Mikulski Archive for Space Telescopes and initially processed them on-the-fly with CalACS version 8.2.1 (14-Nov-2013, LEGUS ACS data), CalWF3 version 3.1.4 (09-Sept-2013, LEGUS WFC3 data), and 3.1.6 (15-Nov-2013, H$\alpha$ LEGUS WFC3 data). These images have gone through the following corrections: bias-shift, crosstalk, bias-stripe, charge transfer efficiency, dark. We then ran {\tt Astrodrizzle}\footnote{We used the following {\tt Astrodrizzle} command: astrodrizzle.AstroDrizzle('i*flc.fits',output=\\'ngc7793w\_f657n',driz\_sep\_bits='64,32',driz\_cr\_corr=yes,\\final\_bits='64,32',final\_wcs=True,final\_scale=0.05,final\_rot=0.)} from {\tt DrizzlePac} \citep{DrizzlePac2012} on the flc images, one filter at a time. {\tt Astrodrizzle} creates mask files for bad pixels and cosmic rays and drizzle-combines the input images using the mask files, while applying geometric distortion corrections, to create a "clean" distortion-free combined image.  For our purposes, it was important to remove cosmic rays from the F658N and F547M flc files, prior to running dolphot.
    
\begin{figure*}
\includegraphics[width=\columnwidth]{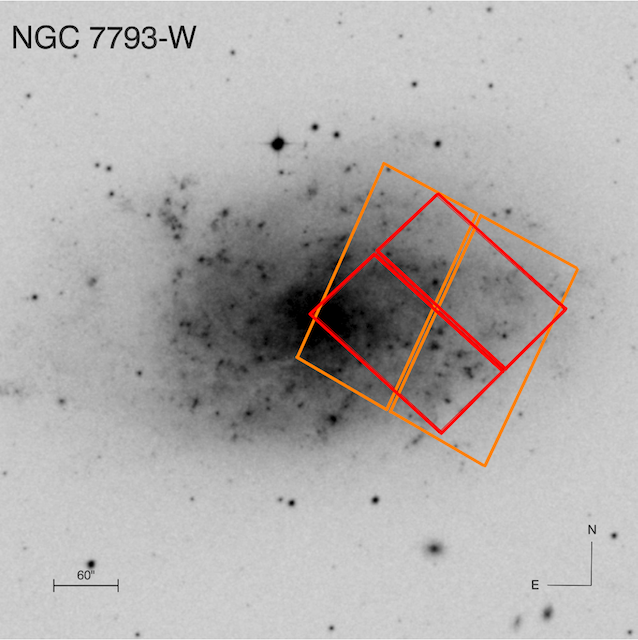}
\includegraphics[width=\columnwidth]{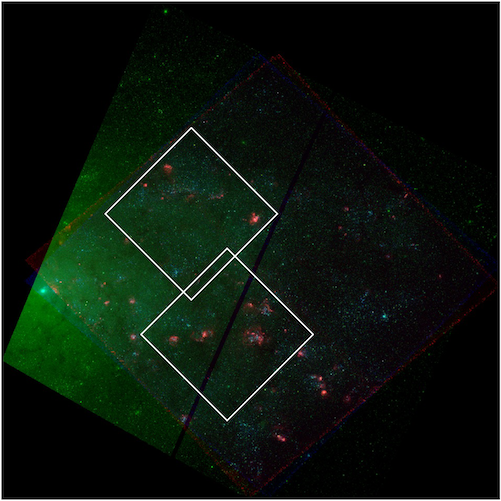}
    \caption{\emph{Left.} DSS image of NGC 7793 with footprints of WFC3/UVIS and ACS/WFC {\it HST} observations of NGC 7793W overlaid (red and orange polygon, respectively). The image dimensions are 10'x10'. \emph{Right.} Composite {\it HST} image of NGC 7793W (red, continuum subtracted F657N; green, F555W; and blue, F275W). We overlay the {\it VLT} MUSE footprints (white polygons), which have dimensions of 1'$\times$1'.}
    \label{fig:footprints}
\end{figure*}

In Table~\ref{tab:obs}, we provide a summary of the LEGUS and H$\alpha$ LEGUS observations. The left panel of Figure~\ref{fig:footprints} shows a  Digital Sky Survey (DSS) image of NGC 7793 with the WFC3 and ACS footprints overlaid. The right panel of Figure~\ref{fig:footprints} shows an \hst~composite image of NGC 7793W with the \vlt~\muse~footprints overlaid. 


\begin{table*}
\centering
\caption{Summary of \hst~observations of NGC 7793W. The F547M and F657N data are from the H$\alpha$ LEGUS program (PI Chandar, PID 13773), and the rest from the LEGUS program (PI Calzetti, PID 13364). Note that the F657N filter is 94 \AA~wide and includes H$\alpha$ + [N\,{\sc iii}].}
\label{tab:obs}
\begin{tabular}{lllllll}
\hline
{Instrument/Camera} & {Pixel Size}  & {Field of View} & {Filter} & {Pivot Wavelength} & {Exposure Time} & {Date Obs.} \\
\hfill &  (") & (") & \hfill & (\AA) & (s) & \hfill \\ 
\hline
WFC3/UVIS & 0.040   & 162 $\times$ 162 & F275W & 2710.1 & 2349.0  & 2014 Jan 18\\   
& &  & F336W & 3354.8 & 1101.0 (367$\times$ 3) & 2014 Jan 18  \\
& & &  F438W & 4326.5 & 947.0 & 2014 Jan 18 \\
& & &  F547M & 5447.4 & 550.0 & 2014 Dec. 07 \\
& & &  F657N & 6566.6 & 1545.0 (515$\times$ 3) & 2014 Dec. 07 \vspace{0.2cm} \\ 
ACS/WFC & 0.05 &  202 $\times$ 202 & F555W & 5359.6 & 680.0 (340$\times$2)  & 2003 Dec 10\\
 & & &  F814W & 8059.8 & 430.0 & 2003 Dec 10 \\
 \hline
\end{tabular}
\end{table*}


\subsection{MUSE spectroscopy}\label{sec:spectroscopy}

The MUSE data reduction is extensively presented in Della Bruna et al submitted. Here we provide a summary for the reader. NGC 7793W was observed on August 15$^{\rm th}$ 2017 as part of the MUSE AO Science Verification run, with programme ID 60.A\-9188(A). The data were acquired in wide field mode (WFM), with a $1$ arcmin$^2$ FOV and a spatial resolution of 0.2 arcsec, and over the extended wavelength range 4650 -- 9300 \AA, sampled in steps of 1.25\AA. The spectral resolution of the MUSE spectra is $FWHM\sim2.6$\,\AA~at 4650 \AA~and $FWHM\sim2.5$\,\AA~at 9300 \AA. The two MUSE pointings are centred at RA \& Dec coordinates (23:57:42.3070, -32:35:48.150) and  (23:57:43.8894,-32:34:49.353).

Four exposures (each rotated 90 degrees with respect to the previous one), with a total of $2100~\rm s$ on source, were acquired for each pointing. As the target fills the entire MUSE field of view, separate sky frames of $120$ s exposure were also acquired in between science exposures (in an object-sky-object sequence).

The data were reduced with the MUSE ESO pipeline \citep[][v2.2]{Weilbacher2014}, which performs standard calibrations (bias subtraction, flat fielding, wavelength- and flux-calibrations) as well as sky subtraction. 
The sky model is constructed on the separate sky frames by running the \texttt{muse\_create\_sky} task, and then subtracted from the science frame using the `subtract-model' option in \texttt{muse\_scipost}. The absolute astrometry of the resulting datacube is matched to the HST LEGUS observations \citep{Calzetti2015}. On the final cube, we measure a seeing of 0.71" full-width at half-maximum at 4900~\AA.

For extracting the spectra of the individual photometric cLBVs, we use the {\tt astropy} library of {\tt python}, and an extraction diameter of 4 pixels or 0.8", i.e., slightly larger than the seeing of the MUSE observations. 

\begin{figure}
\includegraphics[width=1\columnwidth]{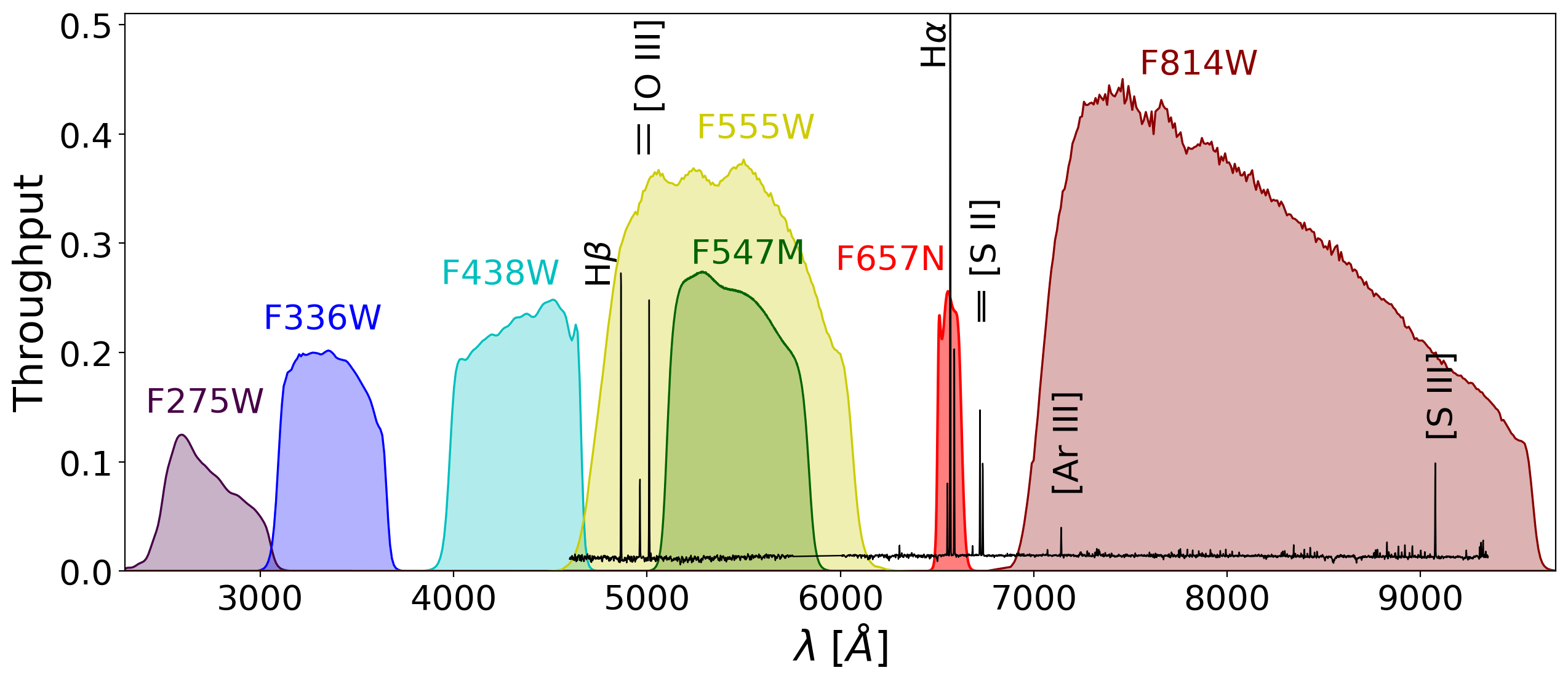}
\caption{System throughput curves of the seven filters which were used in this work (see Table~\ref{tab:obs}). In order to show which strong emission lines are present in the band passes of the filters, we overlay the MUSE spectrum of photometric cLBV \#9 in arbitrary flux units. The F814W filter includes the hydrogen Paschen series.} 
\label{fig:throughputs}
\end{figure}

In Figure~\ref{fig:throughputs}, we show an example of a cLBV MUSE spectrum with the system throughput curves of filters used in this work, overlaid.     


\section{Catalogue of photometric cLBVs}\label{sec:photometric_selection}

We expect cLBVs to have strong Balmer lines (particularly H$\alpha$, e.g., \citealt{King1998}). Such lines can have P-Cygni profiles or be in pure emission if the P-Cygni absorption component is filled in by nebular emission (e.g., \citealt{Grammer2015}).  \cite{King1998} found an efficient method to identify cLBVs in nearby galaxies, based on deep, continuum-subtracted narrow-band H$\alpha$ and [S\2] images. The cLBVs were selected as objects with extremely low [S\2]/H$\alpha$ ratios and with coincident stellar objects in continuum images. Five of their most promising candidates identified by these criteria in M31 were subsequently confirmed by optical spectroscopy to show spectra similar to the previously identified M31 LBV, HS var 15. The latter five candidates also have much in common with B[e] stars. 

In this work, we select potential cLBVs by combining images in the F547M, F657N and F814W filters, with restrictions of roundness, sharpness and H$\alpha$ luminosity. In this section, we describe in detail how we select our sample of potential cLBVs. 

\subsection{Methodology.}

Positions and fluxes for point-like sources were measured via PSF-fitting using the WFC3 and ACS modules of the photometry package dolphot version 2.0, downloaded on 2014 December 12 from the website \url{http://americano.dolphinsim.com/dolphot/}.

{\tt dolphot} is a stellar photometry package that was adapted from HSTphot for general use. We performed the stellar photometry on the flc images with the cosmic rays flagged. In order to run dolphot we used a custom python script which was developed by co-author L. U.. Dolphot was run separately for the five LEGUS filters (F275W, F336W, F438W, F555W, F814W) and the two H$\alpha$ LEGUS filters (F547M and F657N). The steps followed to find the cLBVs follow.

i) Select sources from the {\tt dolphot} output with a signal-to-noise ratio (SNR) > 4 in filters F547M and F657N, and in the intersection of sharpness = [-0.03 to 0.02] and roundness = [0.0 to 0.4]. This avoids selecting larger objects like compact clusters and non-localized H$\alpha$ emission.

ii) Run {\tt TOPCAT} \citep{taylor2011} to find the intersection between F547M, F657N, and F814W point sources using a 2-dimensional Cartesian search, such that the error in the distance between sources is less than 0.5 pixels\footnote{For the relevant datasets, 0.5 pixels is the typical distance error between points which are the same source in different \hst~images.}.

iii) Remove sources which are within 3 pixels from the edges of the images using a routine developed by co-author D. T. 

iv) Use the F547M and F814W images to determine the continuum and subtract it from the F657N image, for each cLBV. Compute the continuum subtracted H$\alpha$ luminosity. For the previous two steps we used a routine developed by co-author J. L..

v) Select all remaining sources with an H$\alpha$ luminosity of log($L$(H$\alpha$)  [erg s$^{-1}$])$\geq35$, which is the 5-$\sigma$ detection threshold for a point source. 

\begin{figure*}
\begin{subfigure}
	\centering
	\includegraphics[width=0.99\columnwidth]{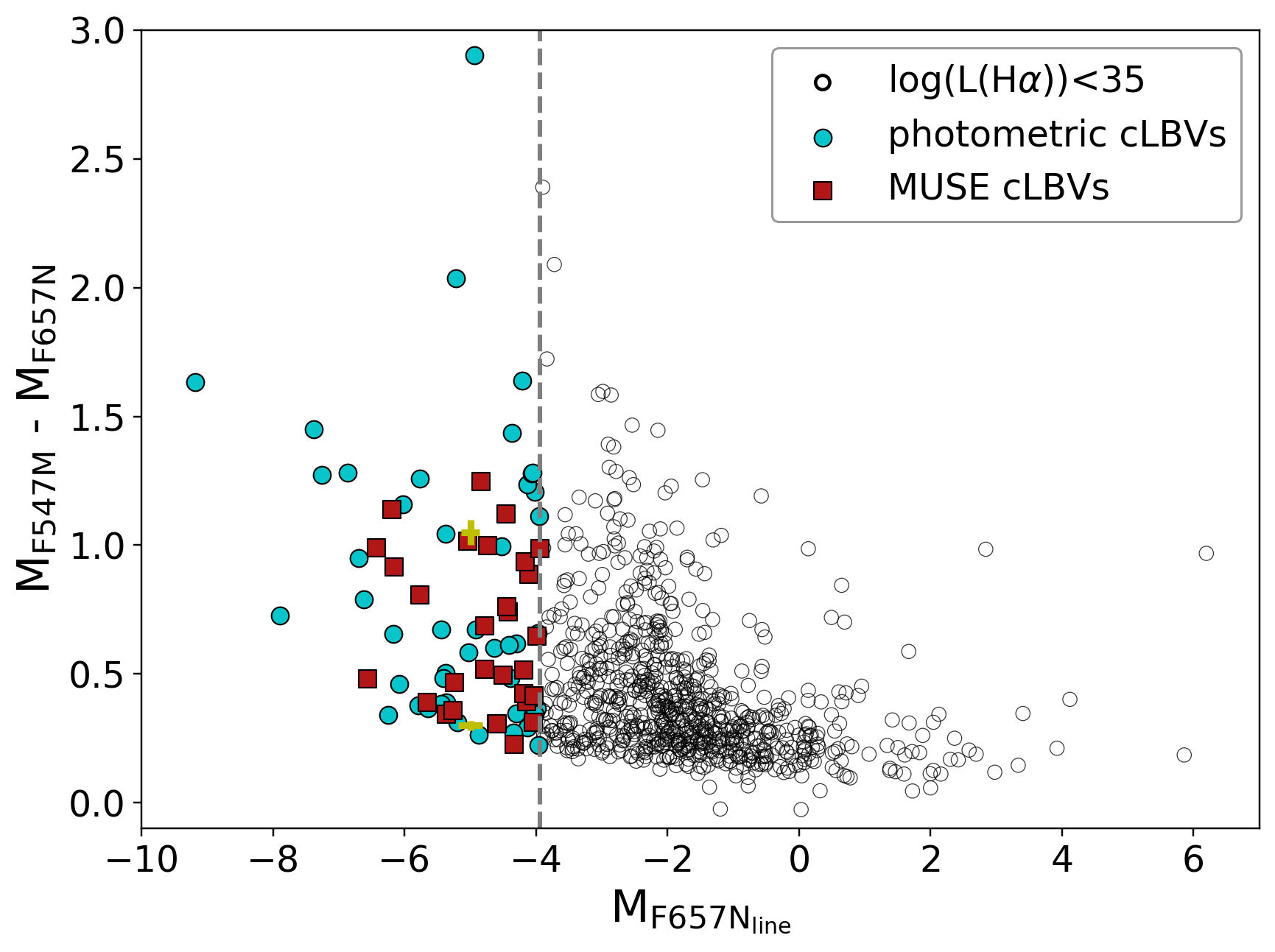}
\end{subfigure}
\begin{subfigure}
	\centering	
	\includegraphics[width=0.99\columnwidth]{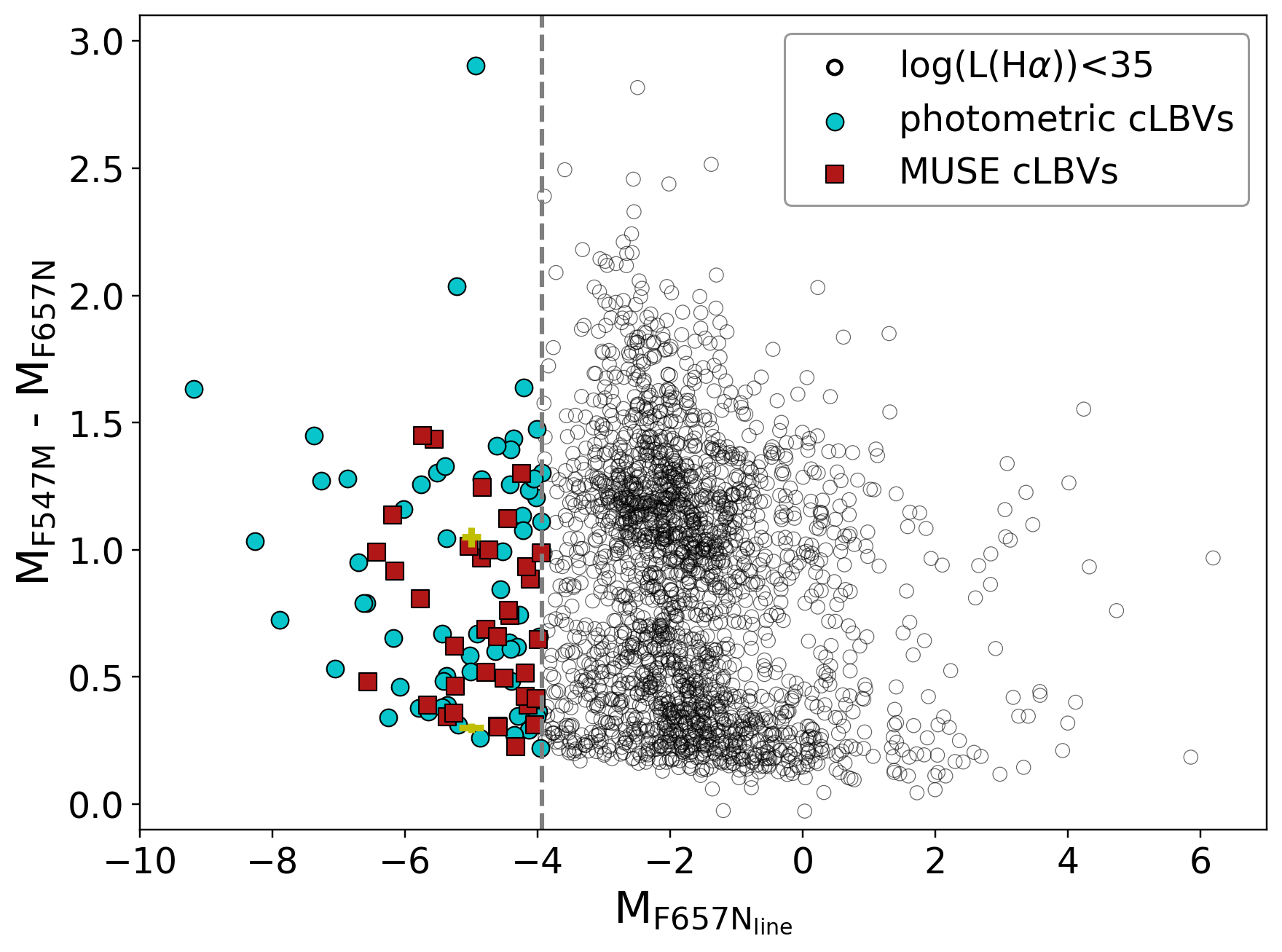} 
\end{subfigure}
    \caption{\emph{Left.} F547M\,$-$\,F657N color versus continuum-subtracted F657N absolute magnitude for sources selected via the conservative approach (all magnitudes are extinction uncorrected). The black unfilled circles are sources with lower H$\alpha$ luminosities than the photometric cLBVs. We mark with a vertical dashed line the minimum H$\alpha$ luminosity of the cLBVs. The filled circles are the photometric cLBVs which are outside of the MUSE fields. The filled squares are the MUSE cLBVs. We show two mean photometric error bars in yellow for points around (-5, 0.25) and (-5, 1), where the first and second numbers are the absolute magnitude in F547M and F547M\,$-$\,F657N. \emph{Right.} Similar to the left panel but we use the less conservative approach, which is the one adopted in this work.}
    \label{fig:selection}
\end{figure*}

Following these steps, we find $\sim$100 potential cLBVs (hereafter, less conservative approach). We can also apply a more conservative approach by requiring in step (ii) a SNR>4 in all five LEGUS and two H$\alpha$-LEGUS filters, which yields 76 potential cLBVs. In Figure~\ref{fig:selection}, we compare the samples of potential cLBVs obtained with the conservative approach (left panel) and less-conservative approach (right panel) in magnitude-color diagrams. We show F547M\,-\,F657N, where F657N is not continuum subtracted, on the y-axis, and F657N$_{\rm line}$, which is continuum-subtracted, on the x-axis. In both panels, sources which are 5-$\sigma$ above the H$\alpha$ detection threshold for a point source are located to the left of the vertical dashed line. The top cloud of unfilled circles in the right panel (which is not present in the left panel)  corresponds to sources with non-detections in the UV, U, and/or B filters. In this work, we adopt the less conservative approach and call the potential cLBVs selected via the less conservative approach, photometric cLBVs. 
 

Since the selection was not performed using the LEGUS stellar catalogues, we compare our catalogue of photometric cLBVs against the v2 stellar catalogue of \cite{Sabbi2018}. All photometric cLBVs have matches in the catalogue of \cite{Sabbi2018}. In addition, by comparing the F555W-band magnitudes of \cite{Sabbi2018} and this work, we find mean and median percent errors of 0.02 and 0.005, for the entire sample of photometric cLBVs, respectively\footnote{The percent errors were computed as follows: absolute value of ((F555W$_{\rm Sabbi}$ - F555W$_{\rm ThisWork}$) / F555W$_{\rm Sabbi})\times100$, where F555W is the apparent magnitude.}. This indicates that the magnitudes from both catalogues are in very good agreement. 
 
A cLBV must have spectroscopic similarities with bona-fide LBVs. Thirty six of the photometric cLBVs fall in the MUSE fields. The MUSE cLBVs are shown with red squares in Figure~\ref{fig:selection}. The latter figure shows that the H$\alpha$-brightest photometric cLBVs are outside of the MUSE field of view.

\subsection{Color-color diagrams}

\begin{figure*}
\begin{subfigure}
	\centering    
	\includegraphics[width=1.99\columnwidth]{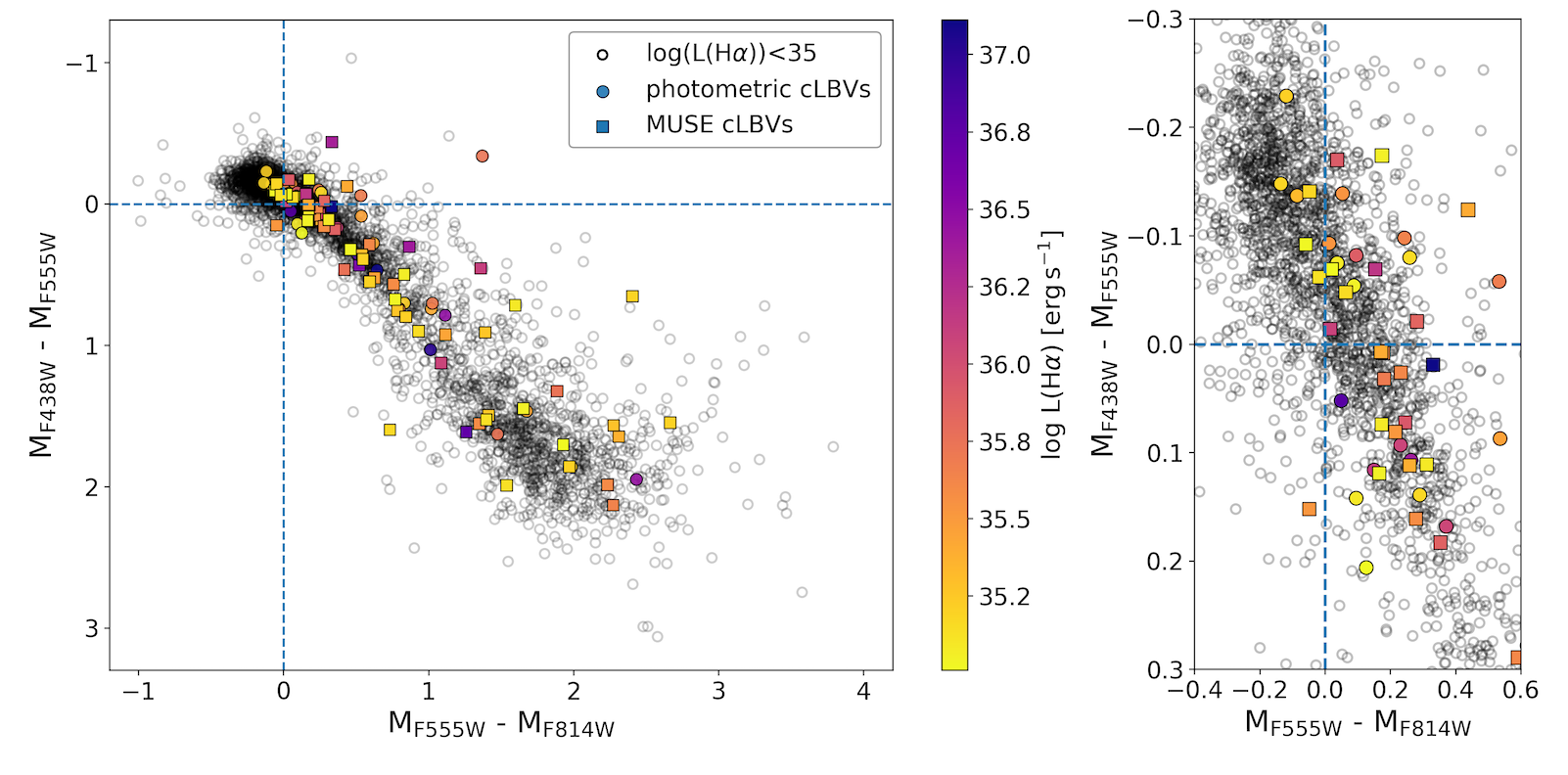}
\end{subfigure}

    \caption{\emph{Left.} F438W\,$-$\,F555W versus F555W\,$-$\,F814W color-color diagram for sources which are selected via the less conservative approach. We plot absolute magnitudes uncorrected for extinction. Not included in this diagram are the photometric cLBVs which are undetected in F438W, F555W and/or F814W; i.e. four photometric cLBVs without MUSE coverage and two MUSE cLBVs. The blue dashed lines mark the positions of zero color in the x and y directions. \emph{Right.} Enlargement of the upper left region of the plot in the left panel.}
    \label{fig:ha_usefulness}
\end{figure*}

Figure~\ref{fig:ha_usefulness} shows the color-color (CCD) diagram, F438W\,$-$\,F555W versus F555W\,$-$\,F814W of our candidate LBVs. The color bar shows the H$\alpha$ luminosity. The right panel of the figure, which is an enlargement of the blue portion of the CCD on the left, shows that two of the most $\alpha$ luminous cLBVs are located in bluest part of the CCD. 


\subsection{Extinction due to dust}

For massive stars, the standard procedure to correct for extinction due to dust intrinsic to the galaxy is to fit a stellar atmosphere model to the observed stellar spectrum in order to find the intrinsic spectrum of the star. It is by comparing the intrinsic spectrum to the observation that one finds the value of the extinction. It is beyond our goal to fit stellar atmosphere models to the MUSE cLBVs and find the intrinsic extinction using this method. Thus, we do not correct the observations for intrinsic extinction. As shown in Section~\ref{sec:classification}, this does not prevent us from finding promising cLBVs in this work. In Table~\ref{tab:photometry} of the appendix, we provide the extinction un-corrected photometry of the MUSE cLBVs. Note that \cite{Kahre2018} provide extinction maps for NGC 7793W. Their extinctions span a range between A(V)$\,=0-1$ peaking around 0.5 mag.  

\subsection{V-band magnitude}

An object more luminous than $M_{\rm{V}}\sim-7.0$ is unusual and therefore a good LBV candidate. However, the latter objects can also be hypergiants \citep{Clark2012}. After the extinction correction, the confirmed LBVs in M31 and M33 have $M_{\rm{V}}\leq-7$ \citep{Humphreys2014}. We found 24 photometric cLBVs with extinction-uncorrected V-band magnitudes such that, $M_{\rm{V}}\leq-7$. These would be even brighter after the extinction correction. Unfortunately, only seven of these are in the MUSE fields, as summarized in Table~\ref{tab:results}, where column 1 gives the ID of the MUSE cLBV and column 2 gives the extinction-uncorrected V-band magnitude. In Table~\ref{tab:results}, the MUSE cLBVs are sorted in order of decreasing luminosity (most luminous at the top). Since the brightest ``normal" blue supergiant stars are at $M_{\rm{V}}\sim-6.0$ or slightly brighter, it is likely that most objects with such magnitudes are normal supergiants. All sources in Table~\ref{tab:results} are spectroscopically classified as described Section~\ref{sec:spectroscopy}.   

\subsection{Color-magnitude diagrams}

\begin{figure*}
\includegraphics[width=2 \columnwidth]{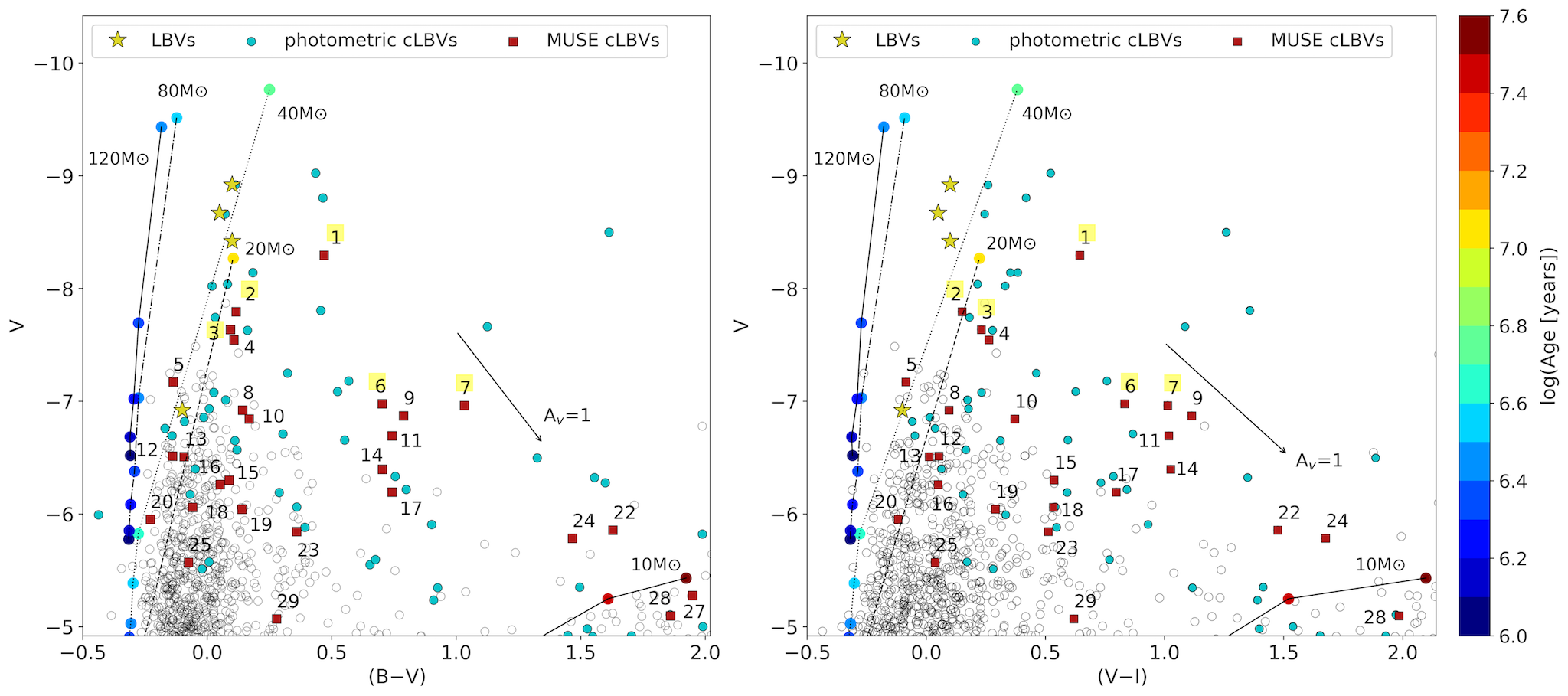}
\caption{Color magnitude diagrams adapted from \protect\cite{Clark2012}. The four yellow stars are bona-fide M33 LBVs, whose V-band magnitudes are provided at the bottom of table 1 of \protect\cite{Clark2012}. To convert to absolute magnitudes we use a distance modulus, $\mu$=24.92 \protect\citep{Bonanos2006}. We overlay our photometric and MUSE cLBVs at the distance of M33. The cyan filled circles represent photometric cLBVs outside of the MUSE fields. The red squares represent MUSE cLBVs. The black unfilled circles represent sources with the same selection criteria as the photometric cLBVs but with lower H$\alpha$ luminosities. The IDs of the five strongest cLBVs are highlighted in yellow. We overlay Padova tracks for various initial stellar masses and a metallicity of $Z=0.005$ (see text for more details). Circles of different colors along the tracks correspond to the stellar ages (years) in logarithmic scale which are given in the colorbar. \emph{Left.} F438W (B) - F555W (V) versus F555W. \emph{Right.} F555W - F814 (I) versus F555W.  The arrow represents the extinction vector corresponding to A(V)=1.}
\label{fig:CMD}
\end{figure*}

We looked at the CMD positions of our photometric cLBVs relative to the bona-fide LBVs of M33 which are plotted in figure 5 of \cite{Clark2012}. The result is shown in Figure~\ref{fig:CMD}, where we put our cLBVs at the distance of M33, i.e., 964 kpc \citep{Clark2012}. We overlay PARSEC v1.2S stellar tracks \citep{Bressan2012, Tang2014} for various initial stellar masses, a metallicity of $Z=0.005$, and a Salpeter Initial Mass Function, using the photometric system of \cite{Maiz-Apellaniz2006} and \cite{Bessell1990}. We attribute the spread of the photometric cLBVs in the diagrams to a combination of extinction and evolutionary phase. The MUSE cLBVs with IDs 1 to 7, i.e., the ones with $M_{\rm{V}}\leq-7$, fall in the region of M33 LBVs. In conclusion, our method for finding potential cLBVs via an H$\alpha$ excess works. We note that a few of the sources with log(L(H$\alpha$))<35 (black open circles) are also close to the bona-fide LBVs of M33 in Figure~\ref{fig:CMD}. This is due to the arbitrary cut at log(L(H$\alpha$))<35.


\section{MATCHES WITH SOURCES IN LEGUS CATALOGUES}\label{sec:other_catalogs}

Prior to the spectral classification of the MUSE cLBVs, we check if their \hst~coordinates match those of other types of sources which are identified in the following LEGUS-collaboration catalogues: candidate O-type stars identified via LEGUS photometry (Lee et al., preliminary catalogue; star clusters of less than 10 Myr classified according to \hst~H$\alpha$ morphology \citep{Hannon2019}; star clusters identified via LEGUS photometry \citep{Adamo2017}; and H\2 regions identified via two-dimensional MUSE spectroscopy (Della Bruna et al., subm.). We also look for members of the above catalogues which are located within the radius used for the spectral extraction and which contaminate the cLBV spectra.   

\subsection{Search radii}

The coordinates of a given source in the different \hst~images can be off by a distance of up to 0.5~pix or 0.02", which at a distance of 3.44 Mpc yields 0.3 pc. We use 0.5~pix as the search radius to find matches between the coordinates of the MUSE cLBVs and candidate O-type stars. The radius used to extract the photometry of LEGUS star clusters in NGC 7793 is 5~pix or 0.2" (3 pc). We use 5~pix as the search radius to find matches between MUSE cLBVs and star clusters. The comparison is performed with {\tt TOPCAT}. The seeing of the MUSE observations is 0.8". We use 0.4" as the search radius to find contaminants to the cLBV spectra. 

\subsection{Candidate O-type stars}

In their preliminary catalogue, Lee et al. (in prep.) identify candidate main sequence O-stars (M $\geq$ 20M$_{\odot}$) by their positions in color-magnitude (I vs B-I) and Q-magnitude (reddening-free index computed from NUV, B, I photometry vs I) diagrams. The definition of Q is given in \cite{Johnson1953}. Lee et al. use simulated HST photometry which is generated using stellar evolutionary tracks along with assumptions on the IMF, star formation history, and dust attenuation, to estimate the completeness and contamination rates, and to optimize the selection. The photometry incorporates accurate photometric errors based on artificial star tests.  With the simulated photometry, they examine their selection based on the 6 three-filter combinations that include the bluest filter available (NUV: F275W), as well as the standard UBV set for reference.  They find that the yield is maximized with NUV, B, and I photometry, with estimated completeness of $\sim$70\% and contamination rate of $\sim$30\%. Comparing against their catalogue, we find one candidate O-type star per MUSE cLBV within the search radius of 0.02", i.e., all MUSE cLBVs are also classified as candidate O-type stars. In addition, we find several candidate O-type star contaminants per cLBV, i.e., beyond 0.02" and within a search radius of 0.4", as summarized in column 6 of Table~\ref{tab:results}. 

\subsection{Star clusters classified by H$\alpha$ morphology}

\cite{Hannon2019} use the  H$\alpha-\,$LEGUS images of three galaxies, including NGC 7793 to classify star-clusters younger than 10 Myr according to their H-alpha morphology. They use the classification scheme of \cite{Hollyhead2015}, who defined three classes based on the presence and shape of the H$\alpha$ emission, which is similar to that of \cite{Whitmore2011}: 1) concentrated, where the target star cluster has a compact H\2 region and where there are no discernable bubbles or areas around the cluster which lack H$\alpha$ emission (these are expected to be about 3 Myr old); 2) partially exposed, where the target cluster shows bubble like/filamentary morphology covering part of the cluster; and 3) no emission, where the target cluster does not appear to be associated with H$\alpha$ emission. We check the overlap between our MUSE cLBVs and the concentrated clusters of \cite{Hannon2019}. We find no matches with concentrated clusters within a radius of 0.2", but we do find 5 matches with clusters of morphological type 2 and 3 within that radius. In addition, we find that no concentrated clusters contaminate the spectra of the MUSE cLBVs, although some clusters of morphological type 2 and 3 do contaminate them. This is summarized in columns 7 and 8 of Table~\ref{tab:results}, which give the morphological type of the star clusters found within a radius of 0.2" form the MUSE cLBVs. The additional clusters found within a radius of 0.4" are marked with asterisk. Note that two independent sets of evolutionary tracks, Padova and Geneva, were used in \cite{Hannon2019} to find clusters with ages of less than 10 Myr. These yield slightly different numbers of such clusters, which explains the case of cLBV \#27, with no matching clusters according to Padova and one matching cluster according to Geneva.

\subsection{Star clusters}

\cite{Adamo2017} produced comprehensive high-level young star cluster (YSC) catalogues for a significant fraction of LEGUS galaxies, including NGC 7793.  For the cluster catalogues, \cite{Adamo2017} perform a visual inspection in order to minimize contaminants in their final catalogues and add missed clusters. The total number of clusters which were visually classified in NGC 7793W is 299. Of the latter, 135 sources were classified as compact star clusters (Class 1 and 2, Kim et al. in prep). The NGC 7793 catalogues have been used by \cite{Krumholz2015} to determine the ages, masses, and extinctions of clusters in NGC 7793 using cluster-SLUGS; \cite{Grasha2017} to investigate the hierarchical clustering of the YSCs; \cite{Grasha2018} to study the relationship between giant molecular cloud properties and the associated star clusters; and \cite{Hannon2019} to study the H$\alpha$ morphology of star clusters. The LEGUS cluster catalogues can be found here: \url{https://archive.stsci.edu/prepds/legus/dataproducts-public.html}.

We find 11 matches between the MUSE cLBVs and star clusters within a radius of 0.2" and three additional cases where clusters are located within a radius of 0.4", i.e., contaminate the cLBV spectra. In Table~\ref{tab:results}, columns 9 and 10 give the ages of the matching/contaminating star clusters, where the additional clusters found within a radius of 0.4" are marked with asterisk. Those ages are based on two different sets of solar-metallicity stellar evolution tracks (as no other metallicity is currently available for NGC 7793). We remind the reader that according to single-star evolution, clusters of $\sim$5 Myr and older are too old to host LBV stars. Seven of the twelve contaminating star clusters are 5 Myr or older. As will be shown in the next section, the rest of the contaminating clusters are around cLBVs with signatures of the presence of W-R stars. Interestingly, Table~\ref{tab:results} shows that six of the candidates with $M_{\rm V}\leq-7$ have no contaminating star clusters. This is discussed in the context of the binary formation scenario of LBVs in section~\ref{sec:formation}.

\subsection{H\2 regions}\label{sec:hii_reg}

The H\2 region selection is presented in Della Bruna et al. (subm.). We provide below a concise description. Della Bruna et al. identify H\2 regions by constructing a hierarchical tree structure on the MUSE H$\alpha$ flux map. To do this, they use the python package {\tt ASTRODENDRO}\footnote{Hierarchical tree structure visual explanation under \url{https://dendrograms.readthedocs.io}}. As input data for the algorithm, they use a flux map obtained by integration of the continuum-subtracted cube in the rest frame wavelength range $6559-6568\,$\AA. They assume a Gaussian background for the diffuse H$\alpha$ radiation and compute the hierarchical tree down to 10 sigma above the background mean. They select all leaves of the structure and also the branches that enclose the largest H\2 regions, inside of which the algorithm identifies resolved sub-peaks as distinct leaves. Therefore, they do not use any fixed aperture to define the H\2 regions. They also look at BPT \citep{Baldwin1981} diagrams ([S\2]/H$\alpha$ and [N\2]/H$\alpha$) to assess that the candidate regions are consistent with being photoionized and to exclude potential contaminants (e.g planetary nebulae and supernova remnants). 

\begin{figure}
\includegraphics[width=0.99 \columnwidth]{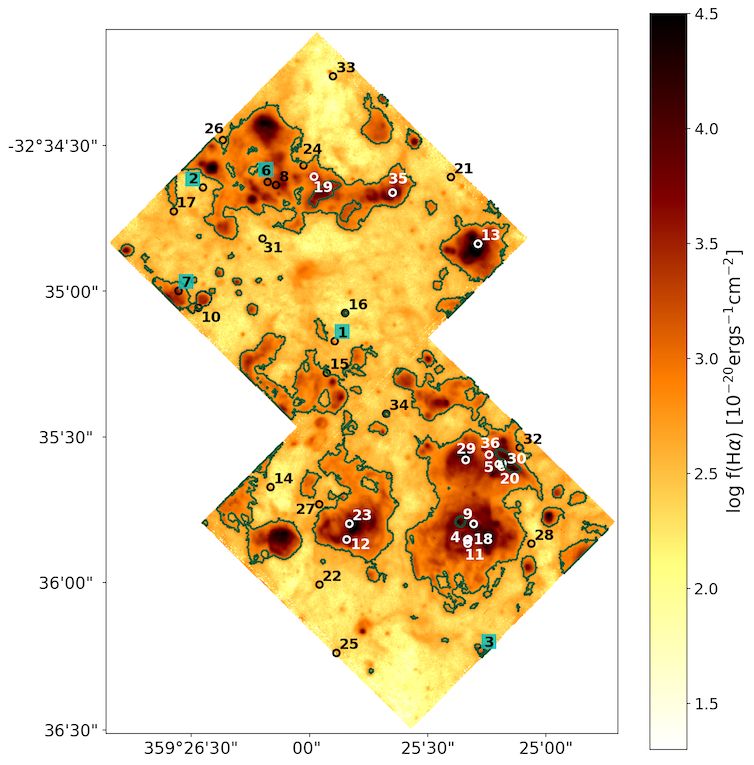}

\caption{MUSE continuum-subtracted H$\alpha$ emission map with positions of MUSE cLBVs indicated with circles with IDs. The IDs of the five strongest cLBVs are highlighted in cyan. The green contours show the H\2 regions from the catalogue of Della Bruna et al. (subm.). Each of the two MUSE fields is 1'$\times$1' on the side. At the adopted distance of 3.44 Mpc, 60" represent $\sim$1 kpc.}
\label{fig:hii_reg}
\end{figure}

In Figure~\ref{fig:hii_reg}, we show the H\2 regions map of Della Bruna et al.. We overlay and identify our 36 MUSE cLBVs. We include MUSE cLBVs which are located at the edges of the H$\alpha$ map because their MUSE spectra do not seem to suffer from any artifacts. We visually check if the MUSE cLBVs are located within H\2-region contours in Figure \ref{fig:hii_reg} and find 22 MUSE cLBVs where this is the case. Column 11 of Table~\ref{tab:results} summarizes these results.

  Candidates \#1 and \#2 are located outside H\2 regions and candidate \#3 is in a small H\2 region which is isolated from the large H\2 region complexes which are shown in Figures~\ref{fig:hii_reg} and~\ref{fig:footprints}. The \hst~postage stamps of the latter cLBVs are shown in Figure~\ref{fig:batch1}, which confirms the findings just described.
 

\begin{table*}
\centering
\caption{Photometric properties of MUSE cLBVs.}
\label{tab:results}
\begin{tabular}{@{}lllllcccccc@{}}
\toprule
{ID}	&	 {RA$^{\rm{a}}$} 	&	 {Dec$^{\rm a}$} 	&	 {$M_{\rm V}^{\rm b}$}  	& {L(H$\alpha$)$^{\rm c}$}  & {O}$^{\rm d}$   &  \multicolumn{2}{c}{CC$^{\rm e}$} 	& \multicolumn{2}{c}{SC$^{\rm f}$} 	&	  {H\2}$^{\rm g}$  \\
& & & & & & Padova & Geneva & Padova & Geneva &\\
(1) & (2) & (3) & (4) & (5) & (6) & (7) & (8) & (9) & (10 & (11) \\
\hline														
1  & 23:57:43.617 & -32:35:10.691 & -8.29$\pm$0.05  & 36.06$\pm$0.03 & 8  & 0    & 0    & 0     & 0    & 0 \\
2  & 23:57:45.821 & -32:34:38.168 & -7.79$\pm$0.05  & 35.58$\pm$0.05 & 8  & 0    & 0    & 0     & 0    & 0 \\
3  & 23:57:41.170 & -32:36:16.184 & -7.64$\pm$0.05  & 35.90$\pm$0.03 & 7  & 0    & 0    & 0     & 0    & 1 \\
4  & 23:57:41.400 & -32:35:52.769 & -7.54$\pm$0.05  & 35.09$\pm$0.11 & 16 & 2, 2* & 2, 2* & 4, 2* & 3, 2* & 1 \\
5  & 23:57:40.872 & -32:35:36.706 & -7.17$\pm$0.05  & 35.38$\pm$0.05 & 10 & 0    & 0    & 0     & 0    & 1 \\
6  & 23:57:44.739 & -32:34:37.020 & -6.98$\pm$0.05  & 35.54$\pm$0.06 & 13 & 0    & 0    & 0     & 0    & 1 \\
7  & 23:57:46.228 & -32:35:00.058 & -6.96$\pm$0.05  & 35.70$\pm$0.04 & 17 & 0    & 0    & 0     & 0    & 1 \\
8  & 23:57:44.600 & -32:34:37.653 & -6.92$\pm$0.05  & 35.11$\pm$0.13 & 10 & 3    & 3    & 5    & 5    & 1 \\
9  & 23:57:41.292 & -32:35:49.332 & -6.87$\pm$0.05  & 35.66$\pm$0.04 & 15 & 0    & 0    & 5*     & 5*    & 1 \\
10 & 23:57:45.899 & -32:35:03.626 & -6.84$\pm$0.05  & 35.17$\pm$0.10 & 8  & 0    & 0    & 0     & 0    & 1 \\
11 & 23:57:41.394 & -32:35:53.490 & -6.69$\pm$0.05  & 35.73$\pm$0.03 & 11 & 0    & 0    & 600   & 700  & 1 \\
12 & 23:57:43.416 & -32:35:52.641 & -6.51$\pm$0.05* & 35.91$\pm$0.03 & 11 & 0    & 0    & 0     & 0    & 1 \\
13 & 23:57:41.221 & -32:34:50.083 & -6.51$\pm$0.05  & 35.74$\pm$0.04 & 21 & 0    & 0    & 0     & 0    & 1 \\
14 & 23:57:44.690 & -32:35:41.549 & -6.40$\pm$0.05* & 35.22$\pm$0.09 & 27 & 0    & 0    & 200   & 200  & 0 \\
15 & 23:57:43.749 & -32:35:17.408 & -6.30$\pm$0.05  & 36.01$\pm$0.03 & 8  & 0    & 0    & 200   & 200  & 1 \\
16 & 23:57:43.442 & -32:35:04.726 & -6.26$\pm$0.05  & 35.05$\pm$0.13 & 10 & 0    & 0    & 0     & 0    & 1 \\
17 & 23:57:46.308 & -32:34:43.224 & -6.19$\pm$0.05  & 35.53$\pm$0.05 & 11 & 0    & 0    & 0     & 0    & 0 \\
18 & 23:57:41.383 & -32:35:52.587 & -6.06$\pm$0.05  & 35.11$\pm$0.07 & 17 & 2, 2* & 2, 2* & 2,4*  & 2, 3* & 1 \\
19 & 23:57:43.964 & -32:34:35.874 & -6.04$\pm$0.05  & 35.20$\pm$0.12 & 8  & 0    & 0    & 0     & 0    & 1 \\
20 & 23:57:40.823 & -32:35:37.211 & -5.95$\pm$0.05  & 35.21$\pm$0.06 & 24 & 3    & 3    & 3    & 3    & 1 \\
21 & 23:57:41.672 & -32:34:35.971 & -5.87$\pm$0.05  & 35.28$\pm$0.07 & 5  & 0    & 0    & 0     & 0    & 0 \\
22 & 23:57:43.872 & -32:36:02.178 & -5.86$\pm$0.05  & 35.05$\pm$0.08 & 9  & 0    & 0    & 0     & 0    & 0 \\
23 & 23:57:43.373 & -32:35:49.305 & -5.84$\pm$0.05  & 35.13$\pm$0.08 & 6  & 0    & 0    & 300   & 3000 & 1 \\
24 & 23:57:44.138 & -32:34:33.512 & -5.78$\pm$0.05  & 35.27$\pm$0.08 & 12 & 0    & 0    & 0     & 0    & 1 \\
25 & 23:57:43.589 & -32:36:16.642 & -5.57$\pm$0.05  & 35.45$\pm$0.04 & 11 & 0    & 0    & 0     & 0    & 0 \\
26 & 23:57:45.487 & -32:34:28.146 & -5.34$\pm$0.05  & 35.35$\pm$0.07 & 9  & 0    & 0    & 0     & 0    & 0 \\
27 & 23:57:43.878 & -32:35:45.186 & -5.28$\pm$0.05  & 35.54$\pm$0.03 & 14 & 0    & 3    & 3000 & 7    & 0 \\
28 & 23:57:40.328 & -32:35:53.519 & -5.10$\pm$0.05  & 35.03$\pm$0.07 & 5  & 0    & 0    & 0     & 0    & 0 \\
29 & 23:57:41.425 & -32:35:35.809 & -5.07$\pm$0.05  & 35.33$\pm$0.04 & 9  & 0    & 0    & 5     & 4    & 1 \\
30 & 23:57:40.812 & -32:35:36.533 & -4.88$\pm$0.05  & 35.37$\pm$0.04 & 14 & 0    & 0    & 1     & 2    & 1 \\
31 & 23:57:44.825 & -32:34:48.969 & -4.70$\pm$0.05  & 35.24$\pm$0.07 & 10 & 0    & 0    & 0     & 0    & 0 \\
32 & 23:57:40.525 & -32:35:33.206 & -4.63$\pm$0.05  & 35.08$\pm$0.06 & 11 & 0    & 0    & 0     & 0    & 1 \\
33 & 23:57:43.647 & -32:34:14.648 & -4.43$\pm$0.05  & 35.28$\pm$0.07 & 7  & 0    & 0    & 0     & 0    & 0 \\
34 & 23:57:42.755 & -32:35:26.029 & -4.20$\pm$0.06  & 35.01$\pm$0.06 & 9  & 0    & 0    & 0     & 0    & 1 \\
35 & 23:57:42.648 & -32:34:39.271 & -4.01$\pm$0.06  & 35.35$\pm$0.06 & 11 & 0    & 0    & 0     & 0    & 1 \\
36 & 23:57:41.036 & -32:35:34.696 & -3.42$\pm$0.06  & 35.10$\pm$0.05 & 5  & 0    & 0    & 0     & 0    & 1 \\
\hline
\end{tabular}
\begin{tablenotes}
      \item $^{\rm{a}}$ J2000 coordinate in LEGUS WFC3/UVIS/F657N image. RA=Right ascension in format hours:minutes:seconds. Dec=declination in format degrees:minutes:seconds. 
      \item $^{\rm{b}}$ Absolute F555W magnitude in ascending order (most luminous at the top). The two cases where the source is undetected in F555W but is detected in F547M are marked with asterisks in the column. In those cases, we give the F547M value. The error in the absolute magnitude includes the photometric and distance modulus error. The error in the distance modulus is 27.68 $\pm$ 0.05 mag \citep{Pietrzynski2010}.
     \item $^{\rm{c}}$ H$\alpha$ luminosity in erg s$^{-1}$ (uncorrected for reddening), in logarithmic scale, calculated from the PSF-photometry. The error in the luminosity includes the photometric and distance modulus errors. The F657N filter includes the [N\2] lines.
     \item $^{\rm{d}}$ Number of candidate O-type stars within a radius of 0.4", including the candidate within a radius of 0.02". 
     \item $^{\rm{e}}$ Morphological type of contaminating star cluster as defined by \citealt{Hannon2019} (2=partially exposed. 3=no H$\alpha$). Only clusters with ages of 10 Myr according to the Padova or Geneva models are considered in \cite{Hannon2019}. Asterisk= clusters within a radius of 0.4" of the cLBV and outside a radius of 0.2".
     \item $^{\rm{f}}$ Ages (in Myr) of contaminating star clusters in the catalogue of \cite{Adamo2017}, based on the Padova or Geneva models. Asterisk= star clusters within a radius of 0.4" of the cLBV and outside a radius of 0.2". 
     \item $^{\rm{g}}$ Contamination by H\2 regions. 1=MUSE cLBV is within an H\2 region contour. 0=MUSE cLBV is not within an H\2 region counter (see text for more details).
    \end{tablenotes}
\end{table*}


\section{Spectral classification of MUSE cLBVs}\label{sec:classification}

Here we discuss the spectral classification of the 36 potential cLBVs in the MUSE fields. For all MUSE cLBVs, Figures~\ref{fig:batch1}, ~\ref{fig:batch2}, ~\ref{fig:batch3}, and~\ref{fig:batch4}, show postage stamps and optical spectra in spectral windows of interest. The IDs of the candidates are given in the upper-right corner of the postage stamps. In Figures~\ref{fig:batch1}, ~\ref{fig:batch2}, and~\ref{fig:batch3}, we skip the MUSE cLBVs which show broad He\2$\,\lambda4686$ emission. The latter sources are shown in the last seven rows of Figure~\ref{fig:batch4} and discussed in the context of W-R stars. For each MUSE cLBV, Table~\ref{tab:results2} provides a summary of the spectroscopic measurements which we discuss in this section.

\subsection{Most promising MUSE cLBVs}

There is no consensus on how to spectroscopically classify an object as a candidate LBV. In general, candidate LBVs show strong Balmer and/or optical He\1 emission lines, typically (but not always) with P-Cygni like profiles. The appearance is not unlike what we see in some Type IIn supernovae, with narrow Balmer emission atop a broader base \citep{King1998}. 

As previously shown, candidates LBVs \#1 to \#7 have photometric properties which are compatible with known properties of LBVs. Among the latter, only candidate \#2 shows H$\alpha$ with a P-Cygni like profile. The profile is shown in Figure~\ref{fig:PCygni2}. The other strong candidates show H$\alpha$ in pure emission. 

The spectral resolution of the MUSE observations in the H$\alpha$ region is $FWHM\sim120$\,km\,s$^{-1}$. We do not detect a broad H$\alpha$ component in emission for cLBVs \#1 to \#7. For cLBVs, typical values of the full width at half maximum ($FWHM$) for the narrow component of H$\alpha$ are in the range $100-200$\,km\,s$^{-1}$ \citep{Humphreys2017a}. We measure the $FWHM$(H$\alpha$) values of the MUSE candidates. For this purpose, we fit one or more Gaussians to H$\alpha$, depending on the presence of components in broad emission and/or absorption. The results are shown in column (2) of Table~\ref{tab:results2}. We find values between 119 and 155\,km\,s$^{-1}$, as shown in the min and max rows of column (2). Only the sources with values above 120\,km\,s$^{-1}$ are resolved. The latter sources are within the upper limit of the expected range. 

\begin{figure*}
\begin{subfigure}
  \centering
  \includegraphics[height=0.25\columnwidth]{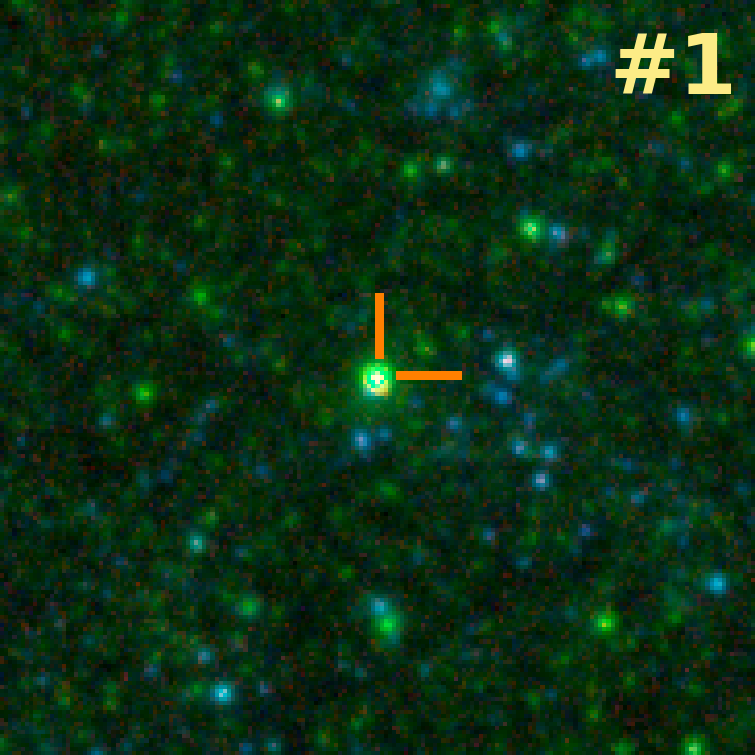}
\end{subfigure}
\begin{subfigure}
  \centering
  \includegraphics[height=0.25\columnwidth]{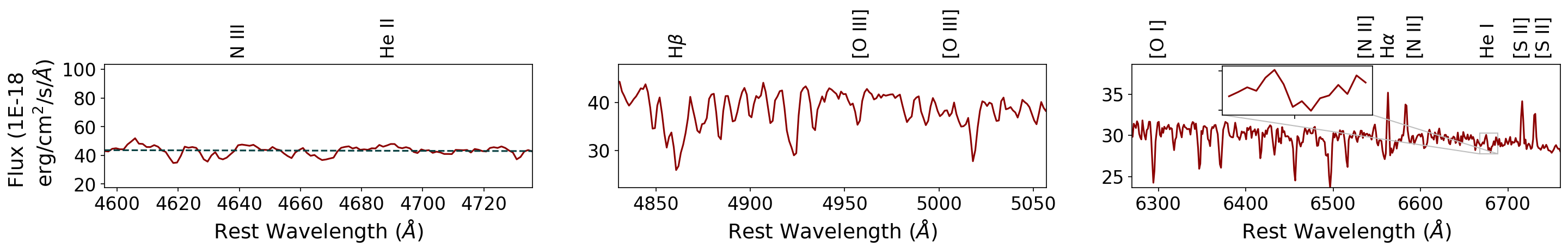}
\end{subfigure}\\
\begin{subfigure}
  \centering
  \includegraphics[height=0.25\columnwidth]{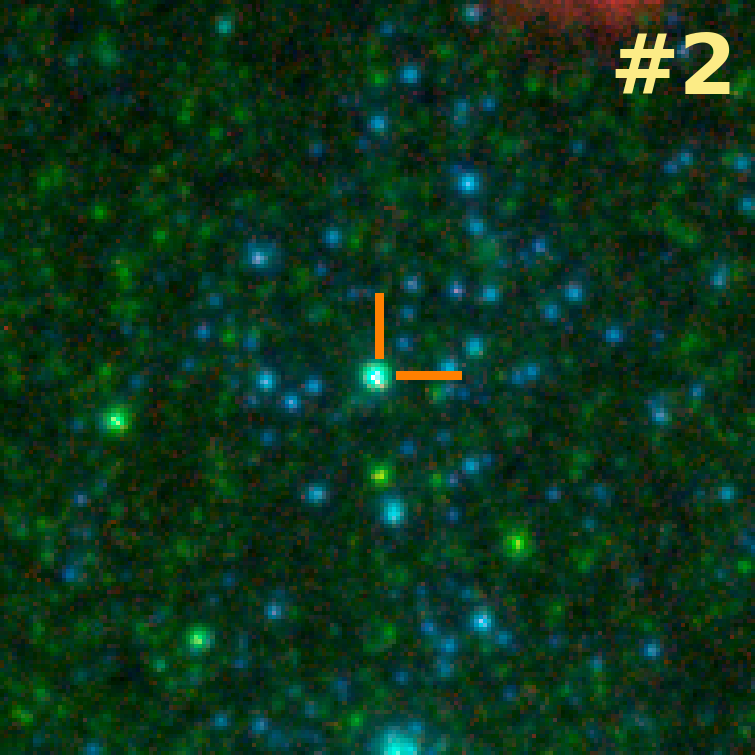}
\end{subfigure}
\begin{subfigure}
  \centering
  \includegraphics[height=0.25\columnwidth]{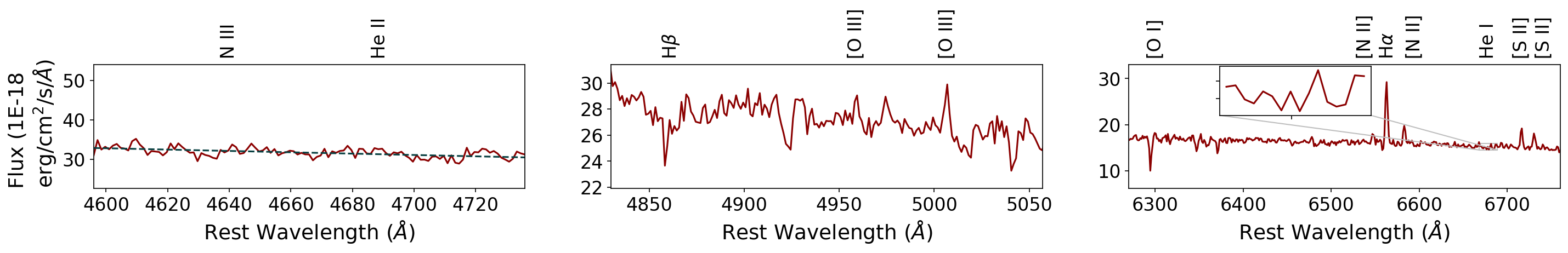}
\end{subfigure}\\
\begin{subfigure}
  \centering
  \includegraphics[height=0.25\columnwidth]{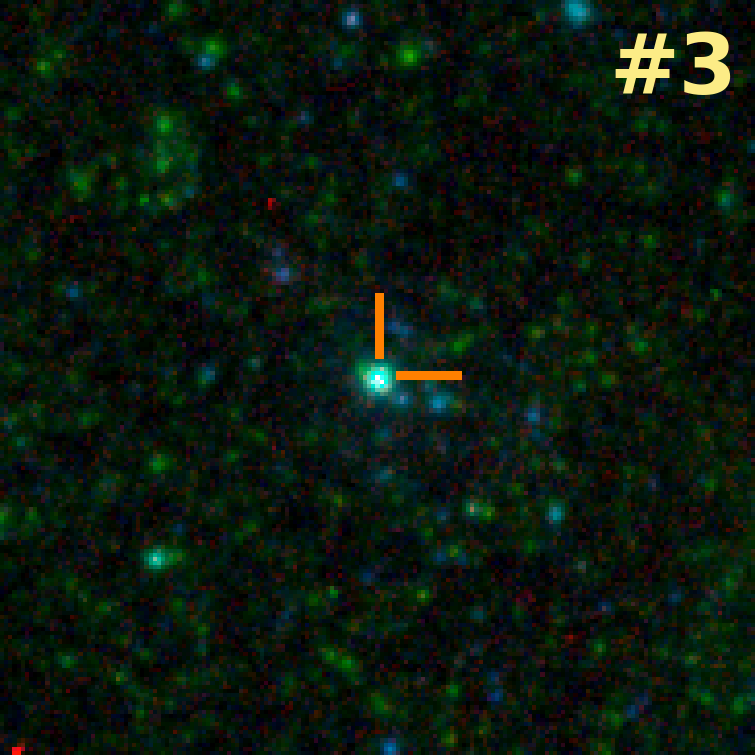}
\end{subfigure}
\begin{subfigure}
  \centering
  \includegraphics[height=0.25\columnwidth]{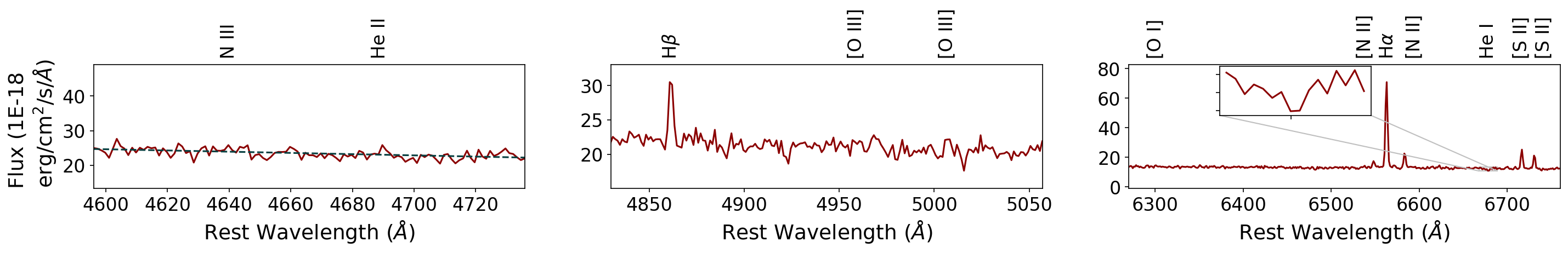}
\end{subfigure}\\
\begin{subfigure}
  \centering
  \includegraphics[height=0.25\columnwidth]{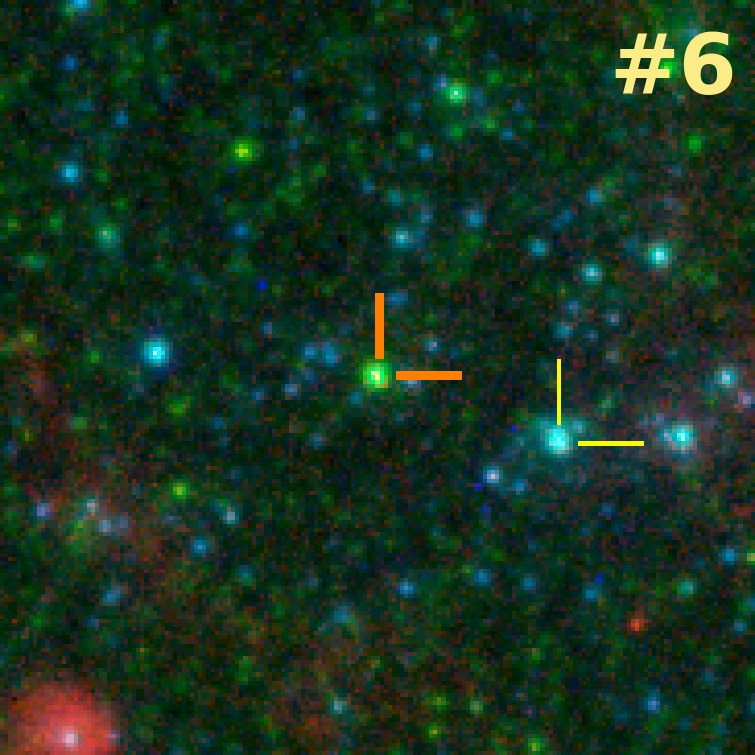}
\end{subfigure}
\begin{subfigure}
  \centering
  \includegraphics[height=0.25\columnwidth]{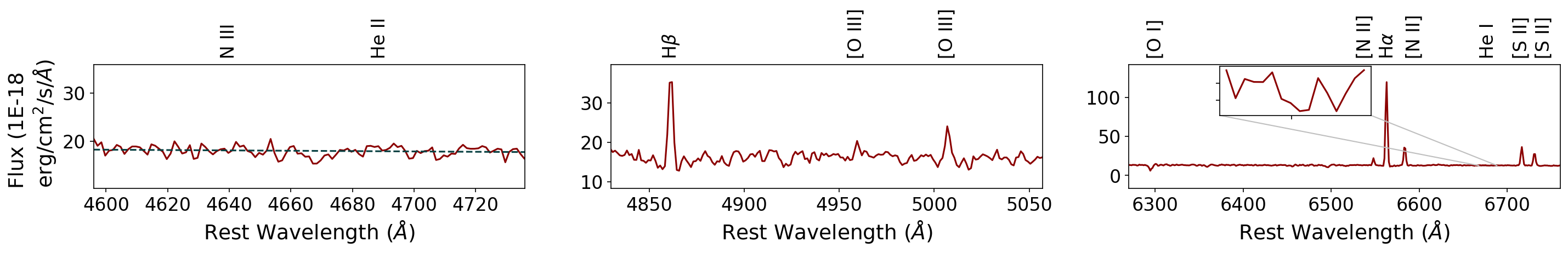}
\end{subfigure}\\
\begin{subfigure}
  \centering
  \includegraphics[height=0.25\columnwidth]{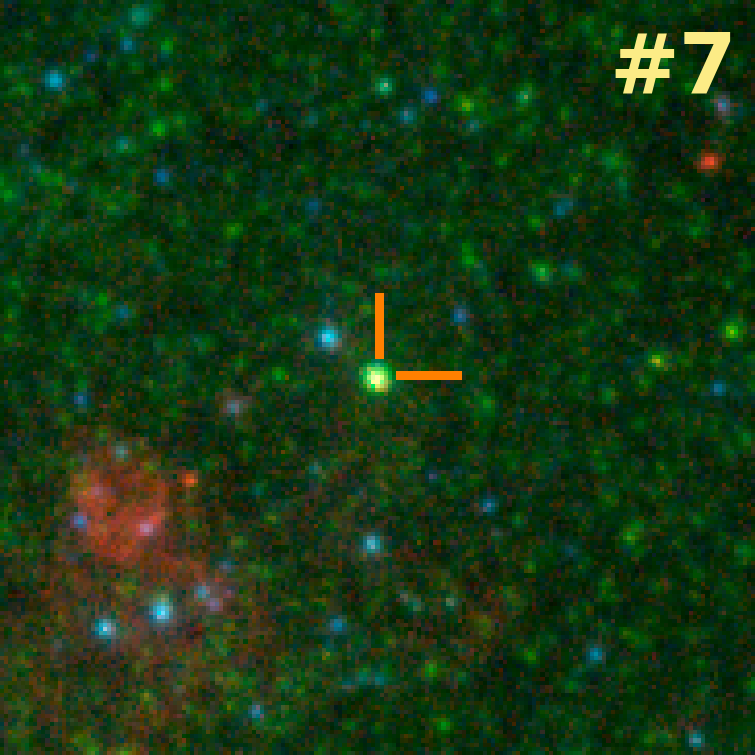}
\end{subfigure}
\begin{subfigure}
  \centering
  \includegraphics[height=0.25\columnwidth]{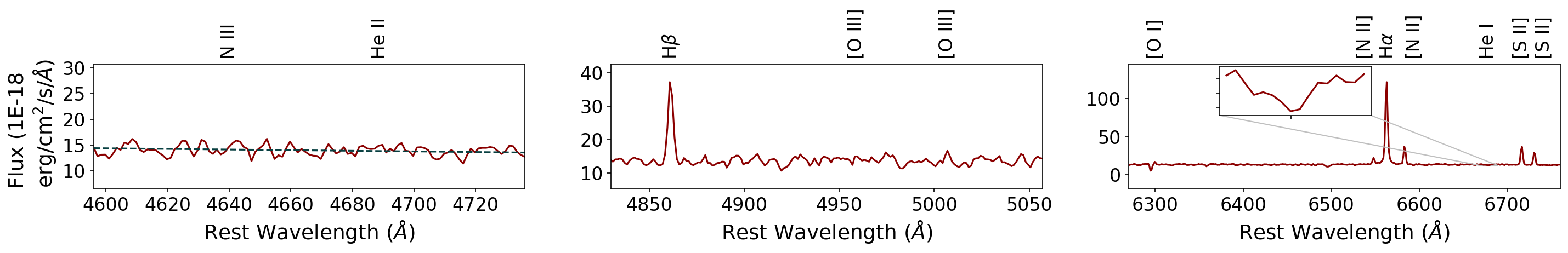}
\end{subfigure}\\
\begin{subfigure}
  \centering
  \includegraphics[height=0.25\columnwidth]{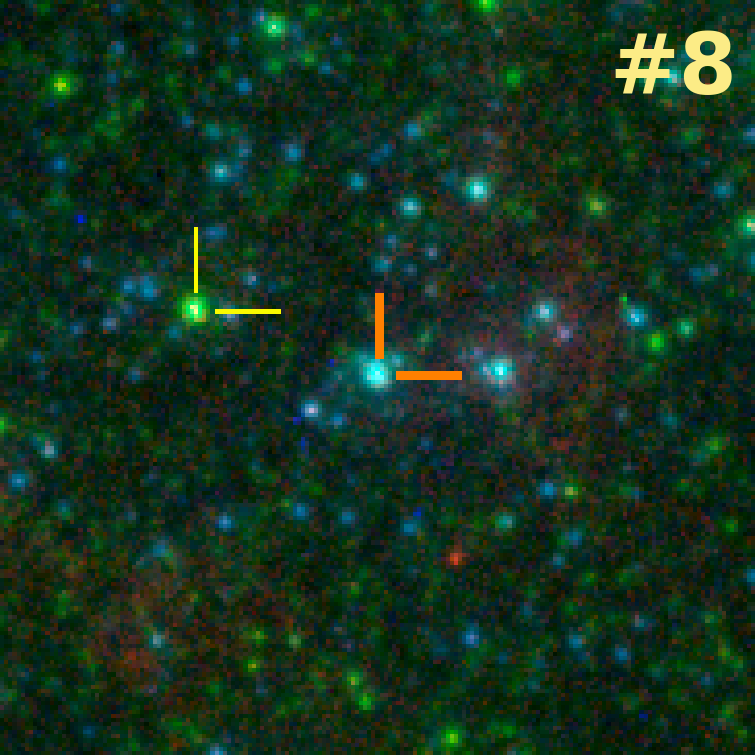}
\end{subfigure}
\begin{subfigure}
  \centering
  \includegraphics[height=0.25\columnwidth]{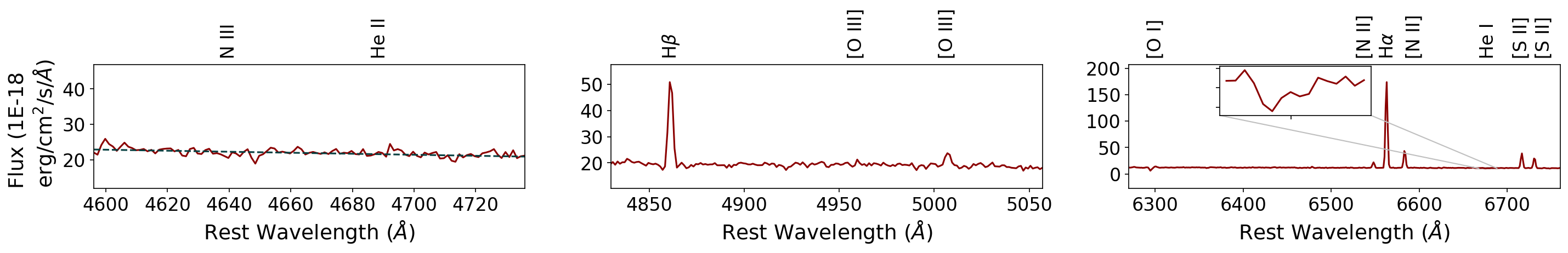}
\end{subfigure}\\
\begin{subfigure}
  \centering
  \includegraphics[height=0.25\columnwidth]{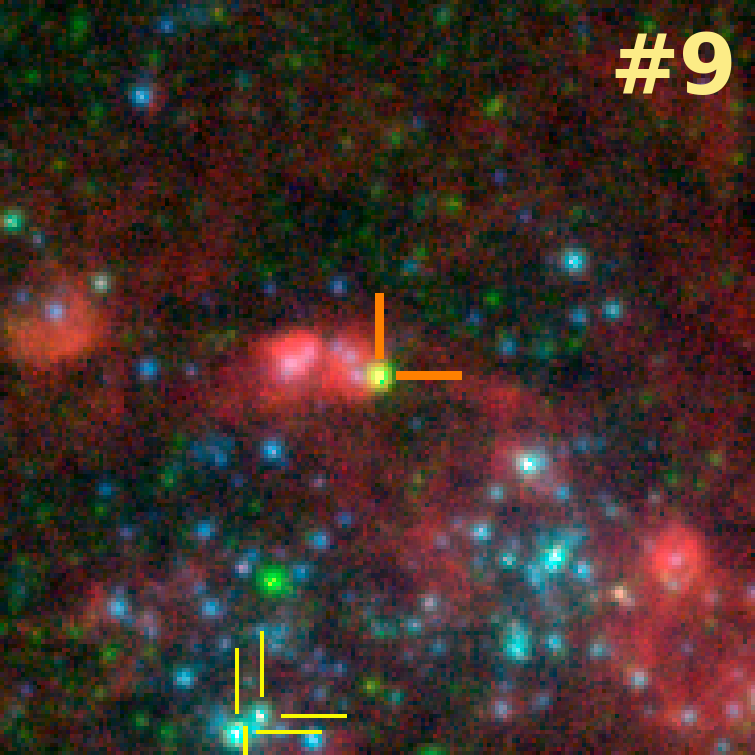}
\end{subfigure}
\begin{subfigure}
  \centering
  \includegraphics[height=0.25\columnwidth]{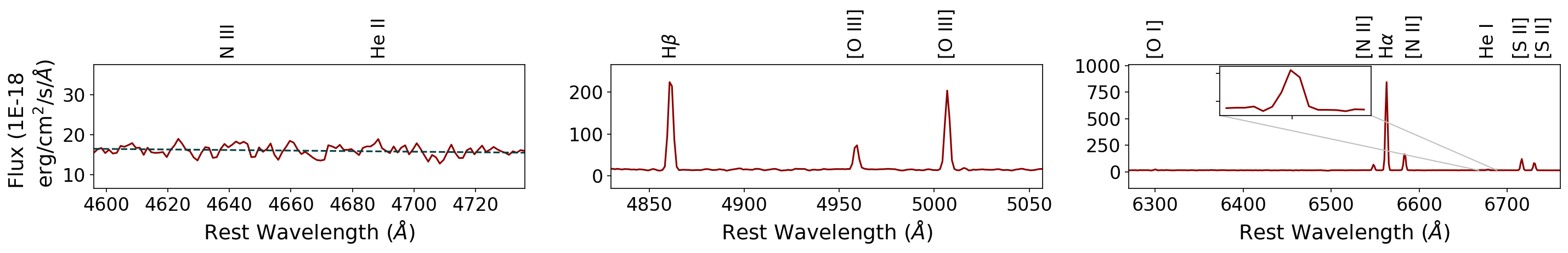}
\end{subfigure}\\
\begin{subfigure}
  \centering
  \includegraphics[height=0.25\columnwidth]{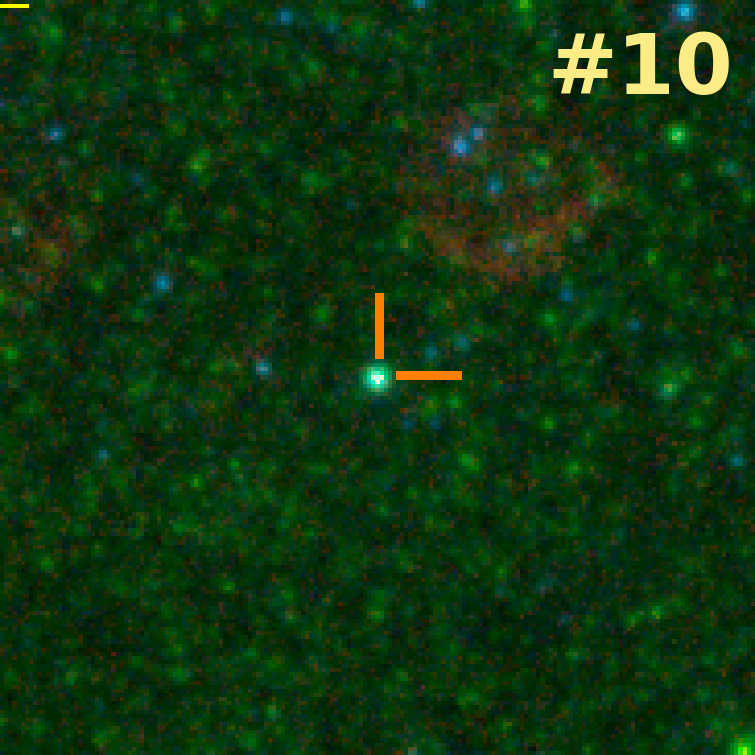}
\end{subfigure}
\begin{subfigure}
  \centering
  \includegraphics[height=0.25\columnwidth]{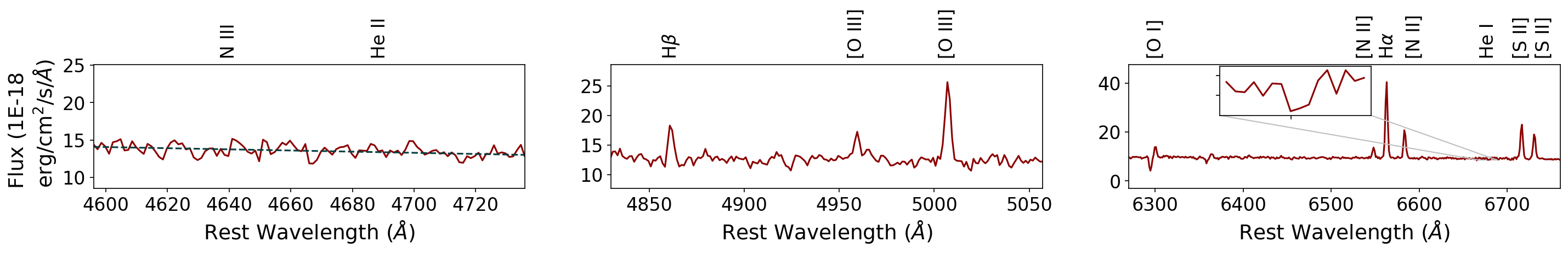}
\end{subfigure}\\
\begin{subfigure}
  \centering
  \includegraphics[height=0.25\columnwidth]{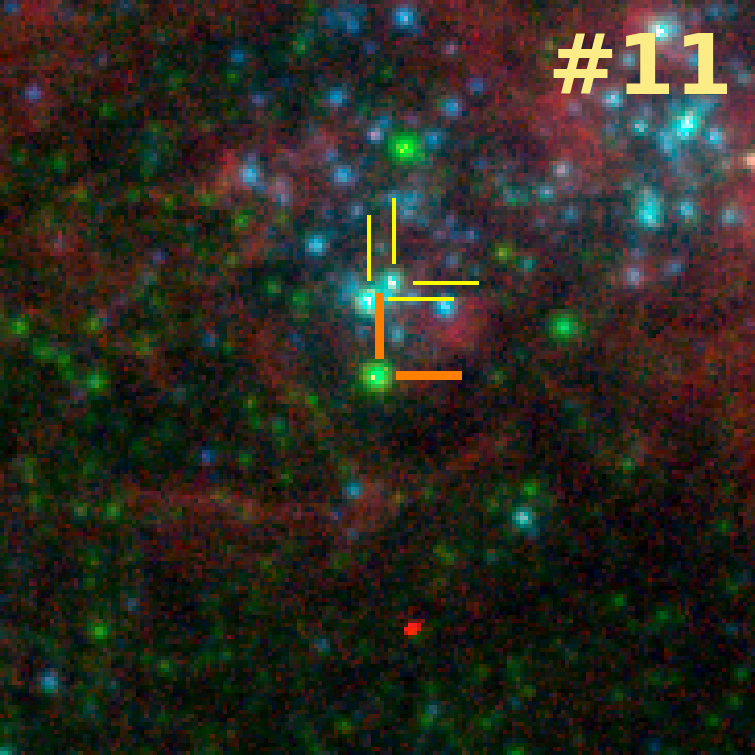}
\end{subfigure}
\begin{subfigure}
  \centering
  \includegraphics[height=0.25\columnwidth]{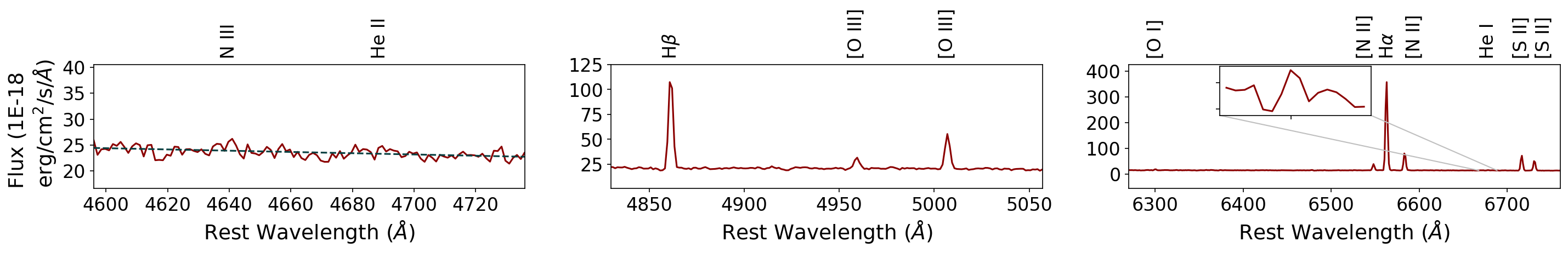}\\
\end{subfigure}
\caption{Postage stamps and portions of MUSE spectra of nine MUSE cLBVs whose ID is given in the upper-right corner of the postage stamp. The postage stamps are RGB composite images where blue corresponds to F275W + F336W, green to F438W + F814W, and red to F657N. At the adopted distance of 3.44 Mpc, 0.6" represent $\sim$10 pc. The markers are 10 pc in length and the images 120 pc on the side. Yellow tick marks correspond to other MUSE cLBVs in the field. The spectra are extracted from circular regions of 0.8" in diameter centred on the cLBVs. We present spectra which are corrected for extinction in the Milky Way and redshift. In the panel showing He\2, the dashed curve corresponds to a fit to the He\2 continuum. The inset on the right-most panel shows an enlargement of the He\1\,$\lambda6678$ line. The tick mark on the x-axis corresponds to 6678 \AA.}
\label{fig:batch1}
\end{figure*}

\begin{figure*}
\begin{subfigure}
  \centering
  \includegraphics[height=0.25\columnwidth]{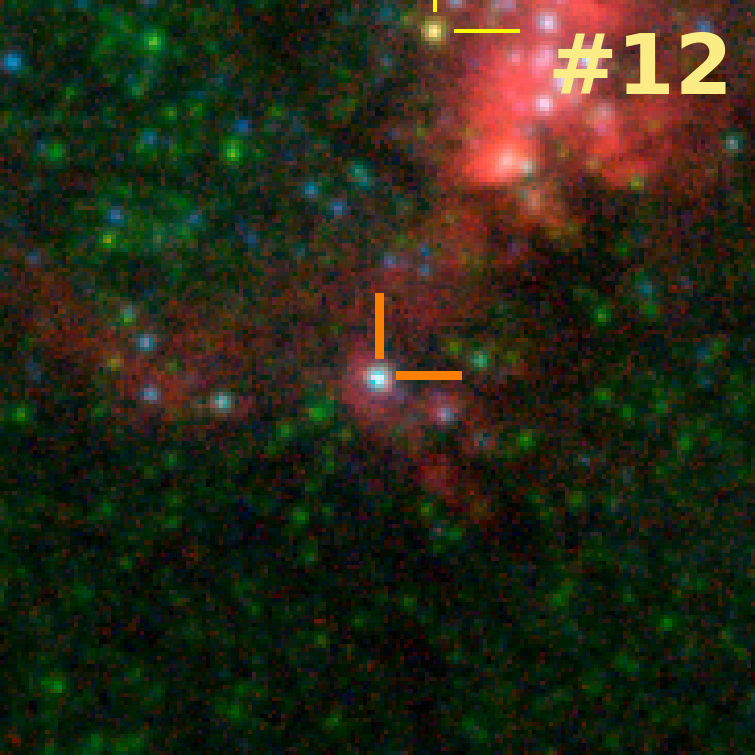}
\end{subfigure}
\begin{subfigure}
  \centering
  \includegraphics[height=0.25\columnwidth]{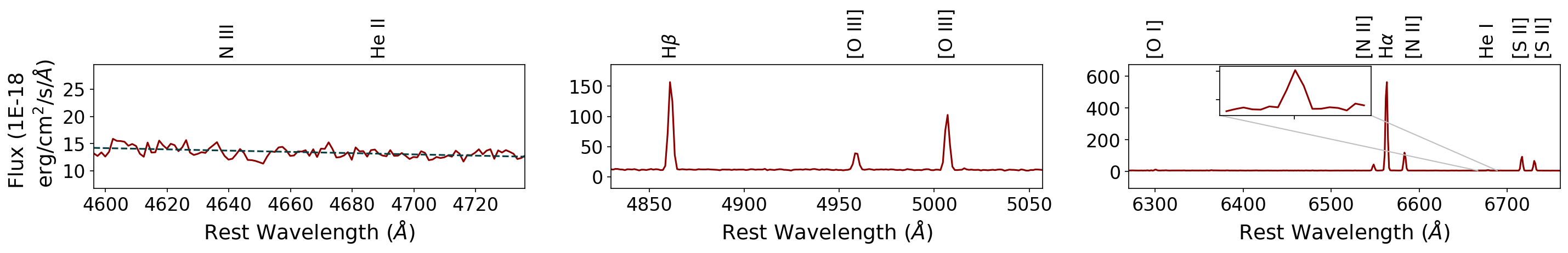}
\end{subfigure}\\
\begin{subfigure}
  \centering
  \includegraphics[height=0.25\columnwidth]{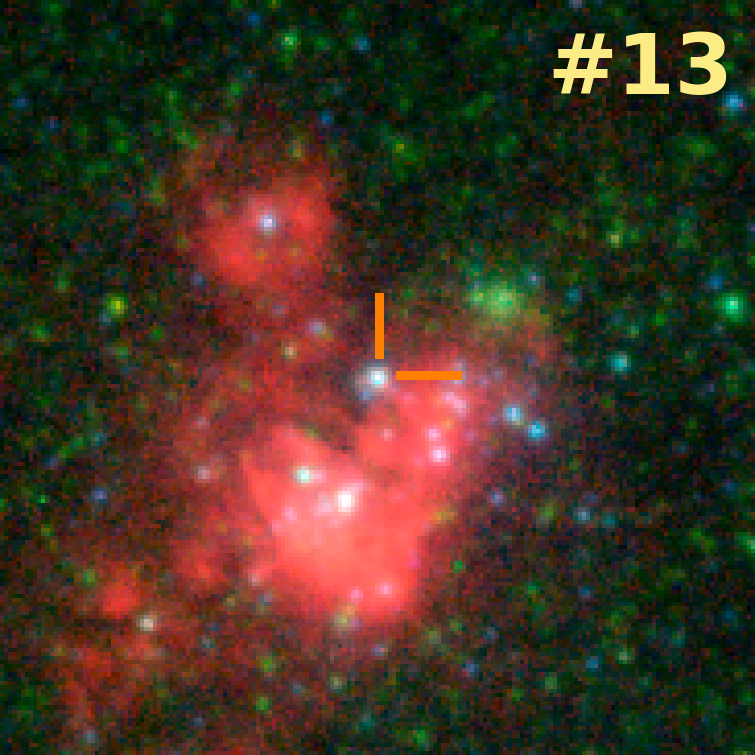}
\end{subfigure}
\begin{subfigure}
  \centering
  \includegraphics[height=0.25\columnwidth]{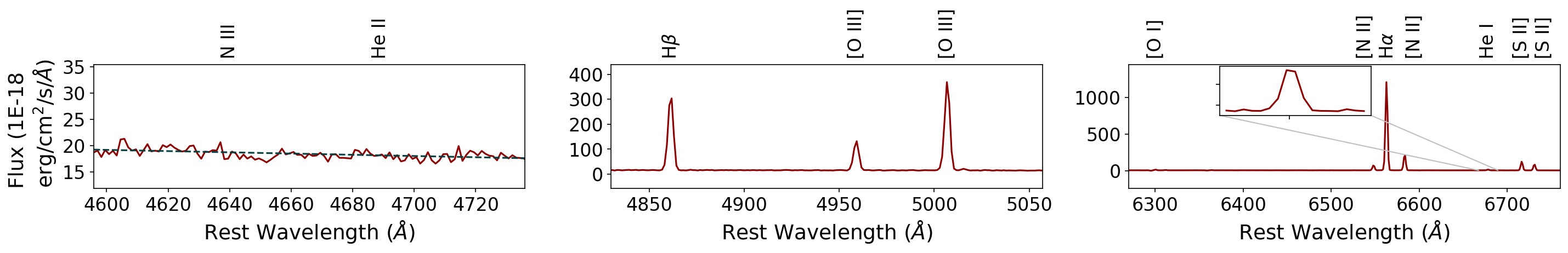}
\end{subfigure}\\
\begin{subfigure}
  \centering
  \includegraphics[height=0.25\columnwidth]{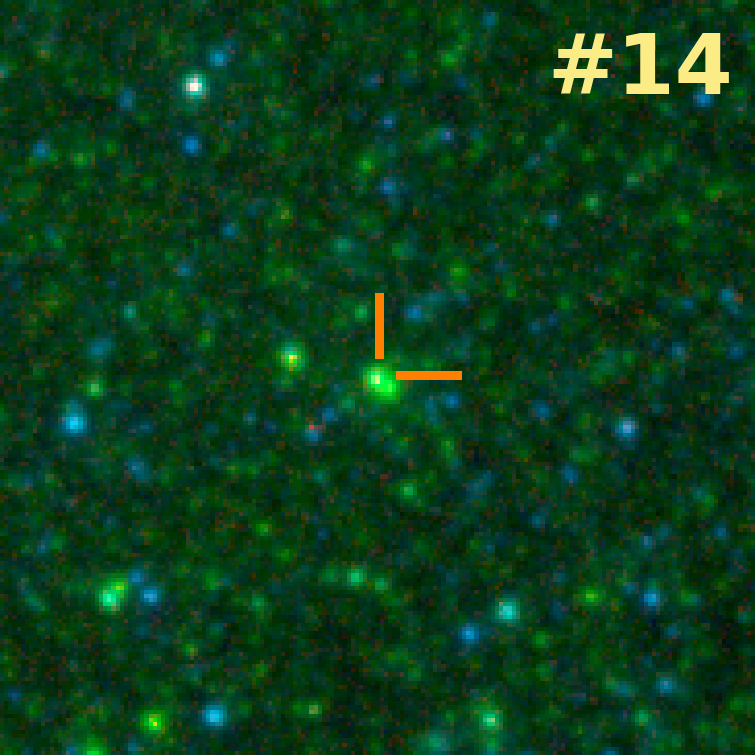}
\end{subfigure}
\begin{subfigure}
  \centering
  \includegraphics[height=0.25\columnwidth]{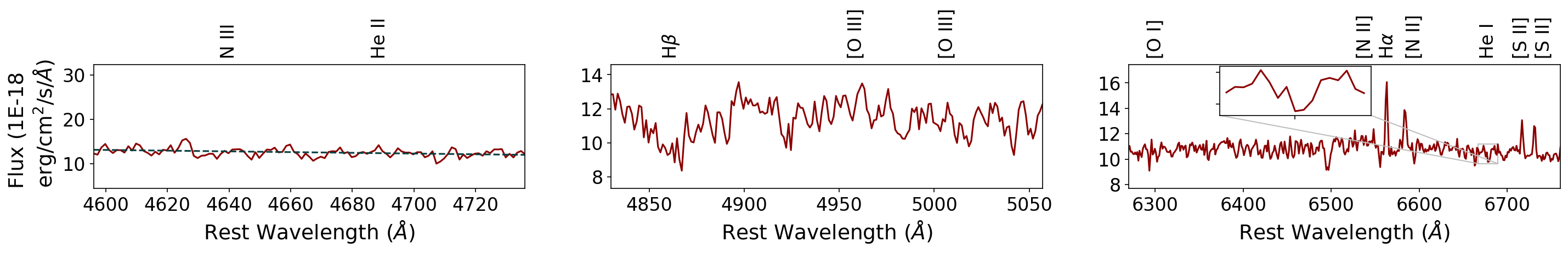}
\end{subfigure}\\
\begin{subfigure}
  \centering
  \includegraphics[height=0.25\columnwidth]{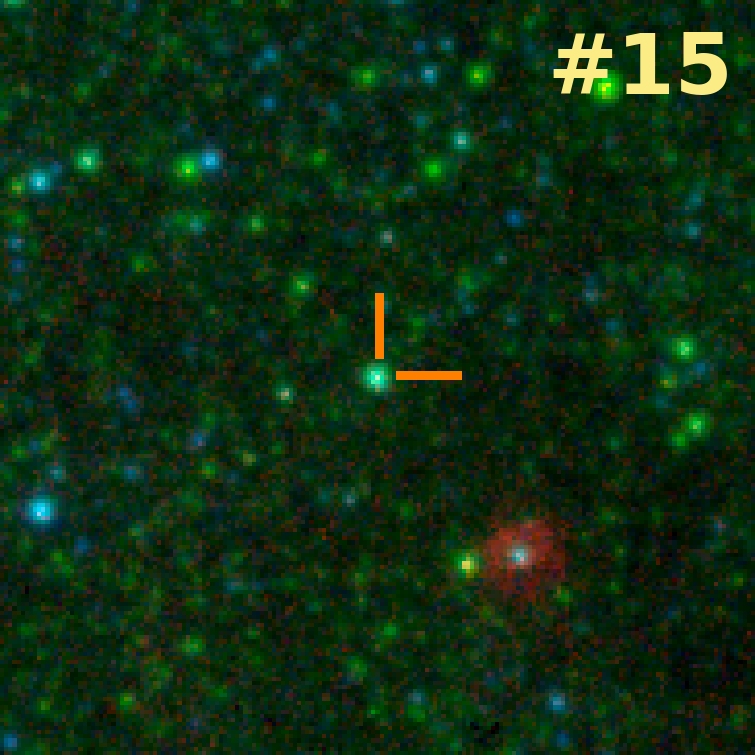}
\end{subfigure}
\begin{subfigure}
  \centering
  \includegraphics[height=0.25\columnwidth]{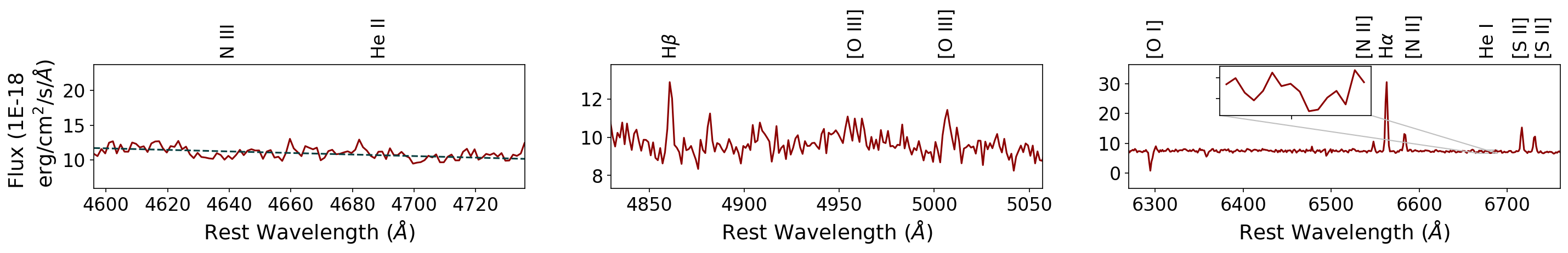}
\end{subfigure}\\
\begin{subfigure}
  \centering
  \includegraphics[height=0.25\columnwidth]{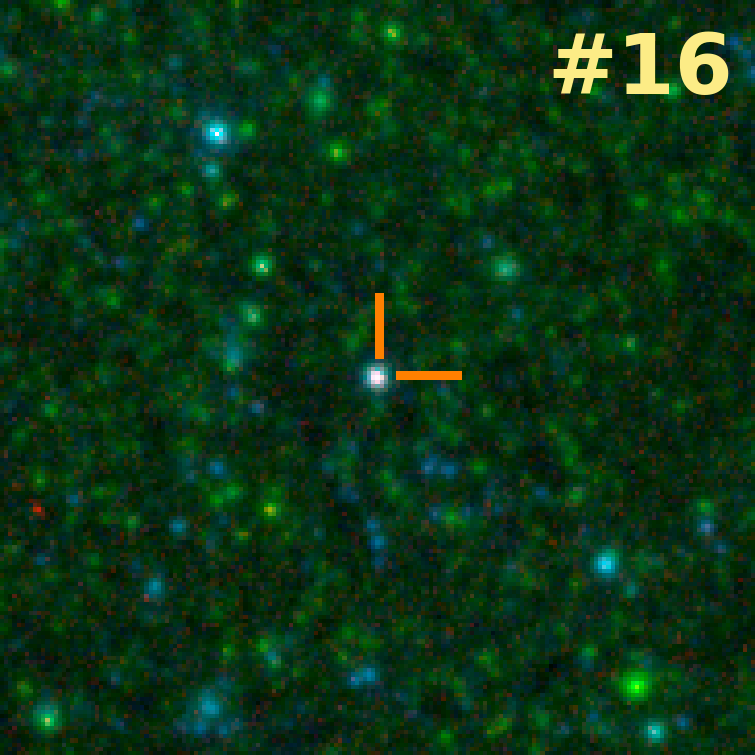}
\end{subfigure}
\begin{subfigure}
  \centering
  \includegraphics[height=0.25\columnwidth]{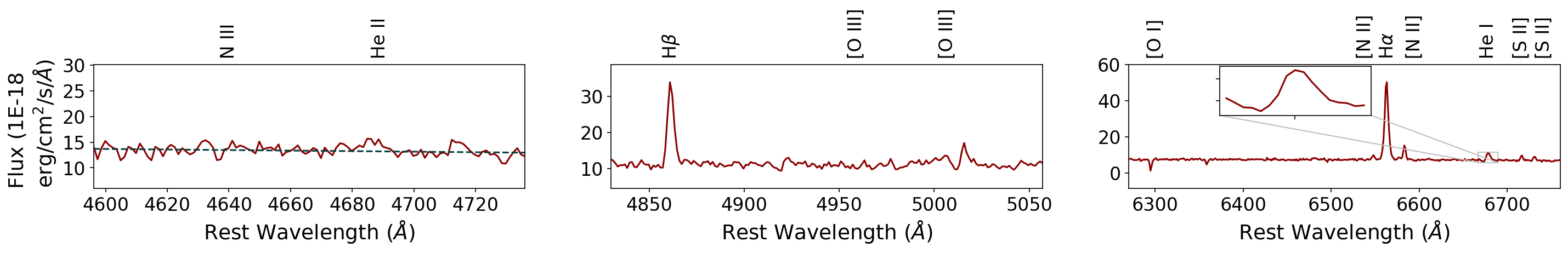}
\end{subfigure}\\
\begin{subfigure}
  \centering
  \includegraphics[height=0.25\columnwidth]{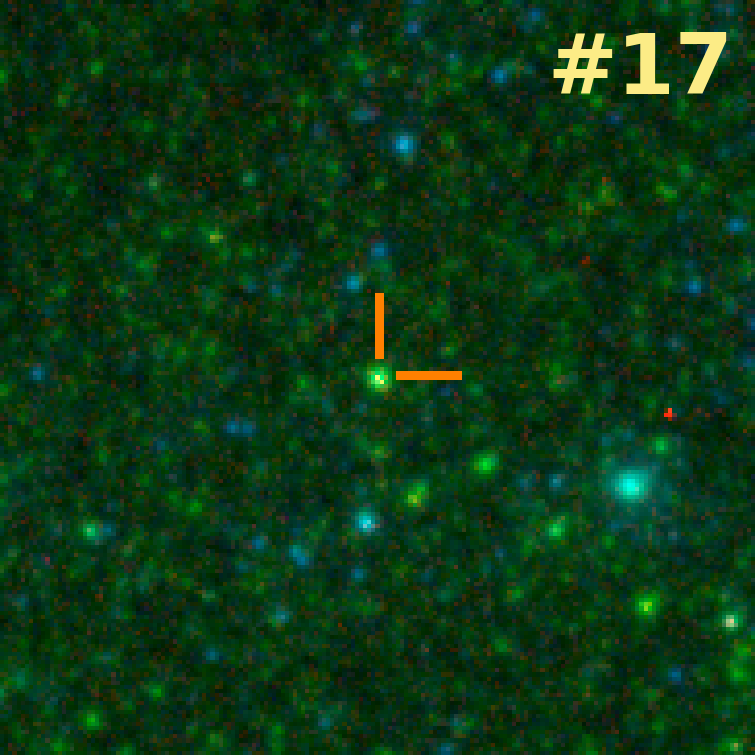}
\end{subfigure}
\begin{subfigure}
  \centering
  \includegraphics[height=0.25\columnwidth]{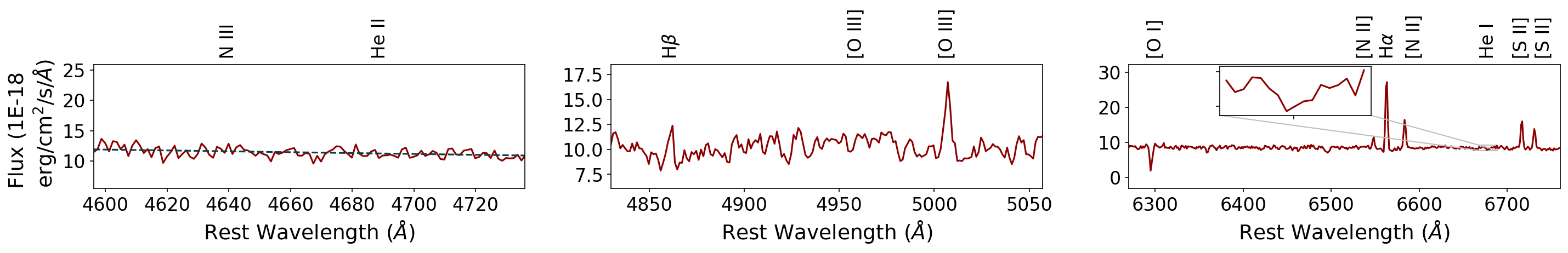}
\end{subfigure}\\
\begin{subfigure}
  \centering
  \includegraphics[height=0.25\columnwidth]{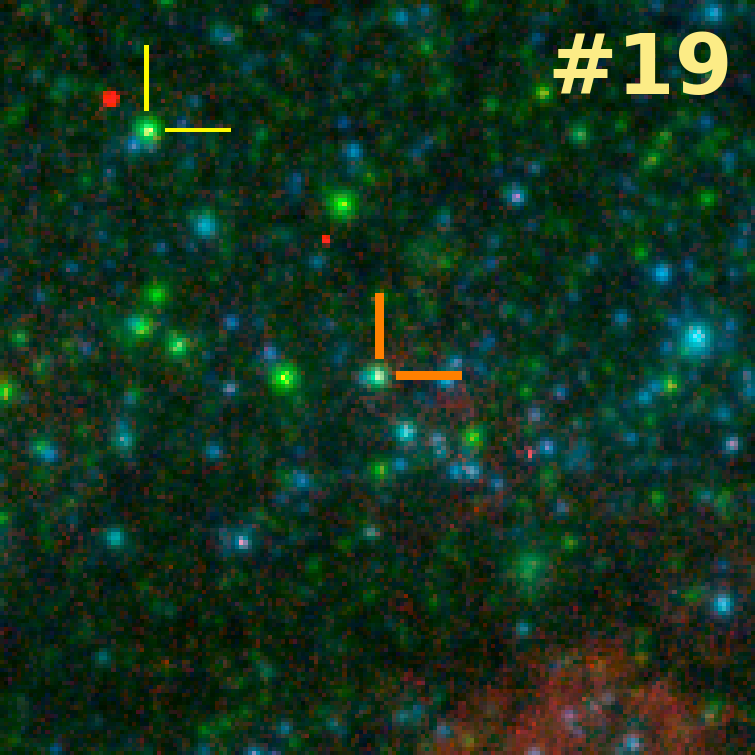}
\end{subfigure}
\begin{subfigure}
  \centering
  \includegraphics[height=0.25\columnwidth]{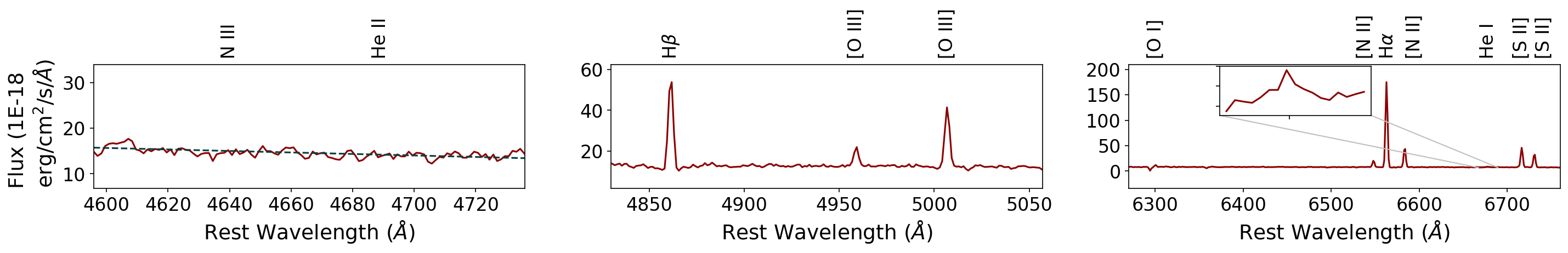}
\end{subfigure}\\
\begin{subfigure}
  \centering
  \includegraphics[height=0.25\columnwidth]{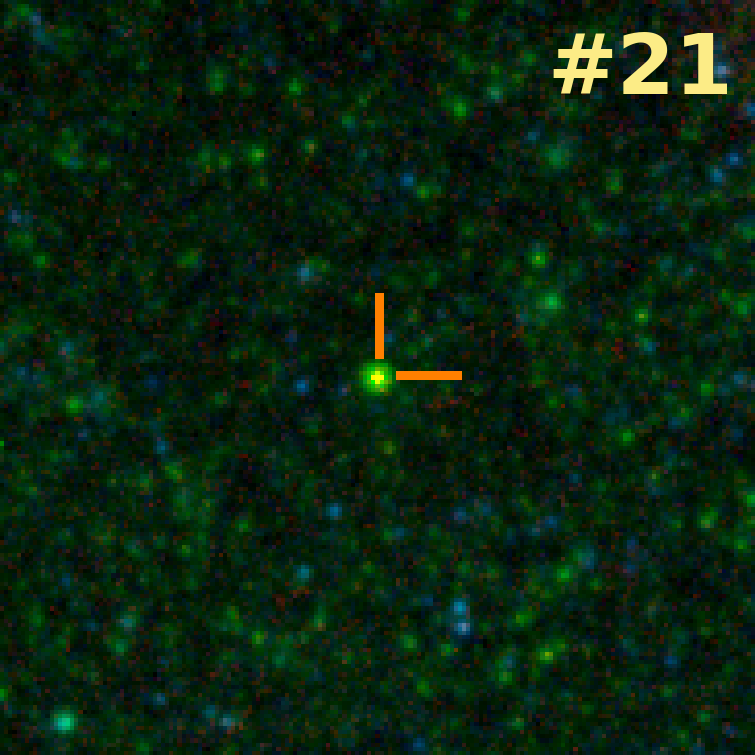}
\end{subfigure}
\begin{subfigure}
  \centering
  \includegraphics[height=0.25\columnwidth]{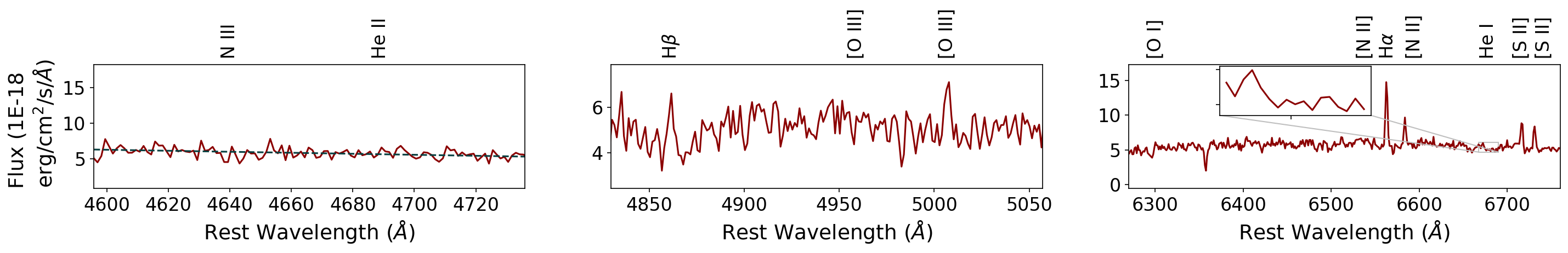}
\end{subfigure}\\
\begin{subfigure}
  \centering
  \includegraphics[height=0.25\columnwidth]{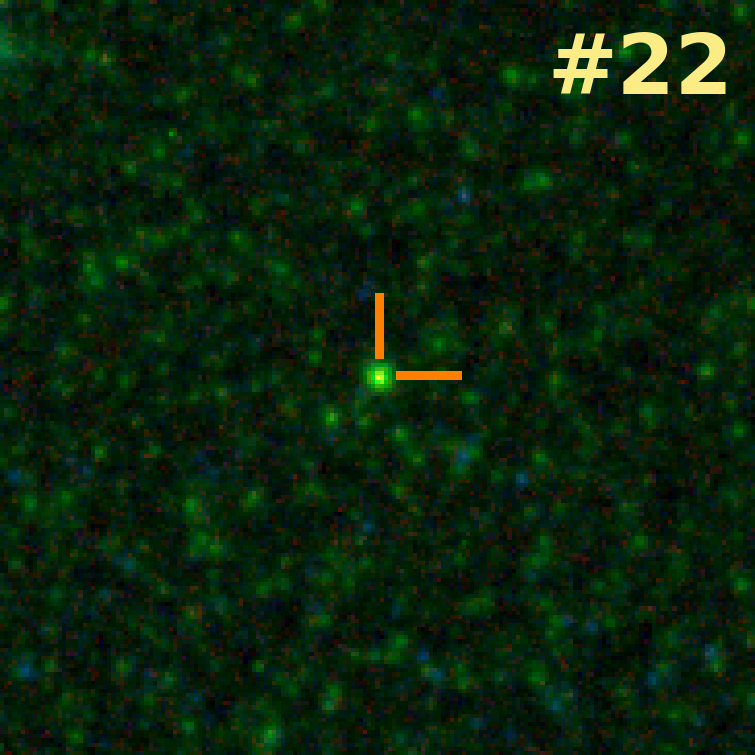}
\end{subfigure}
\begin{subfigure}
  \centering
  \includegraphics[height=0.25\columnwidth]{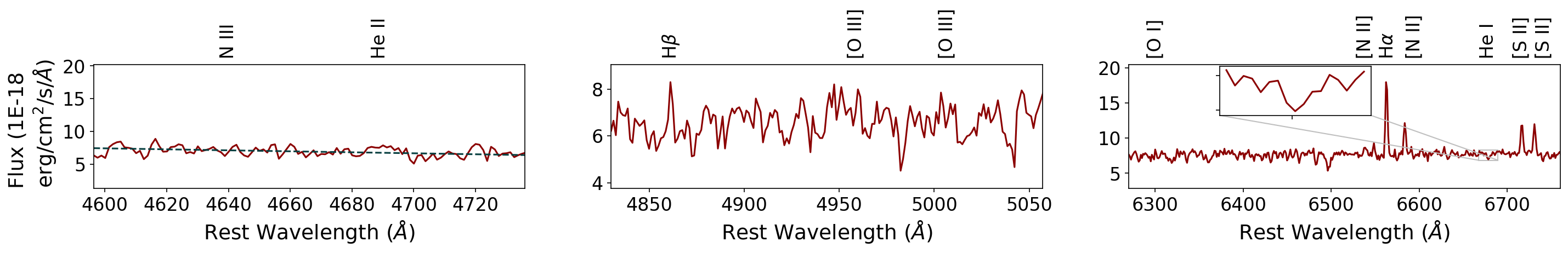}\\
\end{subfigure}
\caption{Same as Figure~\ref{fig:batch1} but for a different set of MUSE cLBVs.}
\label{fig:batch2}
\end{figure*}

\begin{figure*}
\begin{subfigure}
  \centering
  \includegraphics[height=0.25\columnwidth]{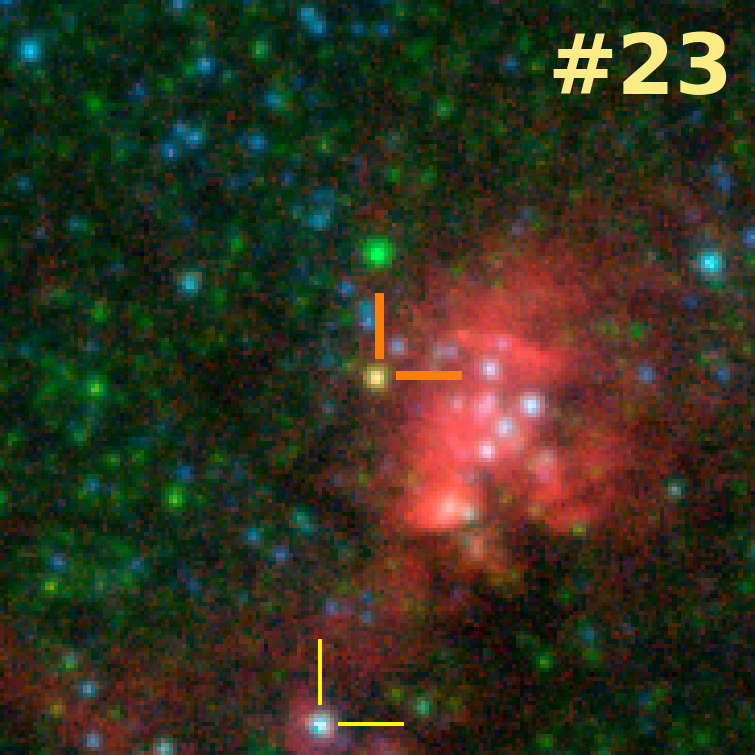}
\end{subfigure}
\begin{subfigure}
  \centering
  \includegraphics[height=0.25\columnwidth]{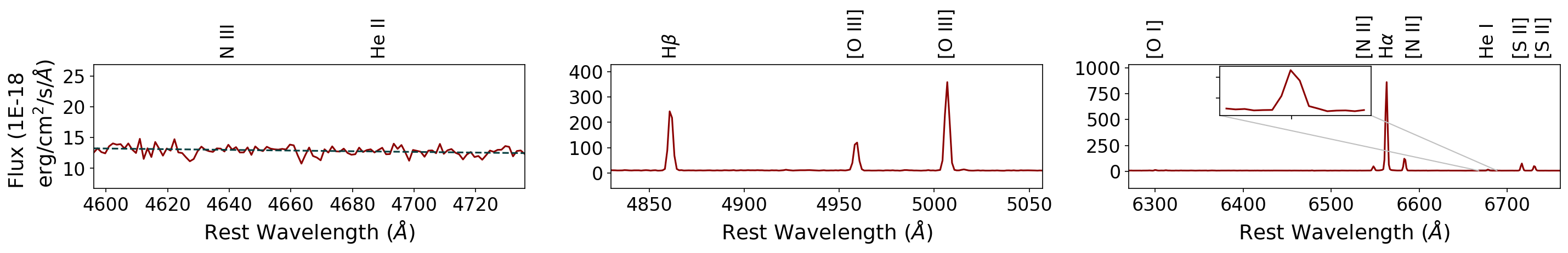}
\end{subfigure}\\
\begin{subfigure}
  \centering
  \includegraphics[height=0.25\columnwidth]{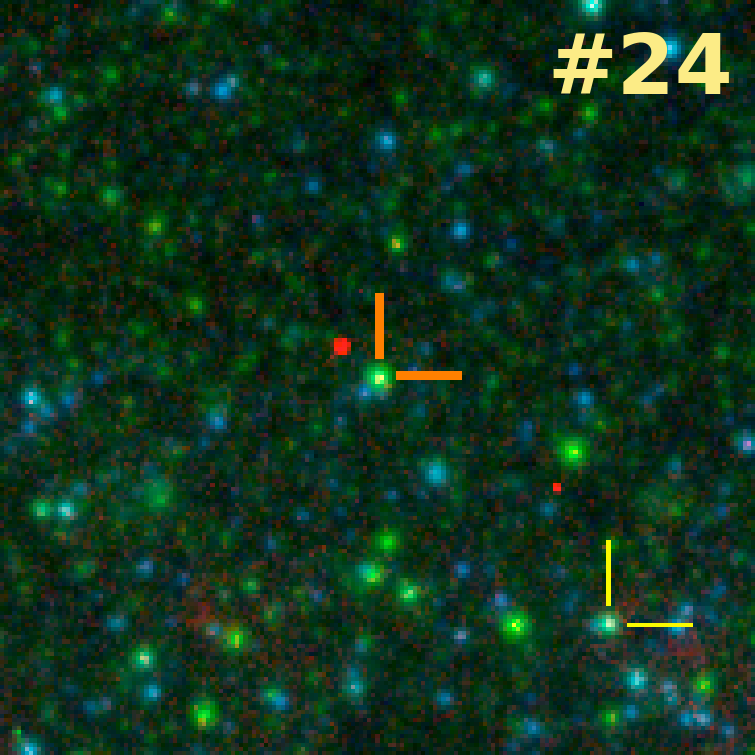}
\end{subfigure}
\begin{subfigure}
  \centering
  \includegraphics[height=0.25\columnwidth]{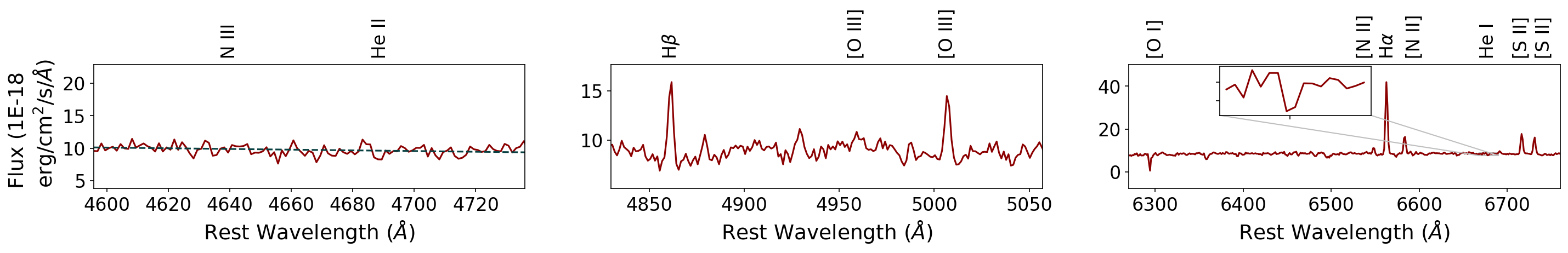}
\end{subfigure}\\
\begin{subfigure}
  \centering
  \includegraphics[height=0.25\columnwidth]{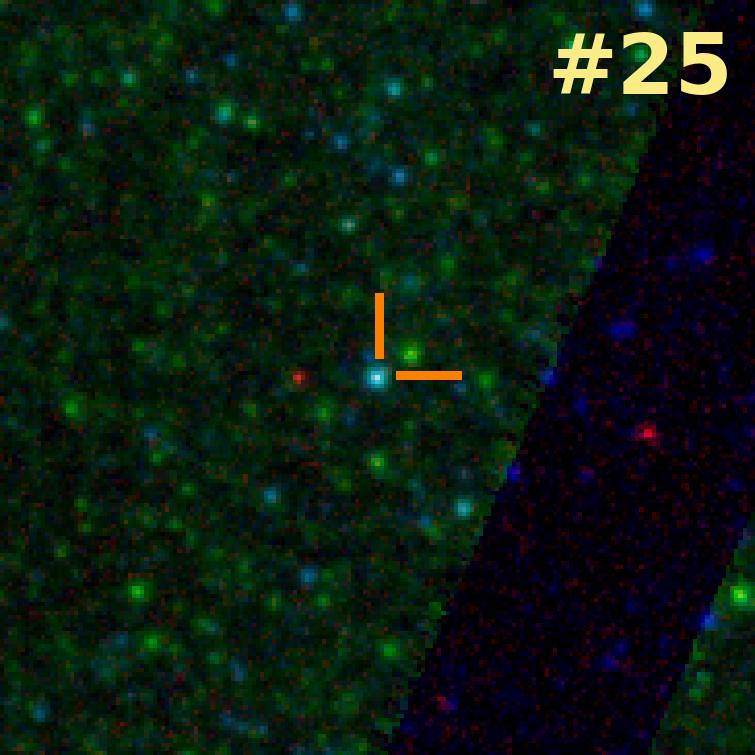}
\end{subfigure}
\begin{subfigure}
  \centering
  \includegraphics[height=0.25\columnwidth]{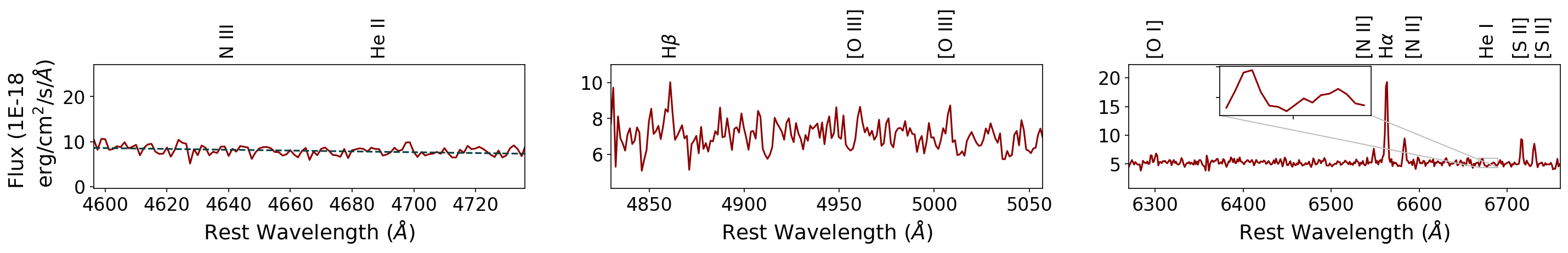}
\end{subfigure}\\
\begin{subfigure}
  \centering
  \includegraphics[height=0.25\columnwidth]{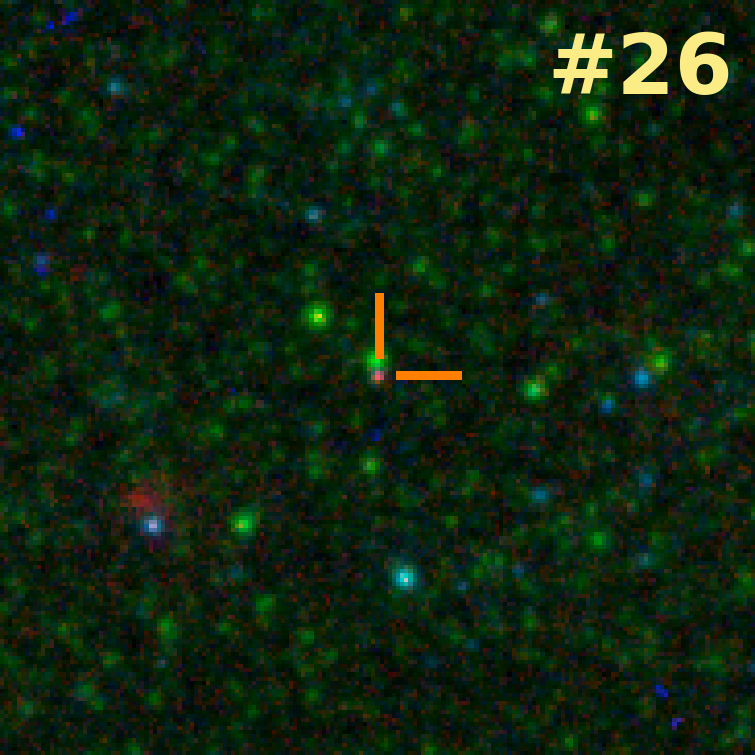}
\end{subfigure}
\begin{subfigure}
  \centering
  \includegraphics[height=0.25\columnwidth]{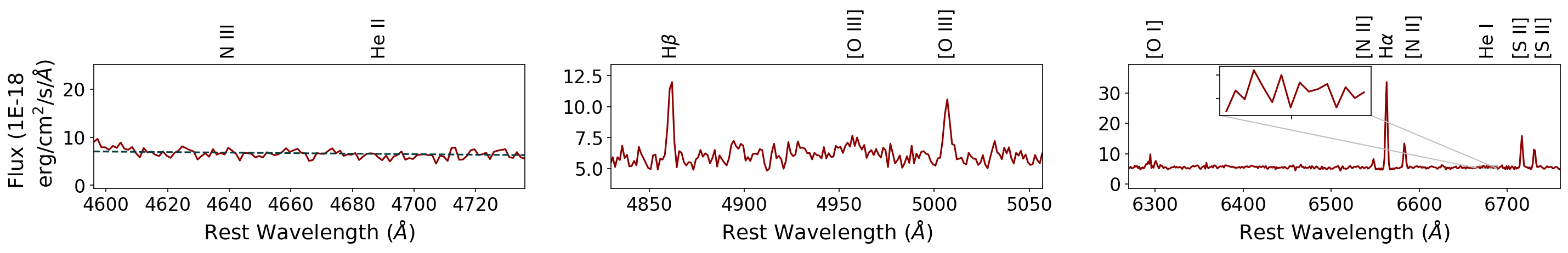}
\end{subfigure}\\
\begin{subfigure}
  \centering
  \includegraphics[height=0.25\columnwidth]{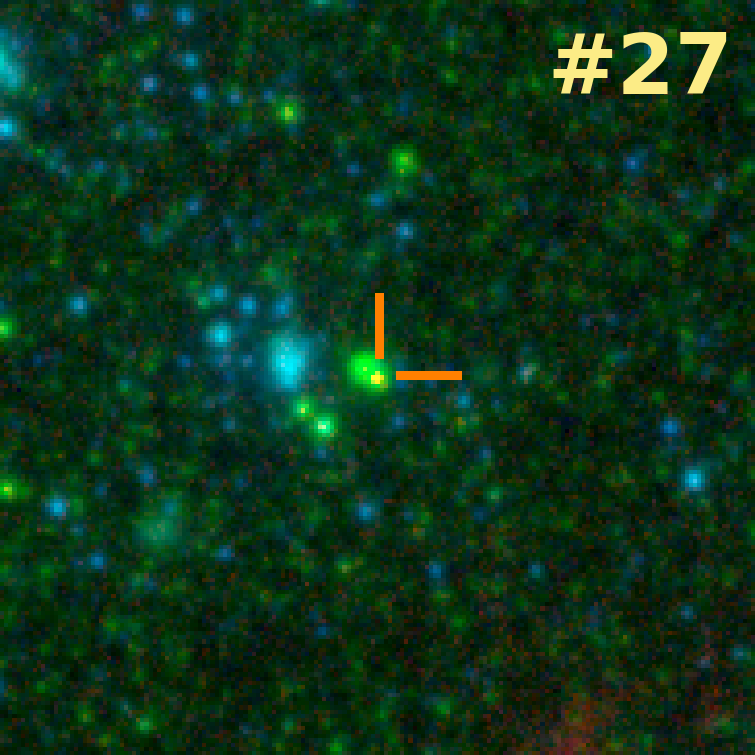}
\end{subfigure}
\begin{subfigure}
  \centering
  \includegraphics[height=0.25\columnwidth]{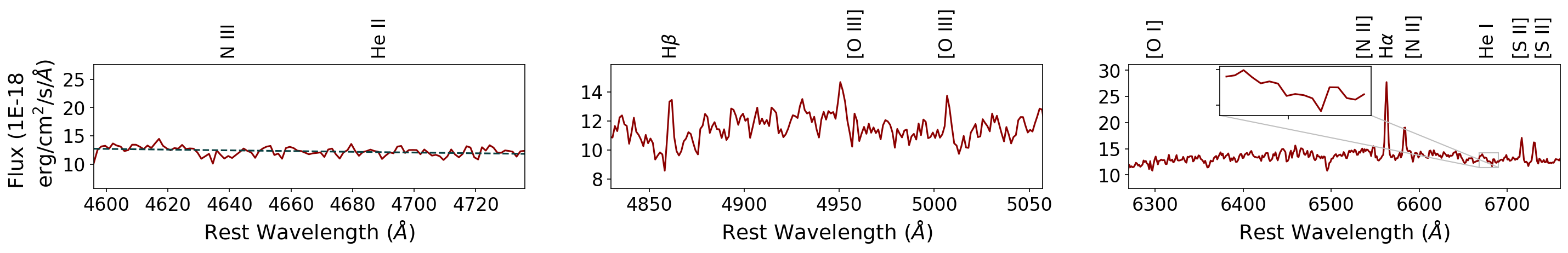}
\end{subfigure}\\
\begin{subfigure}
  \centering
  \includegraphics[height=0.25\columnwidth]{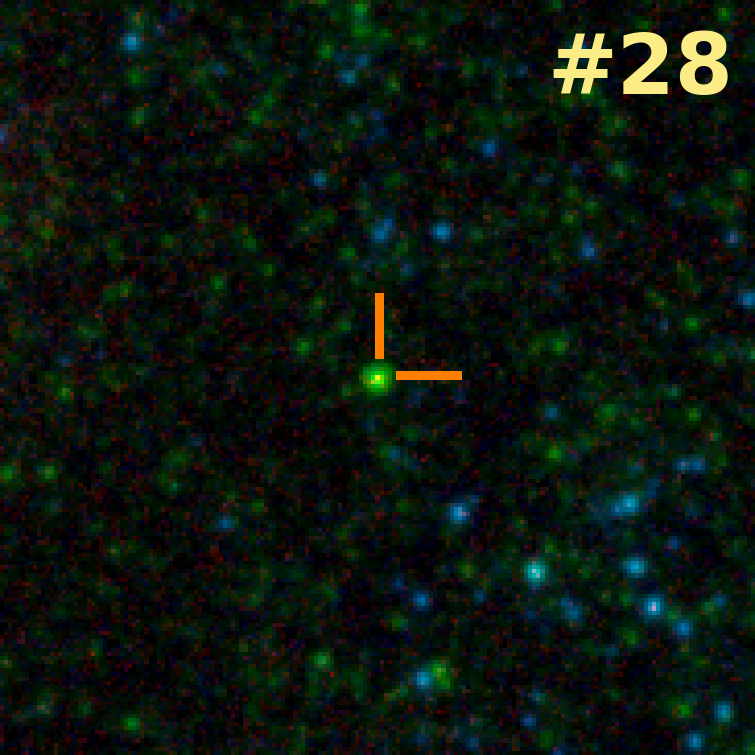}
\end{subfigure}
\begin{subfigure}
  \centering
  \includegraphics[height=0.25\columnwidth]{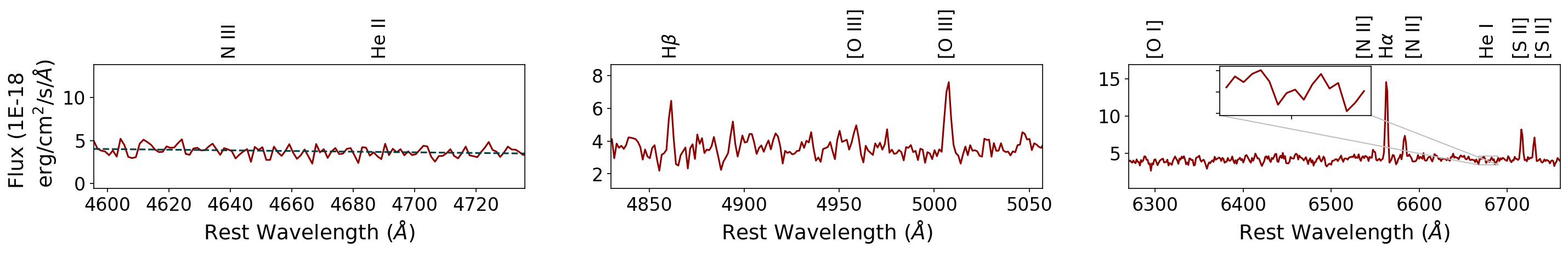}
\end{subfigure}\\
\begin{subfigure}
  \centering
  \includegraphics[height=0.25\columnwidth]{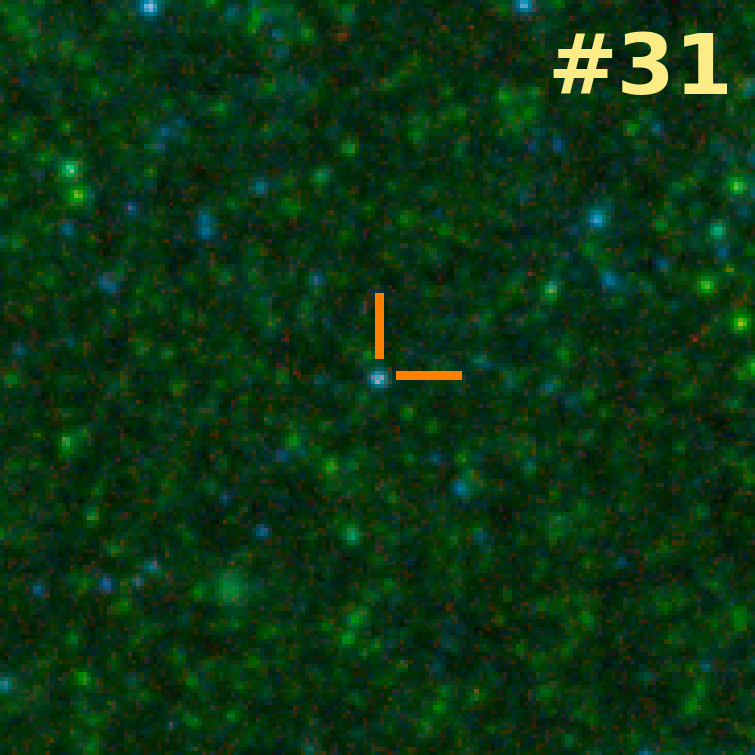}
\end{subfigure}
\begin{subfigure}
  \centering
  \includegraphics[height=0.25\columnwidth]{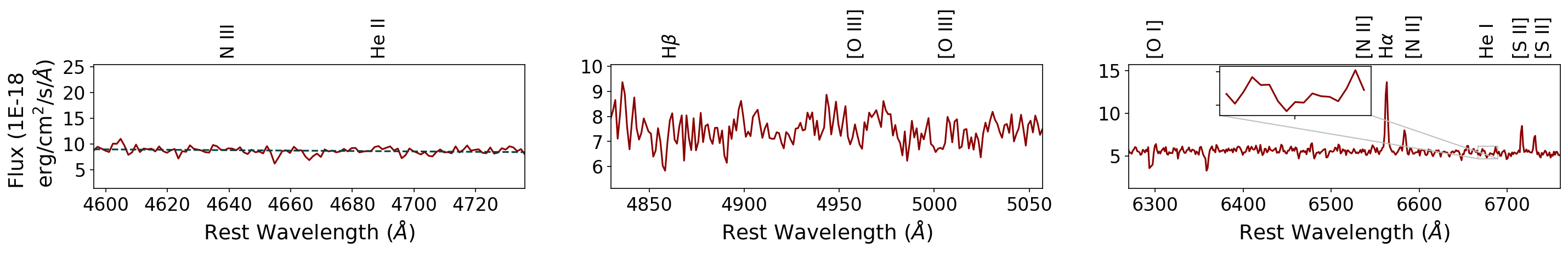} 
\end{subfigure}\\
\begin{subfigure}
  \centering
  \includegraphics[height=0.25\columnwidth]{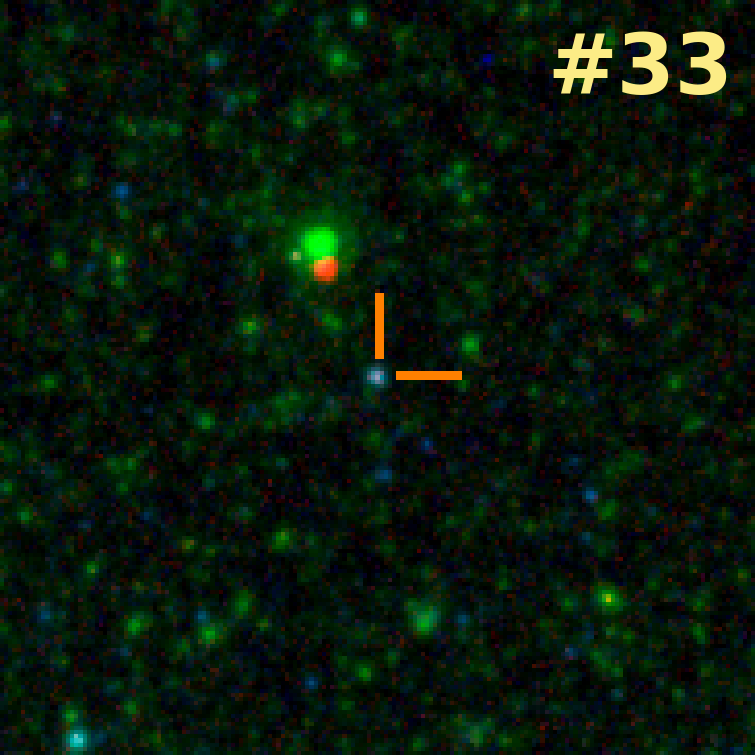}
\end{subfigure}
\begin{subfigure}
  \centering
  \includegraphics[height=0.25\columnwidth]{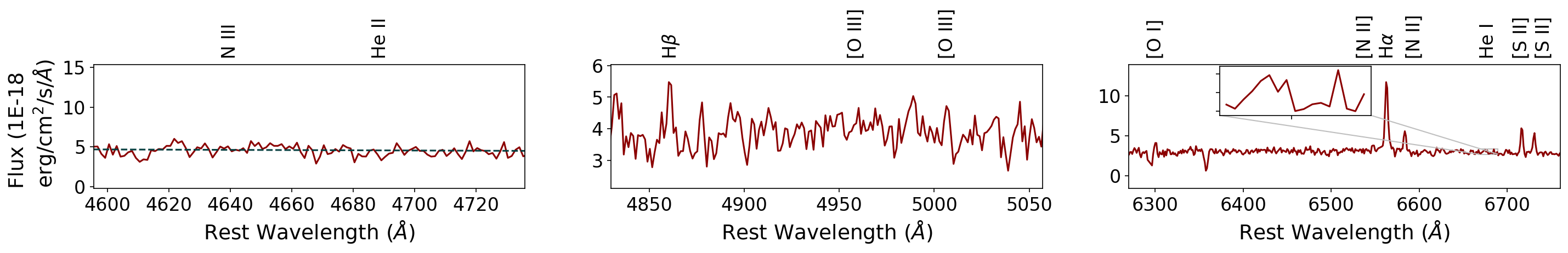}
\end{subfigure}\\
\begin{subfigure}
  \centering
  \includegraphics[height=0.25\columnwidth]{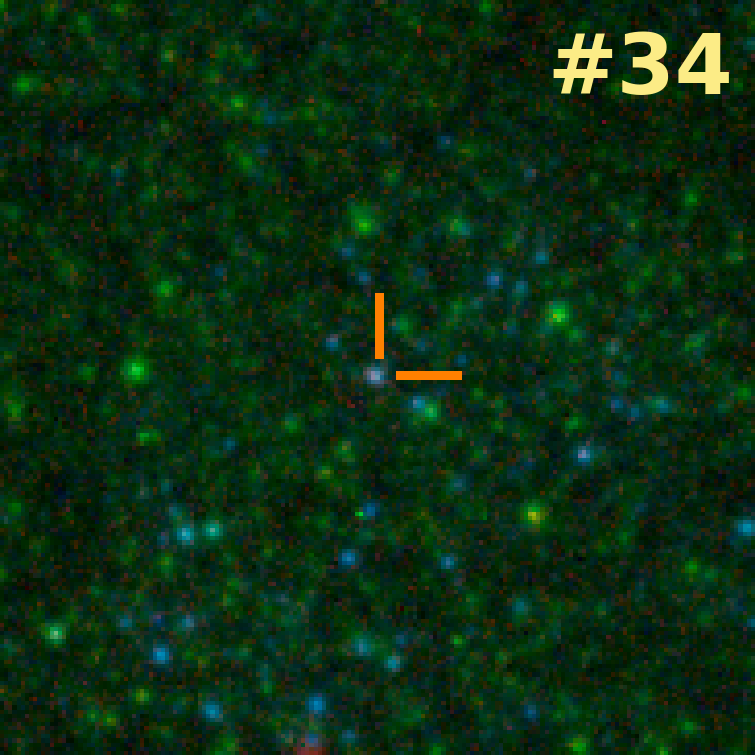}
\end{subfigure}
\begin{subfigure}
  \centering
  \includegraphics[height=0.25\columnwidth]{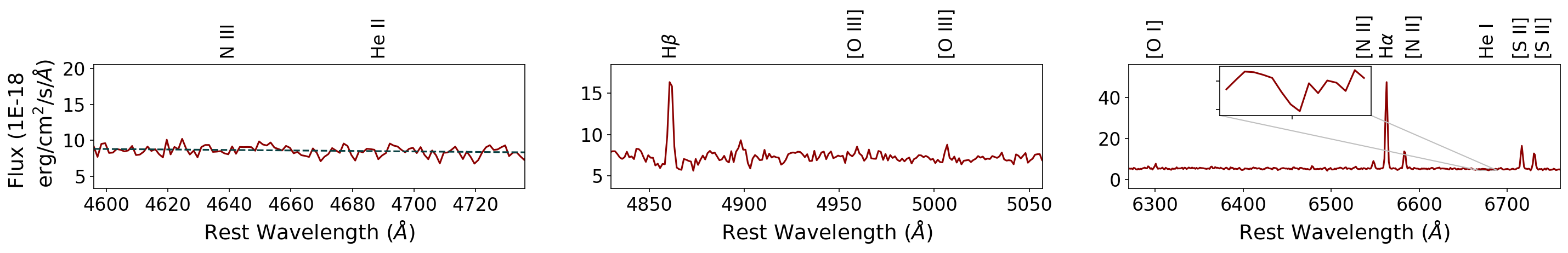}\\
\end{subfigure}
\caption{Same as Figures~\ref{fig:batch1} and~\ref{fig:batch2} but for a different set of MUSE cLBVs.}
\label{fig:batch3}
\end{figure*}

Unfortunately, the strongest He\1 line in the optical, He\1$\,\lambda$5876, falls in the AO laser gap of the MUSE observations and thus was not observed. For this reason, we look at He\1$\,\lambda$6678. As shown in column 3 of Table~\ref{tab:results2}, among MUSE cLBVs \#1 to \#7, He\1$\,\lambda$6678 emission is only observed from \#4 and \#5. The He\1$\,\lambda$6678 line is shown as an inset in the rightmost panel of Figures~\ref{fig:batch1} to~\ref{fig:batch4}.

Candidate LBVs may also show optical emission lines or P-Cygni like profiles of Fe\2 and Fe\3 and are generally distinguished by these ``unusual" features from other H$\alpha$-bright objects. Examples of these lines in cLBVs can be seen in \cite{Castro2008}. We attempted to remove the background light from the spectra of the candidates to see if we could detect faint features in their spectra. For this purpose, we computed the background from a ring around the candidate, which had an inner radius equal to 3/2 times sigma, where sigma is the standard deviation of the Gaussian which we fitted to the radial profile of the candidate's MUSE continuum-subtracted H$\alpha$ emission. The ring had a width of one pixel. We found the median spectrum of all pixels within the ring and multiplied it by the number of pixels within the area of extraction of the cLBV spectrum. Background subtraction using this method was difficult because the surrounding light is spatially inhomogeneous and can be brighter than the cLBV continuum. In the end, we decided not to do a background subtraction, except for sources \#1 through \#7, where there is not much evidence for H$\alpha$ contamination from the surrounding area, and which seem dominated by very local emission line spectra. We detect no Fe lines in the spectra of candidates \#1 to \#7, even after subtracting the nearby light from the background in order to see fainter features. 

\cite{Humphreys2017a} recently published a comparison of luminous stars in M31 and M33 and how to tell them apart. According to their research, confirmed LBVs: i) do not have [O\1]$\lambda6300$ emission lines in their spectra; ii) have [Fe\2] emission lines sometimes; iii) show free-free emission in the near-infrared but no evidence for warm dust; and iv) show S Dor-type variability (this is the most important and defining characteristic). As can be seen in column 5 of Table~\ref{tab:results2}, only candidates \#4 and \#5 show  [O\1]$\lambda6300$ emission in their spectra. As shown in column 4 of the same table, these two same candidates show He\2 emission. Hereafter, we exclude cLBVs \#4 and \#5 from our list of strong MUSE cLBVs.

\cite{Humphreys2017a} also say that cLBVs should share the spectral characteristics of confirmed LBVs with low outflow velocities (measured from the absorption component of the P-Cygni profile) and the lack of warm circumstellar dust (excess at 3.5 and 8 microns reveal the presence of warm circumstellar dust). Luminous Blue Variables are well known to have low wind speeds of 100 to 200 km\,s$^{-1}$ during their eruptions or maximum light phase and LBVs in quiescence are typically 50 to 100 km\,s$^{-1}$. Candidate LBV \#2 has a P-Cygni profile in H$\alpha$ with a  an absorption minimum which is blueshifted by 222 km\,s$^{-1}$. We do not have either infrared data at the required spatial resolution to check for the lack of circumstellar dust. For something to be classified as an LBV, some people also require that there is an H$\alpha$+[N\2] shell nebula, which is suggestive of a previous giant eruption. We do not have the spatial resolution to observe such shell. 

\cite{Smith1998} used the Faint Object Spectrogaph on \hst~to study the nebulae around the known LBVs, R127 and S119, in the LMC. The spectra of these nebulae show strong [N\2], possibly due to N enrichment, relative to H\2 regions, and that the electron density is high, with the [S\2]6716/[S\2]6731 ratio near $\sim1$. We measured this ratio for the MUSE cLBVs. 

Another characteristic of LBVs is that their [O\3]$\,\lambda$5007 is generally low relative to that of H\2 regions. In order to use this fact as a tracer of cLBVs, one has of course to consider the metallicity gradient of the galaxy being studied and that presumably, the [O\3]$\,\lambda$5007 strength varies as the star cycles between hot and cooler phases.

We report our measurements of the line flux ratios, [O\3]$\,\lambda$5007/H$\beta$, [N\2]6584/H$\alpha$, and [S\2]6716/[S\2]6731 in columns 6, 7, and 8 of Table~\ref{tab:results2}. The last three rows of the table give the mean, median and standard deviation around the mean computed from the values of the entire column. The last column of Table~\ref{tab:results} shows that about 2/3 of the MUSE cLBVs have spectra which are contaminated by H\2 regions. Figure~\ref{fig:batch1} and column 6 of Table~\ref{tab:results2} show that cLBVs \#1, \#2, \#3, \#6, and \#7 have weak or no [O\3]$\,\lambda$5007 emission relative to the mean of the column. Column 7 of Table~\ref{tab:results2} shows that cLBVs \#1 and \#2 have above average [N\2]6584/H$\alpha$ ratios relative to the 36 MUSE cLBVs. This is consistent with the N-enriched shells around LBVs. Finally, column 8 of Table~\ref{tab:results2} shows that cLBVs \#1 and \#2 have below average [S\2]6716/[S\2]6731 ratios of 1.22 and 1.11, respectively, similar to the shells studied by \cite{Smith1998}. In summary, we have found five strong cLBVs (\#1, \#2, \#3, \#6, and \#7) among which \#1 and \#2 are the strongest.

\subsection{A word of caution}

Confirmed LBVs AG Car and R127 are Ofpe/WN9 stars when in their hot state. Thus, in principle, any Ofpe/WN9 star could become an LBV as far as we know. Of(pe)/WNh stars will show He\2 $\lambda$4686 and strong [N\2] emission lines. 

While all Ofpe/WN9 stars could be cLBVs, there are other stars that can be cLBVs.  At the lower luminosity end, the S Doradus instability strip is at lower temperatures, so quiescent/hot LBVs (at visual minimum brightness) are not so hot, and they do not have He\2 emission. Some do not even have He\1 emission, and some have pretty weak Balmer emission.  In other words, having some criteria to select LBVs (like an H$\alpha$ equivalent width stronger than some value, or the presence of an Ofpe/WN9 spectrum) will not produce a complete sample. There are some stars that might be dormant LBVs that will escape our criteria, especially at the lower luminosity end, i.e., at log(L/L$_\odot$)=5.3 to 5.8.

\subsection{Candidate Wolf-Rayet stars}

\begin{figure*}
\begin{subfigure}
  \centering
  \includegraphics[height=0.25\columnwidth]{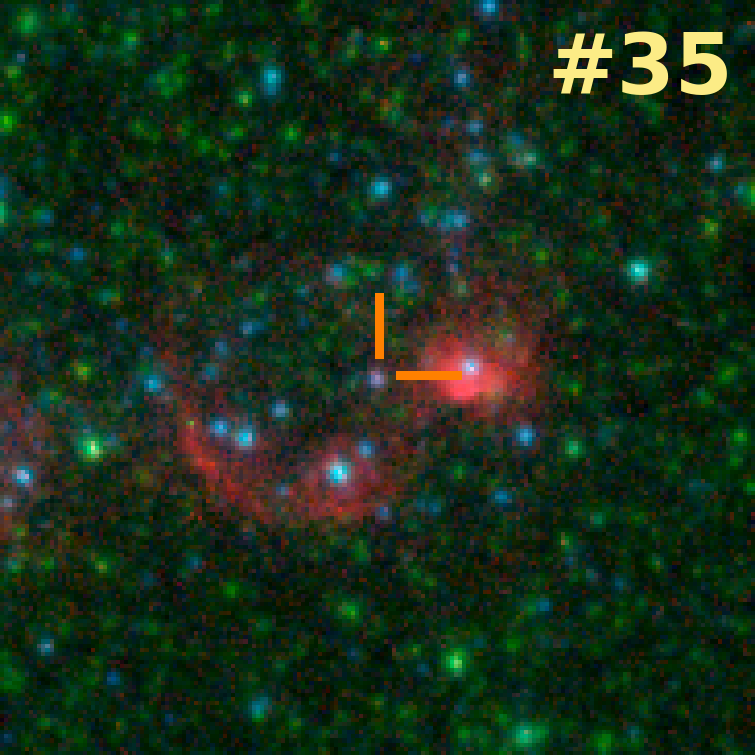}
\end{subfigure}
\begin{subfigure}
  \centering
  \includegraphics[height=0.25\columnwidth]{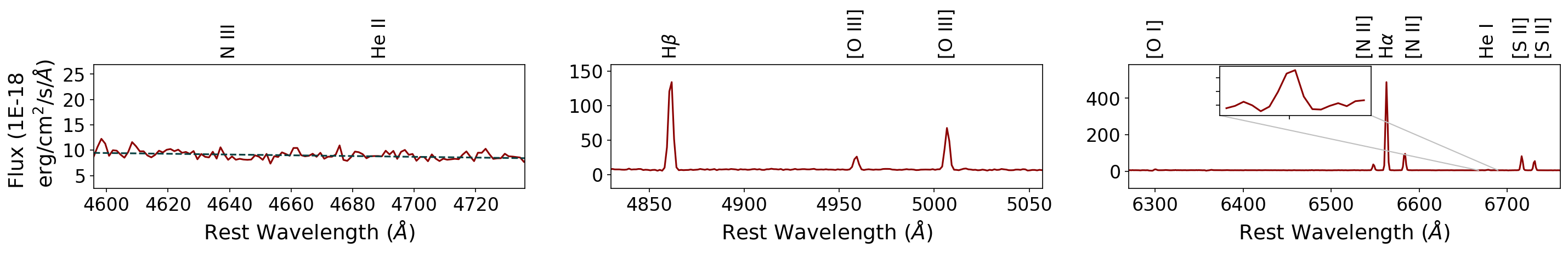}
\end{subfigure}\\
\begin{subfigure}
  \centering
  \includegraphics[height=0.25\columnwidth]{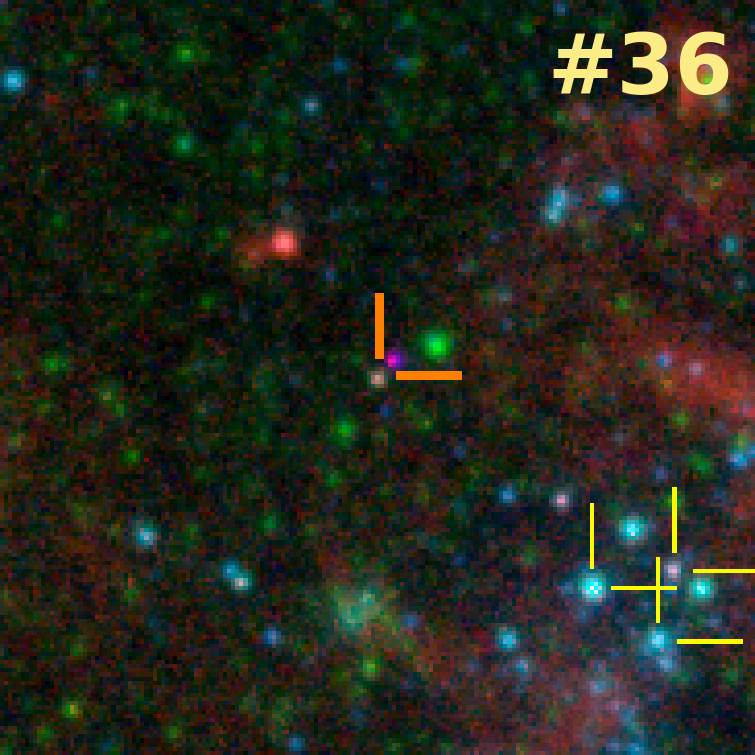}
\end{subfigure}
\begin{subfigure}
  \centering
  \includegraphics[height=0.25\columnwidth]{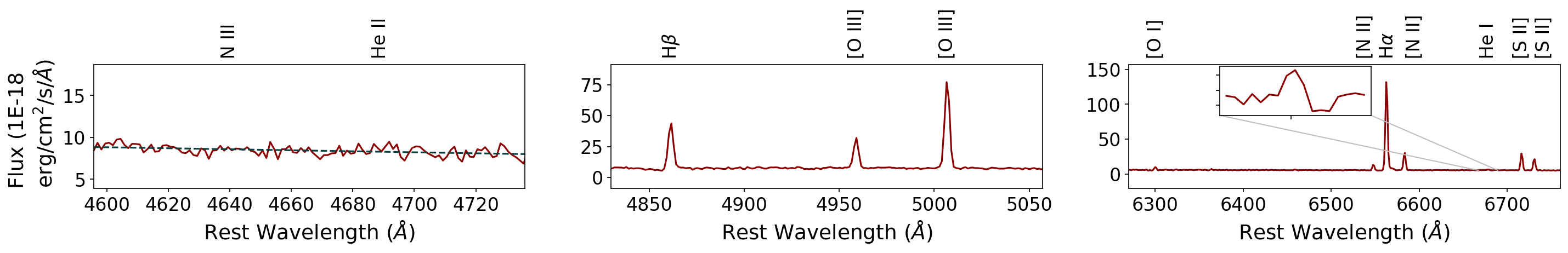}
\end{subfigure}\\
\begin{subfigure}
  \centering
  \includegraphics[height=0.25\columnwidth]{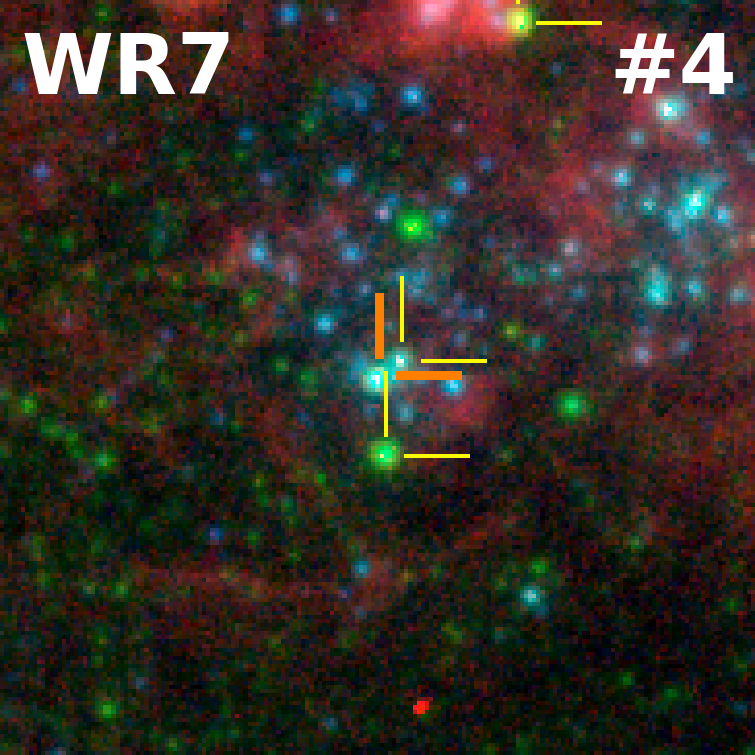}
\end{subfigure}
\begin{subfigure}
  \centering
  \includegraphics[height=0.25\columnwidth]{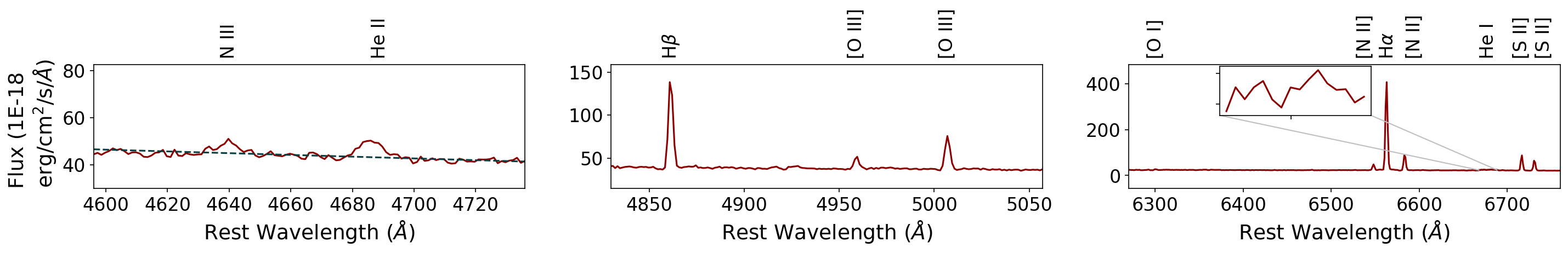}
\end{subfigure}\\
\begin{subfigure}
  \centering
  \includegraphics[height=0.25\columnwidth]{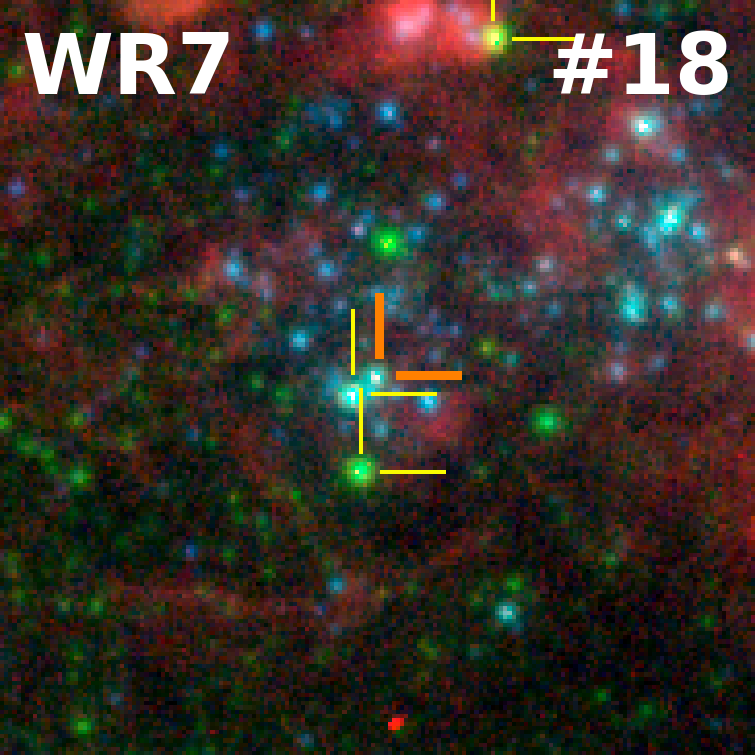}
\end{subfigure}
\begin{subfigure}
  \centering
  \includegraphics[height=0.25\columnwidth]{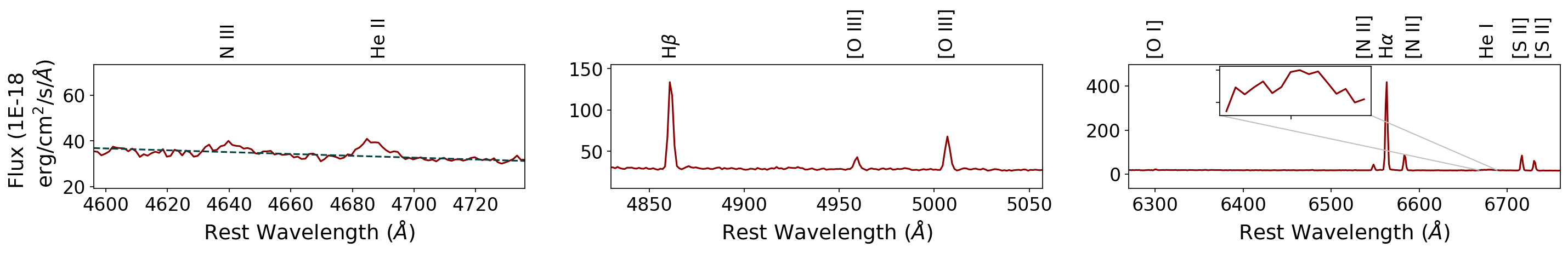}
\end{subfigure}\\
\begin{subfigure}
  \centering
  \includegraphics[height=0.25\columnwidth]{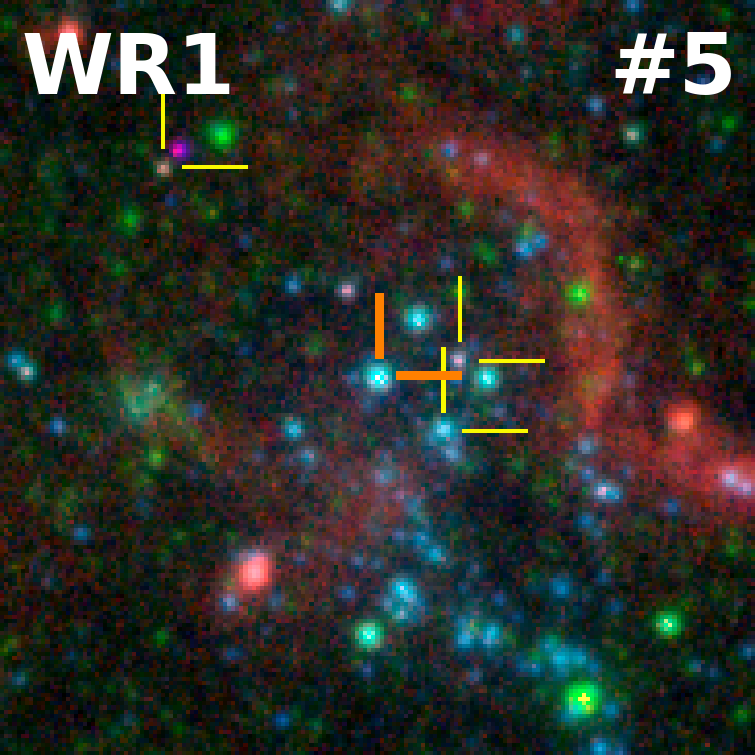}
\end{subfigure}
\begin{subfigure}
  \centering
  \includegraphics[height=0.25\columnwidth]{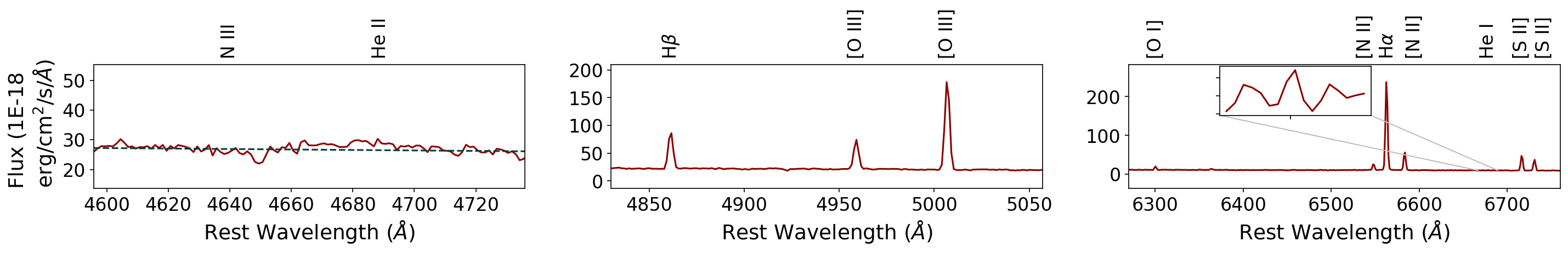}
\end{subfigure}\\
\begin{subfigure}
  \centering
  \includegraphics[height=0.25\columnwidth]{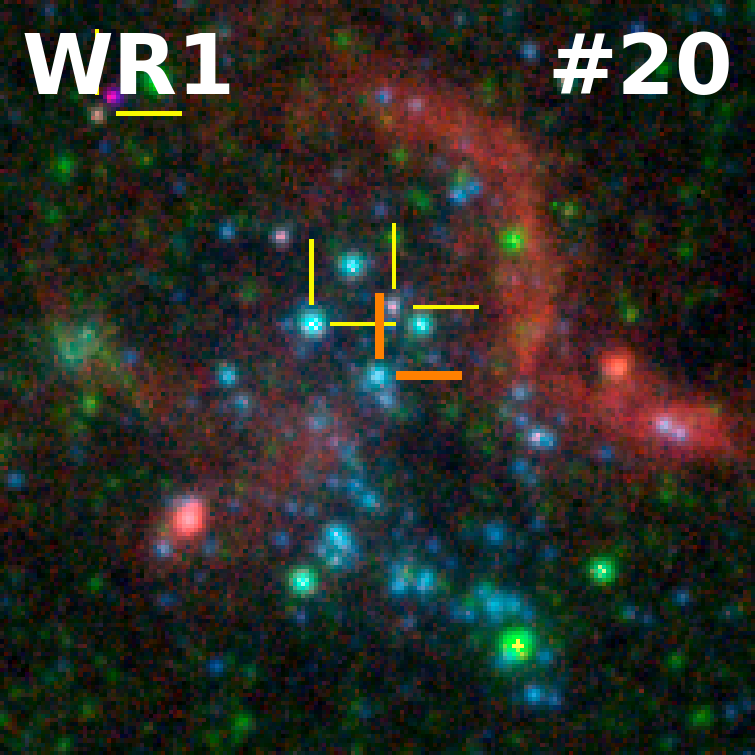}
\end{subfigure}
\begin{subfigure}
  \centering
  \includegraphics[height=0.25\columnwidth]{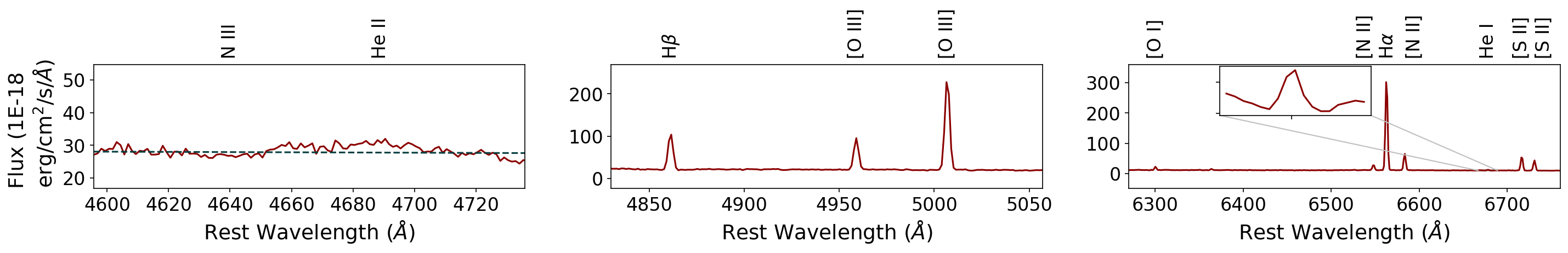}
\end{subfigure}\\
\begin{subfigure}
  \centering
  \includegraphics[height=0.25\columnwidth]{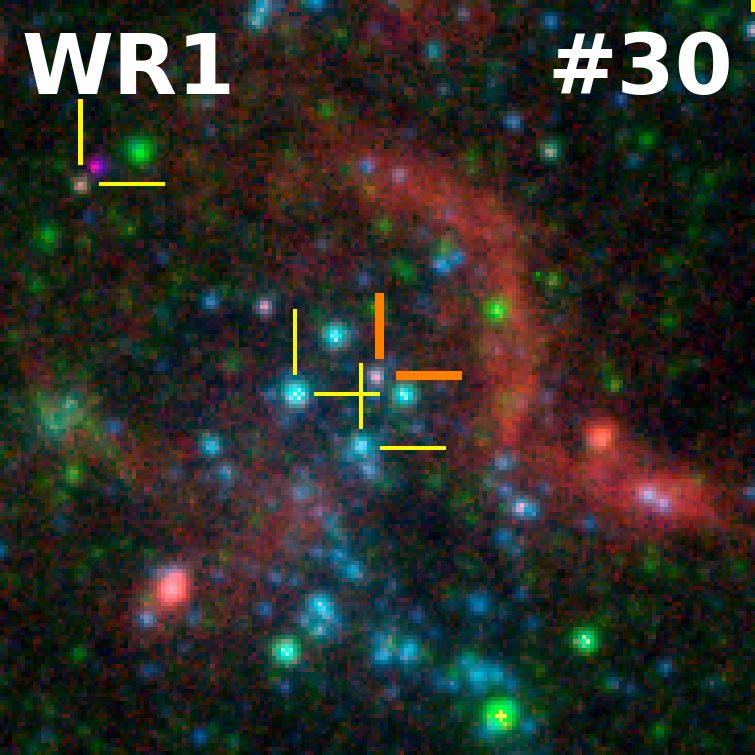}
\end{subfigure}
\begin{subfigure}
  \centering
  \includegraphics[height=0.25\columnwidth]{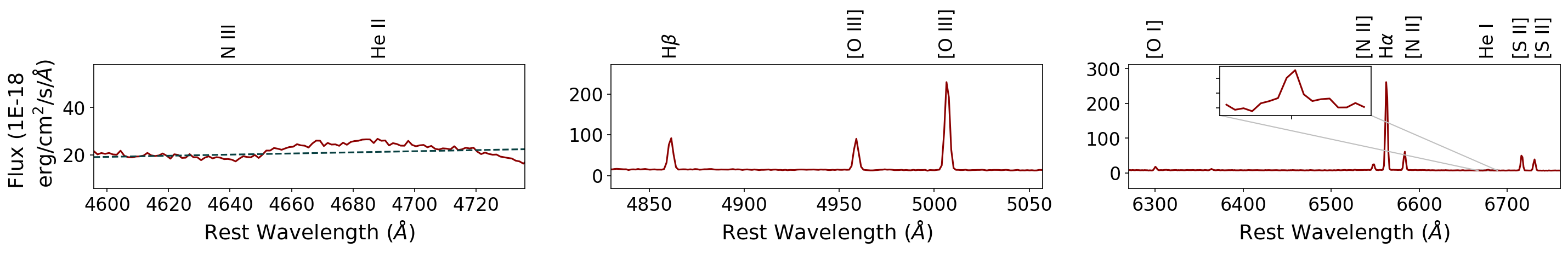}
\end{subfigure}\\
\begin{subfigure}
  \centering
  \includegraphics[height=0.25\columnwidth]{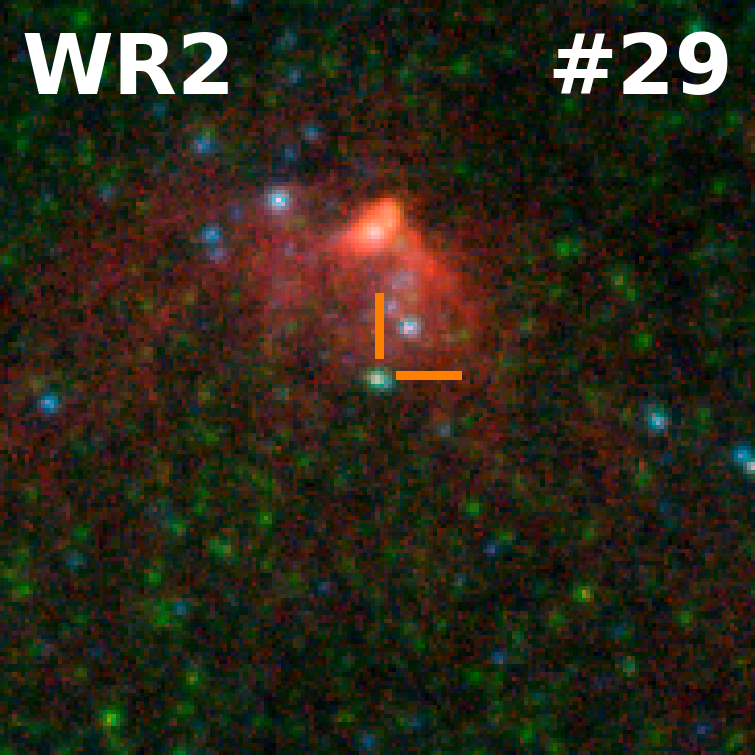}
\end{subfigure}
\begin{subfigure}
  \centering
  \includegraphics[height=0.25\columnwidth]{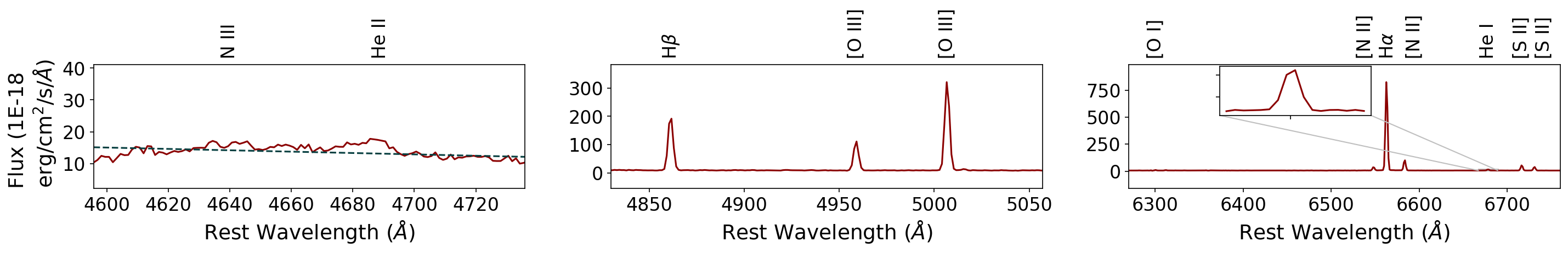}
\end{subfigure}\\
\begin{subfigure}
  \centering
  \includegraphics[height=0.25\columnwidth]{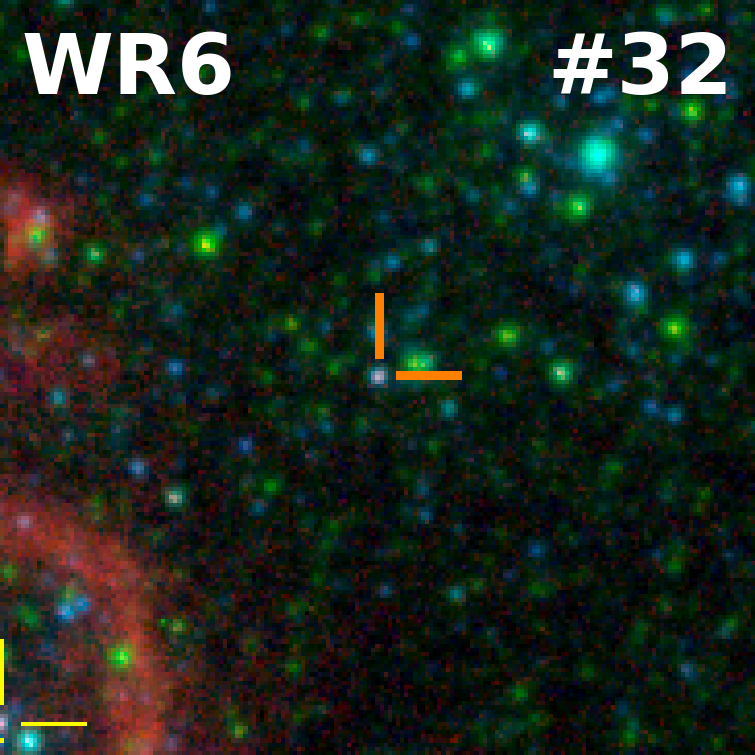}
\end{subfigure}
\begin{subfigure}
  \centering
  \includegraphics[height=0.25\columnwidth]{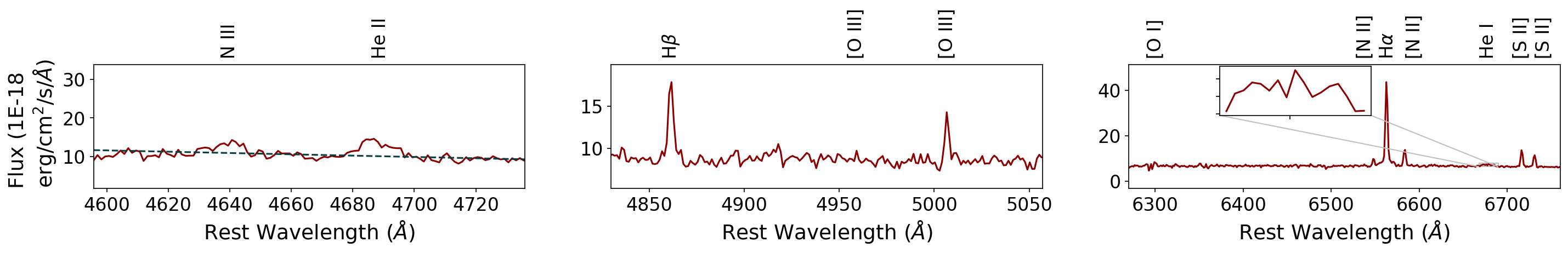}\\
\end{subfigure}
\caption{Same as Figures~\ref{fig:batch1}, ~\ref{fig:batch2}, and ~\ref{fig:batch3}, except for a different set of MUSE cLBVs. The last seven rows show the MUSE cLBVs with detections of the N III$\,\lambda4640$ and/or He II$\,\lambda4686$ bumps.}
\label{fig:batch4}
\end{figure*}

\begin{figure}
\begin{subfigure}
	\centering	
\includegraphics[height=0.49 \columnwidth]{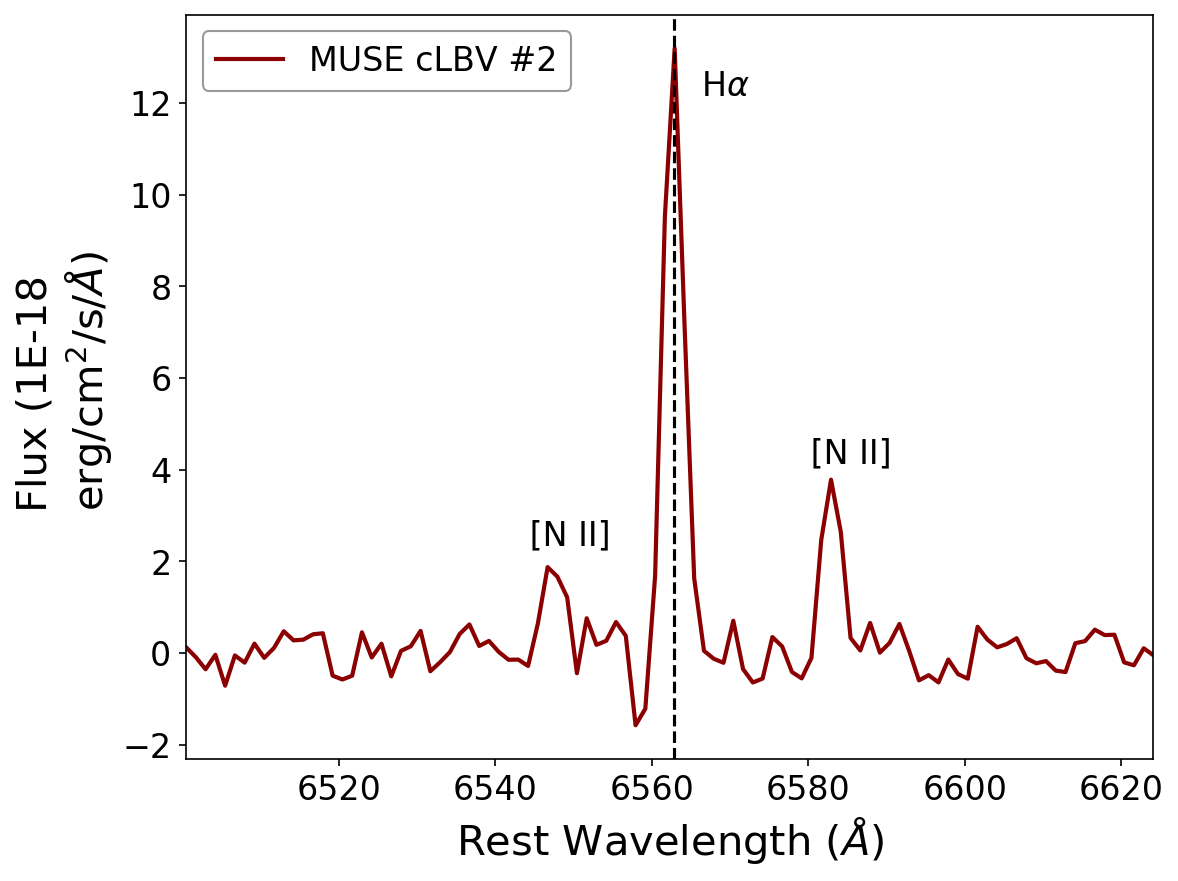}
\end{subfigure}
\caption{H$\alpha$ line with P-Cygni like profile of candidate \#2.}
\label{fig:PCygni2}
\end{figure}

Classic W-R stars are the evolved descendants of massive stars with $\geq25\,M_\odot$ which have completely lost their outer hydrogen, are fusing helium in the core, and are characterized by dense, He-enriched winds, which show prominent broad emission lines of ionised helium and highly ionised nitrogen or carbon. The last seven rows of Figure~\ref{fig:batch4} (rows with IDs in white) show the MUSE cLBVs with broad He\2$\,\lambda$4686 emission detections. Note that candidates \#4 and \#18 are too close to establish which of them is producing the He\2 and N\3 emission bumps which are observed in their spectra. Candidates  \#5, \#10, \#20, and \#30 are also very close to each other. Their spectra, are characterized by an unusually-broad He\2 emission profile. This unusual profile could be explained by mutual contamination of these sources, if several of them emit in He\2. Candidates \#29 and \#32 also show He\2 and N\3 emission bumps. For each MUSE cLBV, column 4 of Table~\ref{tab:results2} says which cLBVs show broad He\2$\,\lambda$4686 emission. In column 4 of Table~\ref{tab:results2} we identify the cLBVs with the unusually broad He\2 profile with an asterisk.

Della Bruna et al. independently identified candidate W-R stars in NGC 7793W based on a MUSE He\2 surface brightness map. We use the coordinates provided by the latter authors to check if we have candidate W-R stars in common. The candidate W-R stars which we have in common are identified with a dagger in column 4 of Table~\ref{tab:results2}. In Figure~\ref{fig:batch4}, the upper-left corner gives the ID of the candidate W-R star given by Della Bruna et al.. Because the spatial resolution of the \hst~images is greater than that of the MUSE data, we find cases where several of our MUSE cLBVs are in the region where Della Bruna et al. found a candidate W-R star based on the MUSE data.

\begin{figure*}
\centering 
\includegraphics[height=0.25\columnwidth]{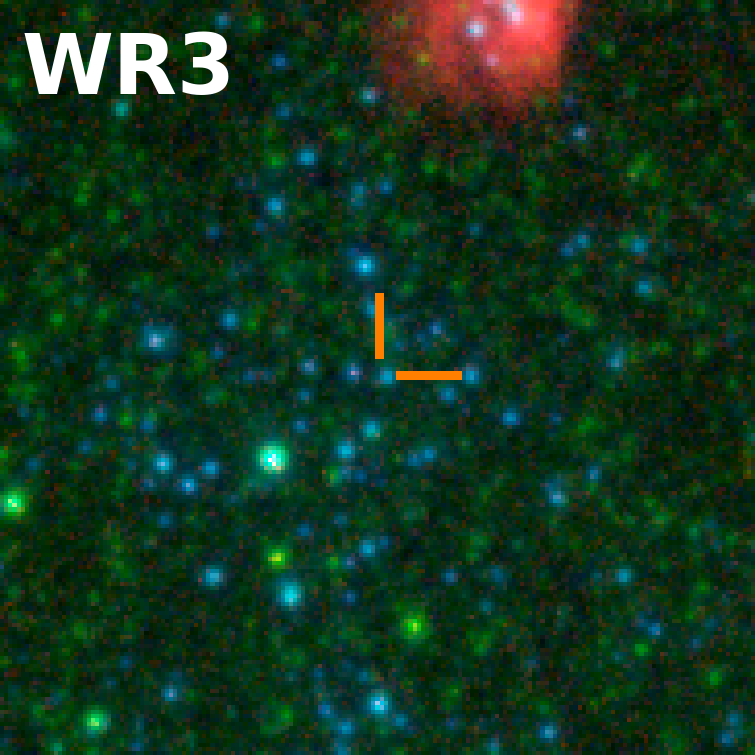}\includegraphics[height=0.25\columnwidth]{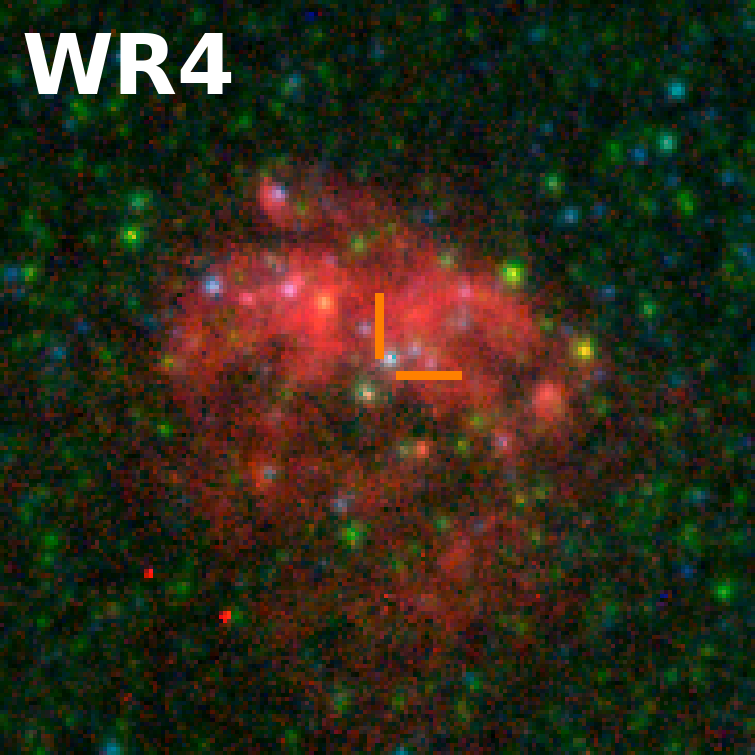}\includegraphics[height=0.25\columnwidth]{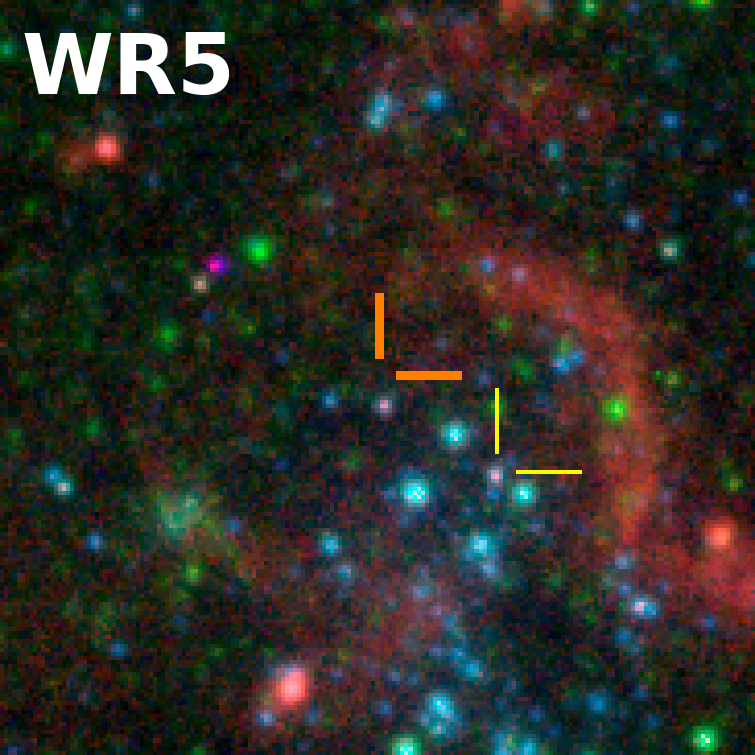}\includegraphics[height=0.25\columnwidth]{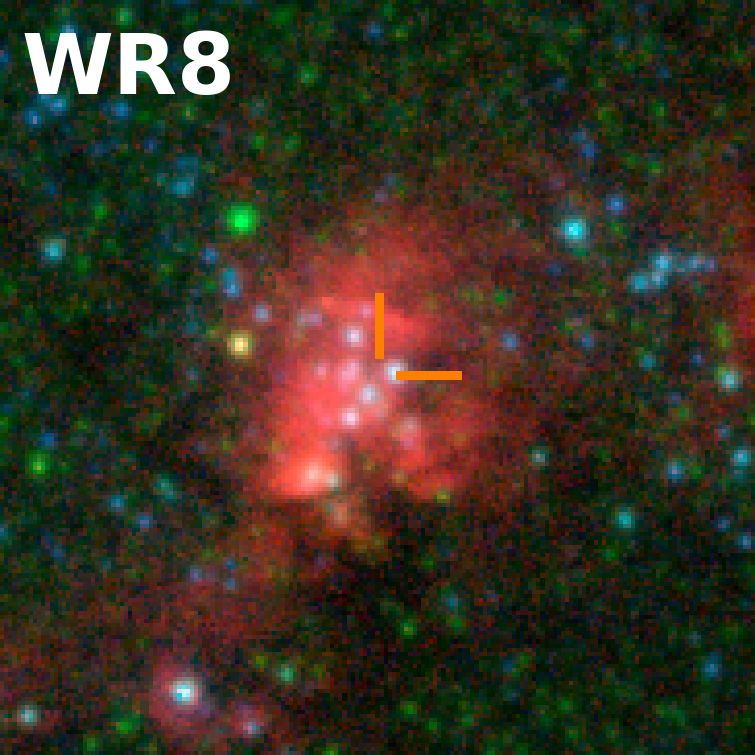}\includegraphics[height=0.25\columnwidth]{figures/ngc7793w_wr8}\includegraphics[height=0.25\columnwidth]{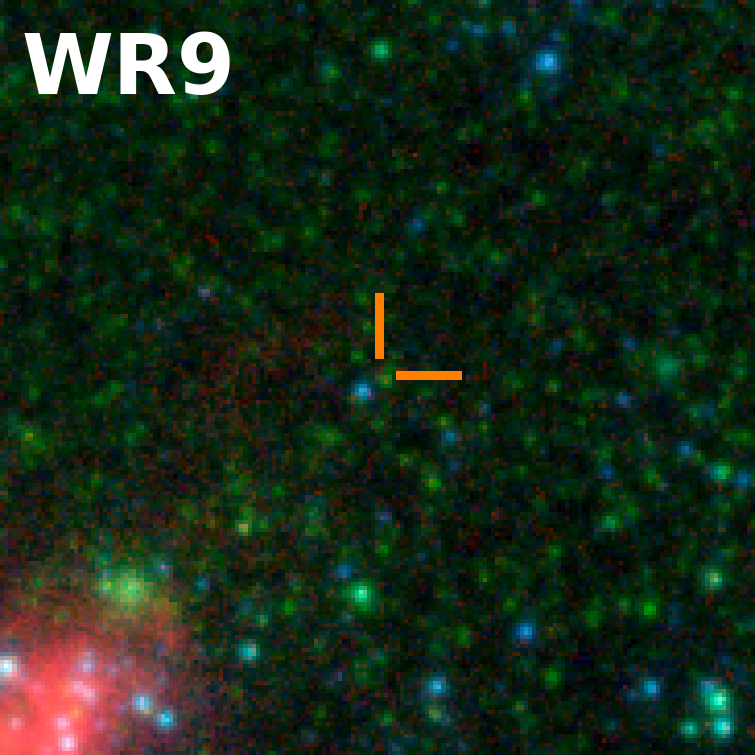}
\caption{LEGUS/H$\alpha$-LEGUS postage stamps of the candidate W-R stars which we do not classify as photometric cLBVs and which were found by Della Bruna et al.. The images are centred on the MUSE He\2 centroid (orange tick marks, see text for explanation of the slight offset between the MUSE and~\hst coordinates. The spectra of these objects are presented in Della Bruna et al..}
\label{fig:undetectedWRs}
\end{figure*}

We noticed that Della Bruna et al. found five candidate W-R stars which are not part of our MUSE cLBV sample. Figure~\ref{fig:undetectedWRs} shows postage stamps of the latter candidates. The postage stamps are centered on the MUSE coordinates provided by Della Bruna et al. (orange markers), which correspond to the MUSE He\2 centroid. The slight offset between the orange markers and point \hst~sources in the figure can be explained by the pixel scale of the MUSE data, which is 0.2”. This corresponds to five \hst~pixels. Thus, the offset is within a distance of five random \hst~pixels from the \hst~centering. In four cases, we did not select the source as a photometric cLBV because the source failed to pass our roundness or sharpness criteria. However, in the case of WR5, we did not select it because after subtracting the continuum from the H$\alpha$ filter, the source did not pass our H$\alpha$ luminosity criterion.


\begin{table*}
\centering
\caption{Spectroscopic properties of MUSE cLBVs.}
\label{tab:results2}
\begin{tabular}{lllllccc}
\hline															
ID	&	FWHM(H$\alpha$)$^{\rm{a}}$	&	He\1$^{\rm{b}}$	&	He\2$^{\rm{b}}$	&	[O\1]$^{\rm{b}}$	&	$\frac{\rm{ F([OIII}]5007)}{\rm{F(H} \beta)}^{\rm{c}}$	&	$\frac{\rm{ F([NII}]6584)}{\rm{ F(H} \alpha)}^{\rm{c}}$	&	$\frac{\rm{ F([SII}]6716)}{\rm{ F([SII} ]6731)}^{\rm{c}}$	\\
(1) & (2) & (3) & (4) & (5) & (6) & (7) & (8) \\
\hline
1	&	120	$	\pm	$	15	&	0	&	0	&	0	&	H$\beta$ in absorption, No [O\3] emission	&	0.56	&	1.22	\\
2	&	132	$	\pm	$	5	&	0	&	0	&	0	&	H$\beta$ in absorption	&	0.29	&	1.11	\\
3	&	130	$	\pm	$	1	&	0	&	0	&	0	&	No [O\3] emission	&	0.15	&	1.40	\\
4	&	126	$	\pm	$	1	&	1	&	1	&	1	&	0.40	&	0.17	&	1.43	\\
5	&	126	$	\pm	$	1	&	1	&	1*	&	1	&	2.46	&	0.19	&	1.39	\\
6	&	122	$	\pm	$	1	&	0	&	0	&	0	&	0.37	&	0.24	&	1.38	\\
7	&	133	$	\pm	$	1	&	0	&	0	&	1	& No [O\3] emission	&	0.19	&	1.41	\\
8	&	122	$	\pm	$	1	&	0	&	0	&	1	&	0.19	&	0.20	&	1.41	\\
9	&	128	$	\pm	$	1	&	1	&	0	&	1	&	0.81	&	0.19	&	1.43	\\
10	&	135	$	\pm	$	1	&	0	&	0	&	1	&	2.81	&	0.38	&	1.36	\\
11	&	126	$	\pm	$	1	&	1	&	0	&	1	&	0.41	&	0.20	&	1.49	\\
12	&	120	$	\pm	$	1	&	1	&	0	&	1	&	0.62	&	0.19	&	1.44	\\
13	&	129	$	\pm	$	1	&	1	&	0	&	1	&	1.17	&	0.17	&	1.43	\\
14	&	133	$	\pm	$	12	&	0	&	0	&	0	&	H$\beta$ in absorption, No [O\3] emission	&	0.86	&	0.96	\\
15	&	129	$	\pm	$	2	&	0	&	0	&	0	&	0.94	&	0.26	&	1.40	\\
16	&	138	$	\pm	$	9	&	1	&	1*	&	0	&	0.58	&	0.12	&	1.08	\\
17	&	128	$	\pm	$	2	&	0	&	0	&	0	&	1.96	&	0.37	&	1.32	\\
18	&	124	$	\pm	$	1	&	1	&	1$^{\dagger}$	&	1	&	0.39	&	0.18	&	1.45	\\
19	&	125	$	\pm	$	1	&	1	&	0	&	1	&	0.69	&	0.23	&	1.46	\\
20	&	126	$	\pm	$	1	&	1	&	1	&	1	&	2.57	&	0.17	&	1.39	\\
21	&	119	$	\pm	$	7	&	0	&	0	&	0	&	0.59	&	0.36	&	1.31	\\
22	&	119	$	\pm	$	5	&	0	&	1*$^{\dagger}$	&	0	&	No [O\3] emission	&	0.38	&	1.05	\\
23	&	123	$	\pm	$	1	&	1	&	0	&	1	&	1.43	&	0.14	&	1.45	\\
24	&	128	$	\pm	$	2	&	0	&	0	&	0	&	0.94	&	0.25	&	1.20	\\
25	&	141	$	\pm	$	4	&	0	&	0	&	1	&	No [O\3]  emission	&	0.26	&	1.21	\\
26	&	123	$	\pm	$	2	&	0	&	0	&	1	&	0.86	&	0.29	&	1.56	\\
27	&	155	$	\pm	$	7	&	0	&	0	&	0	&	0.34	&	0.33	&	0.98	\\
28	&	135	$	\pm	$	4	&	0	&	0	&	0	&	2.33	&	0.34	&	1.45	\\
29	&	124	$	\pm	$	1	&	1	&	1$^{\dagger}$	&	1	&	1.59	&	0.11	&	1.47	\\
30	&	128	$	\pm	$	1	&	1	&	1*$^{\dagger}$	&	1	&	2.80	&	0.19	&	1.40	\\
31	&	128	$	\pm	$	12	&	0	&	0	&	0	&	H$\beta$ in absorption, No [O\3] emission	&	0.29	&	1.36	\\
32	&	127	$	\pm	$	2	&	0	&	1	&	1	&	0.46	&	0.19	&	1.32	\\
33	&	155	$	\pm	$	8	&	0	&	0	&	1	&	No [O\3] emission	&	0.27	&	1.27	\\
34	&	126	$	\pm	$	1	&	0	&	0	&	1	&	0.17	&	0.21	&	1.37	\\
35	&	121	$	\pm	$	1	&	1	&	0	&	1	&	0.45	&	0.20	&	1.46	\\
36	&	131	$	\pm	$	1	&	1	&	0	&	1	&	1.90	&	0.19	&	1.42	\\
min	&	119					&		&		&		&	0.17	&	0.11	&	0.96	\\
max	&	155					&		&		&		& 2.81	&	0.86	&	1.56	\\
mean	&	129					&		&		&		&	1.12	&	0.25	&	1.34	\\
median	&	127					&		&		&		&	0.81	&	0.20	&	1.40	\\
$\sigma$	&						&		&		&		&	0.85	& 0.13  & 	0.15	\\
\hline                 
\end{tabular}
\begin{tablenotes}
      \item \textbf{Notes}
      \item $^{\rm{a}}$ Full width at half maximum of H$\alpha$ line in km\,s$^{-1}$ measured from MUSE spectrum after correction for extinction in the Milky Way. When an underlying stellar absorption component is present, we fit that component. When a broader emission component is present, we also fit that component. We give the FWHM corresponding to the narrow emission component to H$\alpha$. Weaker components to the line may include underlying stellar absorption and/or a broader emission component. 
      \item $^{\rm{b}}$ Presence of He\1$\,\lambda$6678, broad He\2$\,\lambda$4686, and [O\1]$\,\lambda$6300 emission lines in the MUSE cLBV spectra. 1=the line was detected. 0=the line was undetected. Asterisk in column 4=extremely broad He\2$\,\lambda$4686 detection. $\dagger$ in column 4=cLBV identified as candidate W-R candidate in Della Bruna et al. (subm.).
      \item $^{\rm{c}}$ Emission line flux ratio or comment about the ratio.
    \end{tablenotes}
\end{table*}


\section{Discussion}\label{sec:discussion}

\subsection{Isolation of cLBVs}\label{sec:formation}

\cite{Smith2019} argued that LBVs are isolated from locations of star formation. As explained in the introduction, this would favor a binary formation scenario for LBVs. Table~\ref{tab:results} shows that our five strongest cLBVs are at least 7 pc away in projection on the sky from young star clusters; and that three of them are outside of large H\2 region complexes. Table~\ref{tab:results} also shows that several candidate O-type stars are located within 7 pc of our top five cLBVs. Thus, at this point, we cannot conclude if our promising cLBVs are isolated from bona-fide O-type stars or not. Follow-up of these not yet confirmed LBVs could help determine the properties and fraction of LBVs that follow the binary channel.

\subsection{cLBVs in other LEGUS galaxies}

Two other LEGUS galaxies are sufficiently close to search for photometric cLBVs using the method presented in this work, NGC 1313 (4.39 Mpc, \citealt{Calzetti2015}) and NGC 4395(4.3 Mpc, \citealt{Calzetti2015}). For NGC~1313 (southern hemisphere), we already secured optical spectroscopy with Gemini Multi-Object Spectrograph South. We had two masks designed for the NGC1313 field. The first mask got a total of 26 sources in the slits and the second one 19 sources. Both of them were observed in good conditions on Dec 1 and 12, 2018 (first mask) and Jan 1 and 4, 2019 (second one). We used a slit width of 0.75”. The analysis is underway. For NGC~4395 (northern hemisphere), we have failed so far to obtain observations with the OSIRIS MOS on Gran Telescopio Canarias. Since the 0.8" slit is larger than the \hst spatial resolution (0.1 arcsec), contamination from nearby sources in the slit would need to be considered, as we did in the present work.


\section{Summary and conclusions}\label{sec:conclusions}

Bona-fide LBVs are extraordinarily rare, with only $\sim$44 known in 8 galaxies covering a wide range of metallicities and morphologies, including 19 in the Milky Way. Candidate LBVs are also rare, with about 124 known in 12 galaxies. The confirmed LBVs are incompletely characterized, because they require multi-epoch high-spatial resolution imaging and spectroscopy, spanning several decades. This is why searching for cLBVs in new galaxies with good quality data, such as the three nearest galaxies in  LEGUS, is so important. Such observations are seminal in order to establish the maximum luminosity of LBVs and the duration of the LBV phase.

We combine WFC3 images of NGC 7793W in the F547M, F657N and F814W filters, with restrictions of roundness, sharpness and H$\alpha$ luminosity, to produce a catalogue of potential cLBVs (photometric cLBVs). We focus on the subset of the latter sources which have MUSE spectroscopy, finding five strong cLBV candidates. Our five strongest candidates would benefit from follow up observations in order to be confirmed as bona-fide LBVs. In addition, we find three strong candidate Wolf-Rayet stars. Additional Wolf-Rayet stars have been found in the galaxy by other authors but they do not pass our selection criteria for the photometric cLBVs. At this point, we are unable to favor the binary formation scenario for LBVs. We note that the most luminous photometric cLBVs in NGC 7793 were not observed with MUSE AO and are worth following-up spectroscopically, along with photometric cLBVs in other LEGUS galaxies.  


\section*{Acknowledgements}

Based on observations made with the NASA/ESAHubbleSpace Telescope, obtained at the Space Telescope ScienceInstitute, which is operated by the Association of Universitiesfor Research in Astronomy, under NASA Contract NAS5-26555. These observations are associated with Program 13364. Support for Program 13364 was provided by NASA through a grant from the Space Telescope Science Institute. This research has made use of the NASA/IPAC Extragalactic Database(NED), which is operated by the Jet PropulsionLaboratory, California Institute of Technology, under contractwith the National Aeronautics and Space Administration. A. W. and V. R. acknowleges funding from Programa de Apoyo a Proyectos de Investigaci\'on e Innovaci\'on Tecnol\'ogica (PAPIIT) program IA105018.
SdM acknowledges funding from the European Research Council (ERC) grant no. 715063, and the Netherlands Organisation for Scientific Research (NWO) through a VIDI  grant no. 639.042.728.
D. A. G acknowledges support by the German Aerospace Center (DLR) and the Federal Ministry for Economic Affairs and Energy (BMWi) through program 50OR1801 ``MYSST: Mapping Young Stars in Space and Time". We thank the referee for his careful revision of the manuscript.

\bibliographystyle{mnras}
\bibliography{ref} 


\appendix
\section{\hst~photometry of MUSE cLBVs.}
The apparent Vega magnitudes in the seven filters used in this work are presented in Table~\ref{tab:photometry}, uncorrected for foreground or intrinsic extinction.
\begin{table*}
\centering
\caption{Apparent Vega magnitudes of MUSE cLBVs based on \hst~PSF-photometry.}
\label{tab:photometry}
\begin{tabular}{llllllll}
\hline											
ID & F275W & F336W & F438W & F547M & F555W & F657N & F814W \\ 
1 & 22.291$\pm$0.027 & 20.668$\pm$0.012 & 19.857$\pm$0.004 & 19.233$\pm$0.004 & 19.388$\pm$0.003 & 18.751$\pm$0.007 & 18.744$\pm$0.003 \\ 
2 & 19.686$\pm$0.007 & 19.304$\pm$0.006 & 20.004$\pm$0.004 & 19.974$\pm$0.006 & 19.888$\pm$0.003 & 19.631$\pm$0.010 & 19.738$\pm$0.004 \\ 
3 & 19.015$\pm$0.005 & 18.984$\pm$0.005 & 20.140$\pm$0.004 & 20.089$\pm$0.006 & 20.047$\pm$0.004 & 19.731$\pm$0.011 & 19.816$\pm$0.005 \\ 
4 & 19.073$\pm$0.005 & 19.066$\pm$0.005 & 20.245$\pm$0.005 & 20.116$\pm$0.007 & 20.138$\pm$0.004 & 19.727$\pm$0.011 & 19.875$\pm$0.005 \\ 
5 & 18.396$\pm$0.004 & 18.851$\pm$0.005 & 20.375$\pm$0.005 & 20.577$\pm$0.008 & 20.512$\pm$0.004 & 20.350$\pm$0.015 & 20.598$\pm$0.006 \\ 
6 & 23.652$\pm$0.073 & 22.267$\pm$0.031 & 21.407$\pm$0.011 & 20.608$\pm$0.008 & 20.705$\pm$0.005 & 20.092$\pm$0.013 & 19.873$\pm$0.004 \\ 
7 & 22.315$\pm$0.027 & 21.843$\pm$0.020 & 21.752$\pm$0.010 & 20.796$\pm$0.009 & 20.720$\pm$0.005 & 19.806$\pm$0.011 & 19.707$\pm$0.004 \\ 
8 & 19.325$\pm$0.008 & 19.476$\pm$0.008 & 20.904$\pm$0.009 & 20.898$\pm$0.010 & 20.762$\pm$0.005 & 20.585$\pm$0.018 & 20.667$\pm$0.007 \\ 
9 & 24.511$\pm$0.112 & 22.632$\pm$0.032 & 21.599$\pm$0.009 & 20.719$\pm$0.009 & 20.811$\pm$0.005 & 19.912$\pm$0.013 & 19.697$\pm$0.004 \\ 
10 & 20.831$\pm$0.012 & 20.350$\pm$0.010 & 21.008$\pm$0.008 & 20.819$\pm$0.009 & 20.840$\pm$0.005 & 20.354$\pm$0.015 & 20.469$\pm$0.006 \\ 
11 & 24.015$\pm$0.076 & 22.503$\pm$0.029 & 21.731$\pm$0.010 & 21.073$\pm$0.013 & 20.989$\pm$0.005 & 20.332$\pm$0.019 & 19.971$\pm$0.005 \\ 
12 & 19.233$\pm$0.006 & 19.591$\pm$0.006 & 21.031$\pm$0.007 & 21.118$\pm$0.011 & 21.170$\pm$0.006 & 20.811$\pm$0.020 & 21.117$\pm$0.008 \\ 
13 & 19.330$\pm$0.006 & 19.626$\pm$0.007 & 21.081$\pm$0.007 & 21.149$\pm$0.011 & 21.174$\pm$0.006 & 20.844$\pm$0.022 & 21.162$\pm$0.009 \\ 
14 & 24.814$\pm$0.125 & 22.784$\pm$0.036 & 21.989$\pm$0.011 & 21.177$\pm$0.011 & 21.286$\pm$0.006 & 20.490$\pm$0.016 & 20.260$\pm$0.006 \\ 
15 & 21.967$\pm$0.022 & 21.210$\pm$0.014 & 21.469$\pm$0.008 & 21.353$\pm$0.012 & 21.382$\pm$0.007 & 20.857$\pm$0.019 & 20.846$\pm$0.007 \\ 
16 & 19.551$\pm$0.006 & 19.937$\pm$0.008 & 21.473$\pm$0.008 & 21.599$\pm$0.014 & 21.421$\pm$0.007 & 20.683$\pm$0.017 & 21.372$\pm$0.010 \\ 
17 & 0.000$\pm$0.000 & 22.948$\pm$0.037 & 22.230$\pm$0.013 & 21.507$\pm$0.013 & 21.488$\pm$0.007 & 20.849$\pm$0.019 & 20.691$\pm$0.007 \\ 
18 & 20.296$\pm$0.010 & 20.374$\pm$0.010 & 21.565$\pm$0.009 & 21.496$\pm$0.013 & 21.623$\pm$0.008 & 20.978$\pm$0.022 & 21.090$\pm$0.009 \\ 
19 & 22.092$\pm$0.024 & 21.585$\pm$0.018 & 21.780$\pm$0.010 & 21.607$\pm$0.023 & 21.641$\pm$0.008 & 21.216$\pm$0.029 & 21.351$\pm$0.010 \\ 
20 & 19.526$\pm$0.007 & 20.012$\pm$0.008 & 21.501$\pm$0.009 & 22.005$\pm$0.016 & 21.730$\pm$0.008 & 21.582$\pm$0.029 & 21.849$\pm$0.013 \\ 
21 & 0.000$\pm$0.000 & 0.000$\pm$0.000 & 24.273$\pm$0.040 & 21.811$\pm$0.015 & 0.000$\pm$0.000 & 20.376$\pm$0.014 & 19.918$\pm$0.004 \\ 
22 & 0.000$\pm$0.000 & 0.000$\pm$0.000 & 23.455$\pm$0.024 & 21.678$\pm$0.014 & 21.826$\pm$0.008 & 20.710$\pm$0.017 & 20.351$\pm$0.006 \\ 
23 & 22.914$\pm$0.044 & 22.404$\pm$0.030 & 22.198$\pm$0.013 & 21.880$\pm$0.015 & 21.839$\pm$0.009 & 20.743$\pm$0.019 & 21.327$\pm$0.011 \\ 
24 & 22.938$\pm$0.039 & 23.222$\pm$0.046 & 23.364$\pm$0.024 & 21.858$\pm$0.026 & 21.898$\pm$0.009 & 20.736$\pm$0.021 & 20.222$\pm$0.005 \\ 
25 & 20.599$\pm$0.011 & 20.809$\pm$0.012 & 22.037$\pm$0.011 & 22.191$\pm$0.018 & 22.112$\pm$0.010 & 21.776$\pm$0.032 & 22.075$\pm$0.014 \\ 
26 & 0.000$\pm$0.000 & 23.162$\pm$0.073 & 23.069$\pm$0.025 & 22.338$\pm$0.019 & 0.000$\pm$0.000 & 21.717$\pm$0.030 & 24.071$\pm$0.052 \\ 
27 & 0.000$\pm$0.000 & 0.000$\pm$0.000 & 24.354$\pm$0.045 & 21.821$\pm$0.016 & 22.405$\pm$0.013 & 20.371$\pm$0.015 & 19.972$\pm$0.007 \\ 
28 & 0.000$\pm$0.000 & 25.763$\pm$0.257 & 24.446$\pm$0.043 & 22.451$\pm$0.020 & 22.586$\pm$0.012 & 21.152$\pm$0.022 & 20.602$\pm$0.006 \\ 
29 & 22.174$\pm$0.026 & 21.942$\pm$0.023 & 22.891$\pm$0.019 & 22.609$\pm$0.024 & 22.613$\pm$0.013 & 21.847$\pm$0.036 & 21.994$\pm$0.014 \\ 
30 & 20.906$\pm$0.014 & 21.098$\pm$0.014 & 22.723$\pm$0.017 & 22.926$\pm$0.028 & 22.805$\pm$0.014 & 21.912$\pm$0.036 & 22.710$\pm$0.021 \\ 
31 & 21.141$\pm$0.015 & 21.475$\pm$0.020 & 22.933$\pm$0.020 & 23.093$\pm$0.030 & 22.987$\pm$0.016 & 22.447$\pm$0.048 & 22.899$\pm$0.024 \\ 
32 & 21.314$\pm$0.015 & 21.570$\pm$0.017 & 22.957$\pm$0.018 & 23.146$\pm$0.032 & 23.055$\pm$0.016 & 22.147$\pm$0.040 & 22.812$\pm$0.022 \\ 
33 & 21.629$\pm$0.018 & 21.910$\pm$0.021 & 23.456$\pm$0.024 & 23.754$\pm$0.043 & 23.250$\pm$0.018 & 22.767$\pm$0.061 & 23.124$\pm$0.026 \\ 
34 & 21.616$\pm$0.018 & 21.934$\pm$0.021 & 23.398$\pm$0.024 & 23.533$\pm$0.038 & 23.478$\pm$0.021 & 22.647$\pm$0.054 & 23.219$\pm$0.029 \\ 
35 & 21.475$\pm$0.017 & 21.997$\pm$0.022 & 23.521$\pm$0.026 & 23.741$\pm$0.044 & 23.669$\pm$0.024 & 22.806$\pm$0.062 & 23.805$\pm$0.043 \\ 
36 & 23.451$\pm$0.051 & 23.102$\pm$0.042 & 23.924$\pm$0.033 & 23.441$\pm$0.037 & 24.262$\pm$0.033 & 22.194$\pm$0.041 & 22.892$\pm$0.023 \\ 
\hline
\end{tabular}
\end{table*}


\bsp	
\label{lastpage}
\end{document}